\documentclass{aa}  

\usepackage{graphicx}
\usepackage{txfonts}
\usepackage{lipsum}
\usepackage{subcaption}  
\usepackage{hyperref}
\usepackage{color}
\usepackage{lscape}             
\usepackage{placeins}           
\usepackage[dvipsnames]{xcolor}      \definecolor{Peru}{RGB}{205,133,63}
\definecolor{OliveDrab}{RGB}{107,142,35}  
\definecolor{DarkCyan}{RGB}{0,139,139}

\usepackage{multirow,multicol}

\usepackage{soul}
\usepackage{cancel}
\usepackage[normalem]{ulem}
\usepackage[dvipsnames]{xcolor}

\newcommand\bb[1]{\boldsymbol{#1}}

\begin{document}

   \title{Sub-sonic compressible magnetohydrodynamic turbulence}

   \subtitle{I. Alfvénic and fast-magnetosonic injection, amplitude dependence, and compressibility effects}

%
%
%

   \author{E. Puzzoni\inst{1}\corrauth{Eleonora.Puzzoni@oca.eu}        
        \and S. S. Cerri \inst{1}\email{silvio.cerri@oca.eu} 
        \and D. Laveder \inst{1}\email{dimitri.laveder@oca.eu}
        \and T. Passot \inst{1}\email{thierry.passot@oca.eu}
        \and P.-L. Sulem \inst{1}\email{plsulem@oca.eu}
        }

   \institute{Université Côte d'Azur, Observatoire de la Côte d'Azur, CNRS, Laboratoire Lagrange, France}

   \date{\empty}

 
  \abstract
   {The dependence of magnetohydrodynamic (MHD) turbulence on key physical parameters at injection scales remains an open question. 
   This has relevant implications in various astrophysical contexts, in particular for cosmic-ray (CR) transport in the Galaxy.
   }
   {We investigate how properties of sub-sonic compressible MHD turbulence that are relevant for cosmic-ray transport are affected by the nature and amplitude of initial fluctuations, and by the plasma compressibility (via the $\beta$ parameter, i.e., the ratio between the thermal and magnetic pressure). Particular attention is devoted to assess the role of fast-magnetosonic versus Alfv\'enic fluctuations.}
   {We perform 3D simulations of decaying compressible ideal-MHD turbulence at $1024^3$ resolution with the PLUTO code.}
   {The level of density fluctuations $\delta\rho_{\rm rms}/\rho_0$ in fully developed turbulence is insensitive to whether this state is reached starting from Alfvénic or fast-magnetosonic perturbations, being only determined by the initial fluctuation amplitude and plasma $\beta$.  Fast-magnetosonic injection is indeed characterized by an early phase of rapid shock dissipation, followed by a turbulence-dominated decay with a rate comparable to that of the Alfvénic case. In addition, the contribution of fast-magnetosonic fluctuations to the total energy budget in fully developed turbulence remains relevant ($\sim30$-$50\%$) only when the initial injection consists exclusively of fast modes, while becoming negligible ($\lesssim 10\%$) for other types of injection. Large-amplitude turbulence ($\delta B/B_0>1$) is characterized by a nearly isotropic Kolmogorov ($\propto k^{-5/3}$) or Iroshnikov-Kraichnan ($\propto k^{-3/2}$) spectrum for Alfv\'enic or fast-magnetosonic injection, respectively. However, Alfv\'enic turbulence with $\delta B/B_0\sim 1$ already develops a small degree of anisotropy with $\propto k_\perp^{-3/2}$ and $\propto k_z^{-5/3}$ power-law spectra. At low amplitudes ($\delta B/B_0\ll1$), both initial Alfv\'enic and mixed-wave perturbations lead to strongly anisotropic turbulence with a perpendicular spectrum $\propto k_\perp^{-5/3}$ and a parallel spectrum roughly $\propto k_z^{-2}$ (becoming significantly steeper at $\beta\gg1$), while initializing only fast-magnetosonic perturbations produces a turbulent state populated by (weak) shocks and characterized by a nearly isotropic $k^{-2}$ spectrum.
   The statistics of magnetic-field curvature $K_\|= |(\hat{\bb{b}}\cdot\bb{\nabla}) \hat{\bb{b}}|$ and of magnetic-mirror structures (quantified through a {\it``mirroring curvature''} $K_\mathrm{M} = |\hat{\bb{b}} \cdot \bb{\nabla} \ln B|$) is strongly sensitive to the initial fluctuation amplitude and to the plasma $\beta$, exhibiting broader distributions and harder (i.e., shallower) power laws as either parameter increases. The predicted $\mathrm{PDF}(K_\|)\propto K_\|^{-2.5}$ power-law scaling emerges only in the large-amplitude regime and at low plasma compressibility (i.e., high $\beta$).}
  {This work highlights that features of sub-sonic compressible MHD turbulence that may affect CR transport are sensitive to large-scale conditions and to the plasma $\beta$. Their effect on CR diffusion and field-line random walk is the object of Paper II.
  }


   \maketitle
   \nolinenumbers

\section{Introduction}

A wide range of space and astrophysical systems are filled with turbulent plasmas, from the solar wind to accretion disks around compact objects, from the interstellar medium (ISM) of our Galaxy to the intracluster medium (ICM) of galaxy clusters~\citep[e.g.,][]{QuataertGurzinov1999,SchekochihinCowley2006,BrunoCarbone2013,Ferriere2020}.
In particular, turbulence in the ISM and in the Galactic halo has relevant implications for several processes, among which the transport of cosmic rays (CRs) remains an unsolved puzzle more than a century after the discovery of these cosmic particles~\citep[e.g.,][]{Gabici2019,Hopkins2022,Kempski2022}.
The prolonged confinement of CRs in the Milky Way~\citep[$\sim 10$ Myr, see, e.g.,][]{Simpson1988} is due to scattering onto small-scale magnetic fluctuations in the Galactic environment~\citep{Berezinskii1990}. Interstellar turbulence~\citep[][]{Chepurnov2010} is a prime candidate for driving such fluctuations: understanding its properties, such as the spectrum, anisotropy, and compressibility, is thus of central interest. However, the nature and properties of fluctuations and their cascade in MHD turbulence still remain an open question.  

In the incompressible limit, MHD turbulence is dominated by anisotropic Alfvénic fluctuations~\citep{Schekochihin2022}. In the theory by \citet{Goldreich1995}, turbulent eddies (whose perpendicular spectrum scales as $k_\perp^{-5/3}$) are elongated in the direction of the mean magnetic field, with an aspect ratio prescribed by the critical balance condition $\ell_\parallel/\ell_\perp \sim v_A / \delta v_{\ell_\perp}\gg 1$ (where $\delta v_{\ell_\perp}$ is the fluctuation amplitude at scale $\ell_\perp$), yelding an anisotropy $k_\parallel\propto k_\perp^{2/3}$ (corresponding to a spectrum $k_\parallel^{-2}$ along the mean field).
Recent models of Alfv\'enic turbulence also assumed scale-dependent alignment of fluctuations~\citep{Boldyrev2006}, so that the perpendicular spectrum becomes $\propto k_\perp^{-3/2}$ and $k_\parallel\propto k_\perp^{1/2}$.
In general, anisotropy is a fundamental feature of incompressible Alfvénic turbulence, which, among other things, can significantly reduce the efficiency of CR scattering~\citep{Chandran2000}.

Compressible MHD turbulence, however, also includes fast- and slow-magnetosonic fluctuations.
Early numerical studies suggested that fast-magnetosonic turbulence is isotropic with a $\propto k^{-3/2}$ spectrum~\citep{Cho2003}, consistent with weak-turbulence theory~\citep{Zakharov1970,Galtier2023}.
On the other hand, fast-magnetosonic modes tend to form shocks \citep{Suzuki2007}, which may in turn suppress the energy cascade \citep[see, e.g.,][]{Hou2025}. In fact, recent simulations~\citep{Makwana2020} show that $k^{-2}$ spectra can also emerge in fast-magnetosnic turbulence, likely due to weak shocks~\citep{KadomtsevPetviashvili1973}.
Therefore, the development of a cascade of fast-magnetosonic fluctuations can certainly be complicated by a strong damping~\citep[e.g.,][]{GinzburgSyrovatskii1964,Braginskii1965,FooteKulsrud1979}, which limits the extent of the inertial range depending on the plasma environment~\citep[e.g.,][and references therein]{YanLazarian2008,Fornieri2021}, and/or by nonlinear steepening effects developing at a rate $\tau_{\rm steep}^{-1} \sim \delta v_\ell/\ell$~\citep{LandauLifshitz1987} that is faster than the associated (weak) cascade rate $\tau_{\rm casc}^{-1}\sim (\delta v_\ell/\ell)(\delta v_\ell/v_{\rm F})$ with $\delta v_\ell/v_{\rm F}\ll1$ ($v_{\rm F}$ is the fast-magnetosonic speed).

Furthermore, turbulence naturally produces field-line curvature and magnetic mirrors~\citep[e.g.,][]{Yang2019,Bandyopadhyay2020,Yuen2020,LazarianXu2021,Lemoine2023,Kempski2023,ZhangXu2023,Reichherzer2025}. These intermittent, localized bends and magnetic-field magnitude variations represent a feature not captured by the power spectrum alone. 
CR scattering by field-line curvature is an important issue that to date has been explored mainly in the incompressible regime and in the large-amplitude-fluctuation limit ($\delta B/B_0\gg1$) or with $B_0=0$~\citep[][]{Lemoine2023,Kempski2023,Lubke2025}.

In this Paper I, we focus on how properties of compressible MHD turbulence depend on the nature and amplitude of injected fluctuations, and on the plasma compressibility. In particular, we focus on those properties that are considered to be most relevant for cosmic-ray transport, including the spectral properties and the statistics of magnetic-field curvature and magnetic mirrors. This is a necessary preliminary step, as in Paper II we investigate how these features may reflect on the associated CR diffusion and magnetic-field line random walk.

This paper is structured as follows. Section \ref{sec:model} describes the model and numerical setup. Section \ref{sec:results_overview} presents the temporal evolution of the simulations and an overview of the fully developed turbulent state, including statistics of the magnetic-field curvature and magnetic-mirror scales. Section \ref{sec:results_spectra} discusses spectral properties and nature of magnetic-field fluctuations. Additional discussion on mode versus Helmholtz decomposition and frequency--wavevector analysis reported in Appendix~\ref{app:modes_Helm} and Appendix~\ref{app:omegak}, respectively, while Appendix \ref{app:chi} presents a scale-dependent analysis of the turbulence nonlinear parameter. Section \ref{sec:conclusions} summarizes the main findings. 



\begin{table*}[tbh!]
\caption{Summary of relevant simulation parameters. Initial values are denoted by a ``$0$'' subscript, while values at peak activity are denoted by a $^*$ superscript. Injection wavevectors are normalized to $k_0=2\pi/L$, i.e., $\tilde{\bb{k}}_{\rm inj}=\bb{k}_{\rm inj}/k_0$, and $\tilde{\ell}_\mathrm{int}=\ell_\mathrm{int}/L$ is the turbulence integral length $\ell_\mathrm{int}$ normalized to the box size $L$. The root-mean-square fluctuations' amplitude normalized to $B_0$ is denoted with $\delta \tilde{B}=\delta B_{\rm rms}/B_0$, while $M_{\rm A}=\delta v_{\rm rms}/v_{\rm A}$ and $M_{\rm s}=\delta v_{\rm rms}/c_{\rm s}$ are the turbulent Alfv\'enic and sonic Mach numbers, respectively.}              
\label{tab:ic}      
\centering                                      
\begin{tabular}{c c c c c c c c c c c c c c c c}          
\hline
\multirow{2}{*}{initial condition} & \multirow{2}{*}{$\beta$} &  \multicolumn{4}{c}{turbulence injection parameters} & &\multicolumn{4}{c}{values at peak activity} & \multirow{2}{*}{\shortstack{color\\code}} \\    
& & \,\,\,\,$|\tilde{\bb{k}}_{\rm inj}|$ & $\delta\tilde{B}_0$ & $M_{\rm A,0}$ & $M_{\rm s,0}$ & & $\tilde{\ell}_\mathrm{int}^*$ & $\delta\tilde{B}^*$ & $M_{\rm A}^*$ & $M_{\rm s}^*$ \\    
\hline                                   
    Alfv\'en waves (AW) & $1$ & \,\,\,\,$\leq4$ & $0.33$ & $0.3$ & $0.5$ & & 0.17 & 0.3 & 0.3 & 0.4 & $\color{red}\blacksquare$  \\
    fast-magnetosonic waves (FW) & $1$ & \,\,\,\,$\leq4$ & $0.25$ & $0.3$ & $0.5$ & & 0.16 & 0.1 & 0.1 & 0.2 & $\color{black}\blacksquare$\\
    Alfv\'en + fast-magnetosonic waves (MW) & $1$ & \,\,\,\,$\leq4$ & $0.33$ & $0.4$ & $0.6$ & & 0.18 & 0.2 & 0.2 & 0.3 & $\color{brown}\blacksquare$ \\
    Alfv\'en waves (AW) & $200$ & \,\,\,\,$\leq4$ & $0.33$  & $0.3$ & $0.03$ & & 0.17 & 0.3 & 0.3 & 0.03 & $\color{orange}\blacksquare$  \\
    Alfv\'en waves (AW) & $2$ & \,\,\,\,$\leq4$ & $1$  & $1$ & $1$ & & 0.13 & 0.8 & 0.8 & 0.8 & $\color{magenta}\blacksquare$  \\
    Alfv\'en waves (AW) & $200$ & \,\,\,\,$\leq4$ & $1$ & $1$ & $0.1$ & & 0.13 & 0.9 & 0.8 & 0.08 & $\color{OliveDrab}\blacksquare$  \\
    Alfv\'en waves (AW) & $200$ & \,\,\,\,$\leq4$ & $5$ & $5$ & $0.5$ & & 0.20 & 2.3 & 1.3 & 0.13 & $\color{DarkCyan}\blacksquare$   \\
    fast-magnetosonic waves (FW) & $200$ & \,\,\,\,$\leq4$ & $0.14$ & $17$ & $1.7$ & & 0.21 & 1.4 & 1.6 & 0.16 & $\color{Peru}\blacksquare$  \\
\hline                                             
\end{tabular}
\end{table*}

\section{Model, numerical method, and simulations setup}
\label{sec:model}

We describe the plasma using the ideal-MHD equations:
\begin{align}
\frac{\partial \rho}{\partial t} + \bb{\nabla} \cdot (\rho \bb{v}) &= 0, \label{eq:continuity}\\
\frac{\partial (\rho \bb{v})}{\partial t} + \bb{\nabla} \cdot \left[ \rho \bb{v}\bb{v} + \left( p + \frac{B^2}{8\pi} \right)\bb{I} - \frac{\bb{B}\bb{B}}{4\pi} \right] &= 0, \label{eq:momentum}\\
\frac{\partial \bb{B}}{\partial t} - \bb{\nabla} \times (\bb{v} \times \bb{B}) &= 0\label{eq:Faraday},
\end{align}
for the mass density $\rho$, the fluid velocity $\bb{v}$, and the magnetic field $\bb{B}$, complemented by the solenoidal constraint $\bb{\nabla} \cdot \bb{B} = 0$
and an isothermal equation of state for the thermal pressure $p = c_{\rm s}^2 \rho$, where $c_{\rm s}^2=k_{\rm B}T_0/m$ is the isothermal sound speed, $k_{\rm B}$ the Boltzmann constant, and $T_0$ the uniform mean plasma temperature.
The relative importance of thermal and magnetic pressure is quantified by the parameter $\beta=8\pi p/B^2$.

Equations~(\ref{eq:continuity})--(\ref{eq:Faraday}) are solved in a triply periodic cube of size $L = 1$ discretized with $1024^3$ uniformly distributed grid points, using the \textsc{PLUTO} code~\citep[][]{Mignone2007, Mignone2012} involving a fifth-order WENO-Z reconstruction algorithm \citep[e.g.,][]{Mignone2010} in combination with the HLLD Riemann solver of \cite{Miyoshi2005} and the UCT-HLLD electromotive force averaging scheme \citep[][]{Mignone2021}; the solenoidal condition $\bb{\nabla}\cdot\bb{B}=0$ is satisfied to machine precision using a constrained-transport method \citep[e.g.,][]{Londrillo2004, Gardiner2005}. 
Time integration employs a third-order Runge-Kutta scheme.
Despite employing the least dissipative algorithms available in PLUTO \citep{Puzzoni2021}, the code does not specifically mitigate the enhanced numerical diffusivity that can arise in strongly subsonic regimes~\citep[see, e.g.,][]{Watt2025}.
Simulation outputs are processed using PyPLUTO \citep{Mattia2025}.

We consider a homogeneous plasma with background density $\rho_0=1$ embedded in a uniform magnetic field $\bb{B}_0=B_0\hat{\bb{e}}_z$ with $B_0=1$ (so that the Alfvén speed is $v_{\rm A} = 1$ in code units).
Freely-decaying turbulence is initialized as a superposition of forward- and backward-propagating waves (with respect to $\bb{B}_0$) having wavevectors $\bb{k} = (k_x, k_y, k_z)  =(i, j, l)\,k_0$, with $k_0 = 2 \pi/L$ and non-zero integers $i, j, l \in [-4, 4]$, and random phases $\varphi_{ijk}^{\pm}$ drawn with uniform probability in $[0, 2\pi]$ for each propagation direction (``$+$'' for forward, ``$-$'' for backward).

The influence of the compressibility of the injection on the turbulent cascade is investigated by considering an initial condition of Alfv\'enic or fast-magnetosonic fluctuations, as well as a mixture of both.
To further highlight the impact of plasma compressibility, values $\beta=1,2,200$ of the beta parameter have been considered, so that turbulence is never in a super-sonic regime (i.e., $0.01\lesssim M_{\rm s}\lesssim 1$) while allowing to explore sub- and super-Alfv\'enic regimes (i.e., $0.1\lesssim M_{\rm A}\lesssim 2$). 
The $\beta\gg1$ regime is thus explored with the sole purpose of maintaining $M_{\rm s}\ll1$ while exploring large-amplitude turbulence: other well-known collisionless high-$\beta$ effects~\citep[e.g.,][and references therein]{Arzamasskiy2023,Majeski2024} are neglected in our MHD model.
The turbulence integral scale is computed as $\ell_\mathrm{int} = 2 \pi \int k^{-1} E_\mathrm{\delta B}(k)dk/\int E_\mathrm{\delta B}(k) dk$, where $E_\mathrm{\delta B}(k)$ is the magnetic power spectrum~\citep[see, e.g.,][]{Shukurov2021}.
A summary of the performed simulations is given in Table \ref{tab:ic}.

\subsection{Alfvénic injection}\label{subsec:AW_init}

Alfvénic fluctuations are initialized using a superposition of arc-polarized
waves, using a generalized version of the one proposed by \citet{DelZanna2001}, for which magnetic-field fluctuation $\delta \bb{B}$ lies in a plane perpendicular to $\bb{k}$ and each wave ensures a constant modulus of the total magnetic field, $|\bb{B}|=|\bb{B}_0+\delta\bb{B}|=\mathrm{const}$, and no density fluctuations, $\delta\rho=0$. The magnetic-field perturbation reads (see Appendix~\ref{app:arc} for the derivation)
\begin{equation}
{\delta \bb{B}} = 
- \delta B_1
\begin{pmatrix}
\cos\theta\,\sin\alpha \\
\cos\theta\,\cos\alpha \\
-\sin\theta
\end{pmatrix}
 + \delta B_2
 \begin{pmatrix}
 \cos\alpha \\
-\sin\alpha \\
0
\end{pmatrix}
\label{eq:deltaB}
\end{equation}
where $\theta = \tan^{-1}\left(k_\perp/k_z\right)$ is the angle between $\bb{k}$ and $\bb{B}_0$, and $\alpha = \tan^{-1}\left(k_x/k_y\right)$ is the angle between $\bb{k}_\perp=\sqrt{k_x^2+k_y^2}$ and $\hat{\bb{e}}_y$. The fluctuation amplitudes are $\delta B_2 = \eta\, B_0\, \cos\phi$, where $\phi=\bb{k}\cdot\bb{x}+\varphi$ ($\varphi$ is a random phase) and $\eta$ is an amplitude parameter, and $\delta B_1 = \sqrt{B_0^2 \sin^2 \theta + \delta B_{\rm rms}^2 -\delta B_2^2} - B_0 \sin \theta$.
The root-mean-square fluctuation $\delta B_{\rm rms}=b_{\rm rms}\,B_0$ is obtained by numerically solving the relation $E(\gamma) = \frac{\pi}{2}\sin \theta\,(\sin^2 \theta + b_{\rm rms}^2)^{-1/2}$, where $E(\gamma) = \int_0^{\pi/2} \sqrt{1 - \gamma^2 \cos ^2\phi}\,{\rm d} \phi$ is an elliptic integral with $\gamma = \eta\, (\sin^2 \theta + b_\mathrm{rms}^2)^{-1/2}$.
The velocity fluctuations are related to magnetic fluctuations by $\delta \bb{v} = \pm\, (\delta \bb{B}/B_0)\,v_{{\rm A},0}$ \citep{Barnes1974},
with the sign determines the propagation direction.
Note that, while a single wave is an exact nonlinear Alfv\'enic solution with $|\bb{B}|={\rm const}$, $\bb{\nabla}\cdot\delta\bb{v}=0$ and $\delta\rho=0$, a superposition of waves with random phases is not: due to their interference, small initial fluctuations $\delta|\bb{B}|$ are present and drive $\delta \rho$ fluctuations which are, however, much smaller than the initialized magnetic and velocity fluctuations, i.e., $\delta\rho_{\rm rms}/\rho_0\ll\delta B_{\rm rms}/B_0\sim\delta v_{\rm rms}/v_{\rm A}$ (see Section~\ref{subsec:time_evo}).
This remains true also with large-amplitude fluctuations, contrary to simpler initialization with linearly polarized shear-Alfv\'en waves, thus motivating our choice.


\subsection{Fast magnetosonic injection}\label{subsec:FW_init}

Fast-magnetosonic fluctuations are initialized following the usual linear approach, e.g., as in \citet{Cho2003}. The velocity perturbation is
    \[
{\delta \bb{v}} = v_0
 \cos(\bb{k}\cdot\bb{x}+\varphi)
\begin{pmatrix}
(1+a+\sqrt{D})\,\sin\theta\,\sin\alpha \\
(1+a+\sqrt{D})\,\sin\theta\,\cos\alpha \\
(-1+a+\sqrt{D})\,\cos\theta
\end{pmatrix}
\]
where $v_0$ is the amplitude of a single wave, $a = c_{\rm s}^2/v_{\rm A}^2$ and $D = (1 + a)^2 - 4a\cos^2 \theta$.  
The associated density and magnetic field fluctuations are then respectively given by the linear relations, $\delta\rho = \pm\,\omega^{-1}(\bb{k} \cdot \delta \bb{v})\,\rho_0$ and $\delta \bb{B} = \pm\,\omega^{-1}[(\bb{k} \cdot \delta \bb{v})\bb{B}_0 - (\bb{k} \cdot \bb{B}_0)\delta \bb{v}]$,
where $\omega = \{(k^2v_{\rm A}^2/2)(1 + a + \sqrt{D}\,)\}^{1/2}$ is the linear fast-mode frequency, and the $\pm$ sign refers to the two propagation directions.  


\subsection{Mixed injection}\label{subsec:MW_init}
We also consider a mixed-wave (MW) injection case consisting of a superposition of arc-polarized Alfvén (A) and linear fast-magnetosonic (F) waves: $\delta\bb{B} = \delta\bb{B}^{\rm(A)} + \delta\bb{B}^{\rm(F)}$, $\delta\bb{v} = \delta\bb{v}^{\rm(A)} + \delta\bb{v}^{\rm(F)}$, and $\delta\rho = \delta\rho^{\rm(F)}$.
The rms magnetic field fluctuation is set to have equal contributions from the two components, $\delta B_{\rm rms}^{\rm(A)} \simeq \delta B_{\rm rms}^{\rm(F)}$.

\section{Overview of the developed turbulent state}
\label{sec:results_overview}

\begin{figure*}[tbh]
    \centering
    \includegraphics[width=0.9\linewidth]{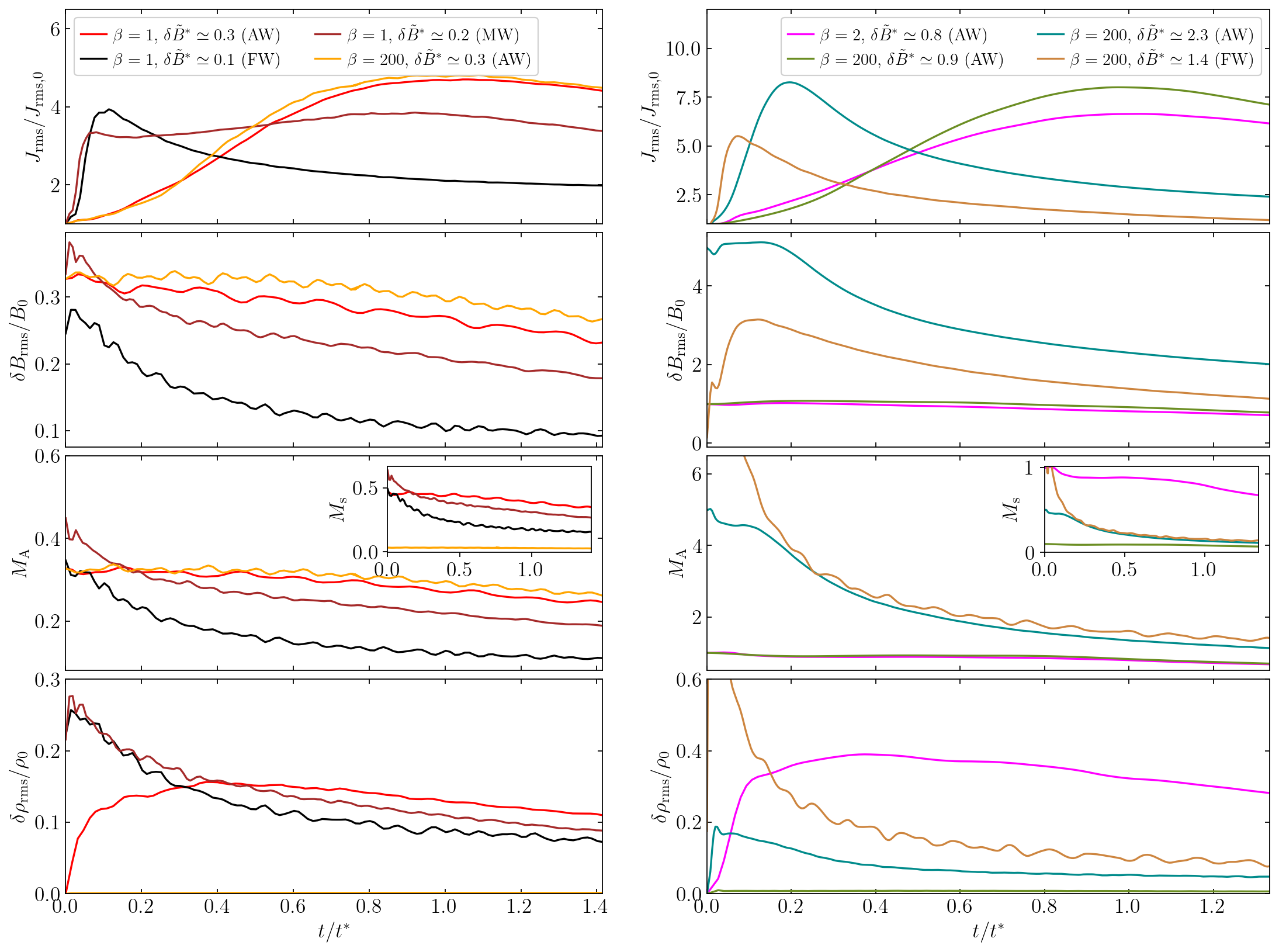}
    \caption{{\it Top to bottom}: Time evolution of the root-mean-square current, magnetic field fluctuations, Alfvénic Mach number (inset: sonic Mach number), and density fluctuations. {\it Left}: small-amplitude injection of Alfvénic (red at $\beta = 1$ and orange at $\beta=200$), fast-magnetosonic (black; $\beta=1$), and mixed waves (brown; $\beta=1$). \textit{Right}: large-amplitude injection of Alfvénic (magenta at $\beta=2$, olive green and dark cyan for different $\delta B/B_0$ at $\beta=200$) and fast-magnetosonic waves (ocher; $\beta=200$).}
    \label{fig:Jrms}
\end{figure*}

\subsection{Global time evolution}
\label{subsec:time_evo}

First, we want to characterize the global properties of the final turbulent state, and in particular analyze whether its main features are sensitive to the nature of large-scale injection.

The temporal evolution of several root-mean-square (rms) quantities in different simulations is shown in Figure \ref{fig:Jrms}, where the time has been normalized to the time $t^*$ at which turbulence is fully developed. This time corresponds to the peak of the current density $J_{\rm rms}$ when Alfv\'enic fluctuations with $\delta B_{\rm rms,0}/B_0\leq1$ are initialized. However, when large-amplitude fluctuations of any kind or small-amplitude fast-magnetosonic perturbations are initialized, this time does not correspond to the peak of $J_{\rm rms}$ anymore, as rapid transients associated with shock dissipation dominate its initial evolution. For these regimes, $t^*$ is associated with  the onset of steady fluctuations' spectra (see Section~\ref{subsec:spec_B}). From top to bottom row, Figure~\ref{fig:Jrms} displays the current density normalized to its initial value, $J_{\rm rms}/J_{\rm rms,0}$, the magnetic-field fluctuations $\delta B_{\rm rms}/B_0$, the Alfvénic Mach number $M_{\rm A} = \delta v_\mathrm{rms}/v_{\rm A}$ (with the sonic Mach number $M_{\rm s} = \delta v_\mathrm{rms}/c_{\rm s}$ in the inset), and the density fluctuations $\delta\rho_{\rm rms}/\rho_0$. The left panels correspond to the $\beta = 1$ regime with small-amplitude initial fluctuations ($\delta B_{\rm rms,0}/B_0 <1$) of different nature (Alfv\'enic, fast-magnetosonic, and mixed waves), while the right panels show runs where $\beta$ and/or the amplitude of the initial fluctuations are varied.
In the fully developed turbulent state, all simulations are sub-sonic ($M_{\rm s}<1$) while ranging from sub- to super-Alfv\'enic ($0.1\lesssim M_{\rm A}\lesssim1.5$).

When fast-magnetosonic fluctuations are initialized, both with low amplitude at $\beta=1$ (left panels, black and brown curves) and with large amplitude at $\beta=200$ (right panel, ocher lines), the system is initially dominated by the formation and dissipation of strong shocks, as highlighted by the sharp increase and successive rapid decay of $J_{\rm rms}$ and $\delta\rho_{\rm rms}$ for $t/t^*\lesssim 0.3$. A similar behavior is found also when Alfv\'enic fluctuations with $\delta B_{\rm rms}/B_0>1$ are initialized (right panels, dark cyan). After this transient, for $t/t^*\gtrsim 0.5$ all these systems become dominated by a gradual turbulent decay at a rate that is the same for simulations with the same $\beta$ and similar fluctuations' amplitude, regardless of the nature in the initial perturbations. The same holds for density fluctuations, eventually achieving similar levels regardless of the nature of initial fluctuations, namely $\delta\rho_{\rm rms}^*/\rho_0\sim 0.1$ for both low-amplitude injection at $\beta=1$ and very large amplitude fluctuations ($\delta B_{\rm rms}/B_0>1$) at $\beta=200$; in this large-$\beta$ regime, density fluctuations become negligibly small ($\delta\rho_{\rm rms}^*/\rho_0\lesssim0.01$) for lower initial amplitudes $\delta B_{\rm rms}/B_0\lesssim1$ (left panels, orange), while significant density fluctuations ($\delta\rho_{\rm rms}^*/\rho_0\sim 0.3$) develop at $\beta=2$ when $\delta B_{\rm rms}/B_0\sim1$ (magenta lines in Figure~\ref{fig:Jrms}).

\begin{figure}[t!]
    \centering
    \includegraphics[width=0.99\linewidth]{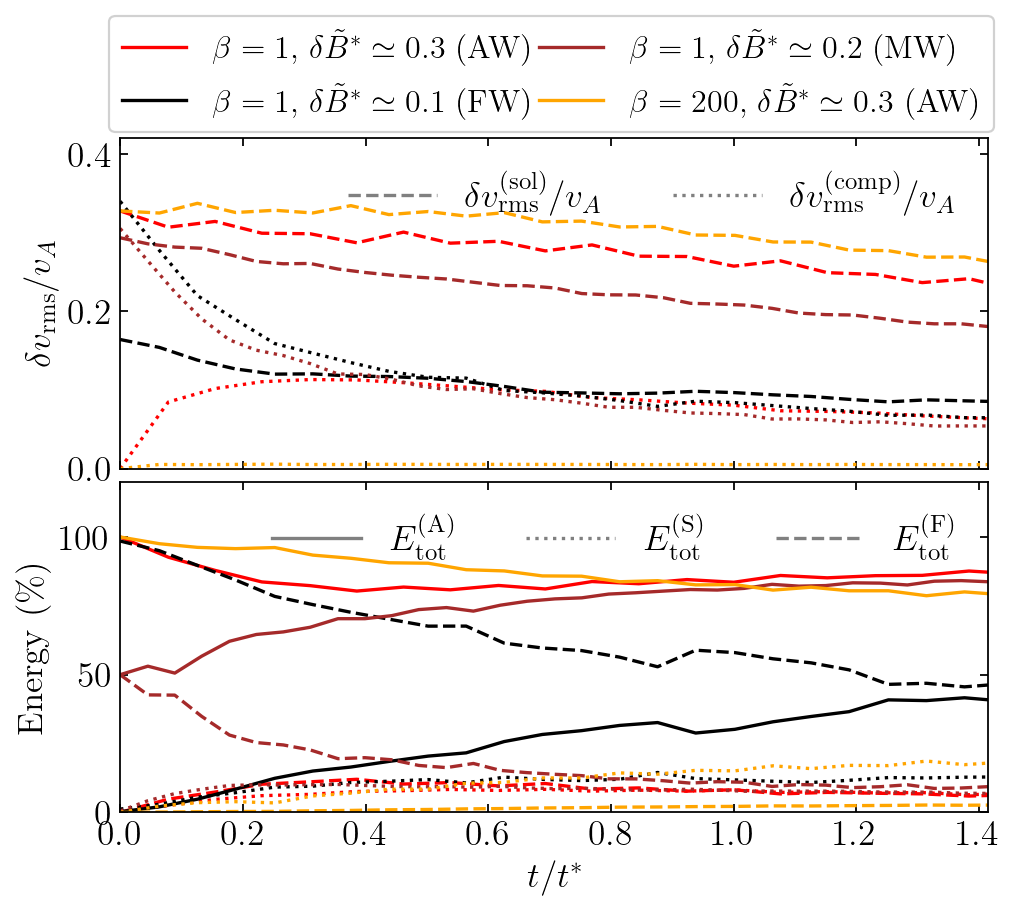}
    \caption{\textit{Top panel}: Time evolution of the compressible (dotted line) and solenoidal (dashed line) velocity fluctuations in small-amplitude turbulence.
    \textit{Bottom panel}: Time evolution of the energy fraction in the Alfvén (solid line), slow (dotted line), and fast (dashed line) modes for the same simulations.}
    \label{fig:vHelm_time}
\end{figure}

\begin{figure}[t!]
    \centering
    \includegraphics[width=0.99\linewidth]{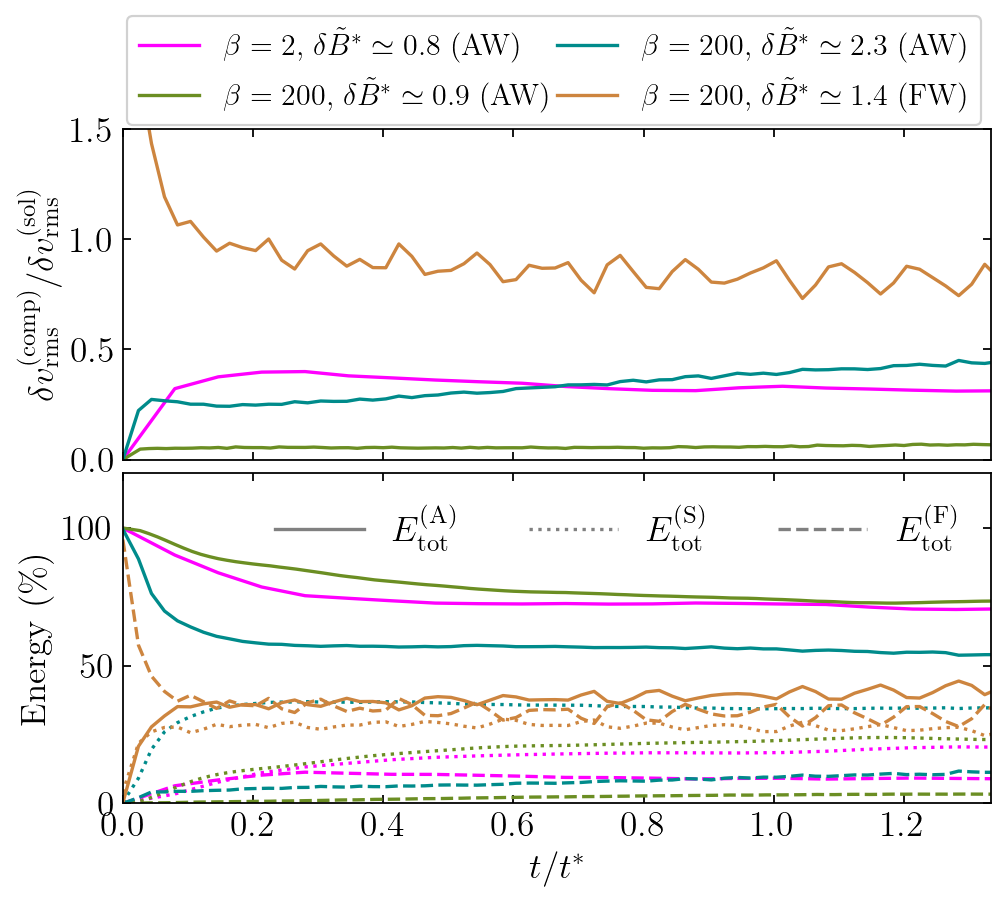}
    \caption{Same as Figure \ref{fig:vHelm_time} but for large-amplitude turbulence.}
    \label{fig:vHelm_time2}
\end{figure}

The time evolution of fluctuations' nature is analyzed in Figures~\ref{fig:vHelm_time} and $\ref{fig:vHelm_time2}$ using two different methods. 
The first method consists of decomposing velocity fluctuations into their compressible and solenoidal contributions through Helmholtz decomposition in Fourier space, i.e., $\delta\bb{v}_{\bb{k}}^{\rm(comp)}=(\bb{k}\cdot\delta\bb{v}_{\bb{k}})\bb{k}/k^2$ and 
$\delta\bb{v}_{\bb{k}}^{\rm(sol)}=\delta\bb{v}_{\bb{k}}-\delta\bb{v}_{\bb{k}}^{\rm(comp)}$.
The second method consists of projecting velocity fluctuations onto the linear eigenvectors corresponding to Alfvén (A), slow-magnetosonic (S) and fast-magnetosonic (F) modes in Fourier space~\citep{Cho2003}: 
\begin{align}
\hat{\bb{\zeta}}_{\bb{k}}^{\rm(A)} &= \hat{\bb{k}}_\perp \times \hat{\bb{k}}_\parallel\,, \label{eq:eigenvect_A}\\
\hat{\bb{\zeta}}_{\bb{k}}^{\rm(F)} &\propto (-1 + a + \sqrt{D})\, k_\parallel \hat{\bb{k}}_\parallel 
+ (1 + a + \sqrt{D})\, k_\perp \hat{\bb{k}}_\perp\,,\label{eq:eigenvect_F} \\
\hat{\bb{\zeta}}_{\bb{k}}^{\rm(S)} &\propto (-1 + a - \sqrt{D})\, k_\parallel \hat{\bb{k}}_\parallel 
+ (1 + a - \sqrt{D})\, k_\perp \hat{\bb{k}}_\perp\,,\label{eq:eigenvect_S}
\end{align}
where the hat stands for unit vector (e.g., $\hat{\bb{k}}_\perp=\bb{k}_\perp/|\bb{k}_\perp|$) and the quantities $a$ and $D$ are defined in Section~\ref{subsec:FW_init}.
With this method one can obtain also the corresponding magnetic and density fluctuations of each mode m = A, F, S as $\delta\bb{B}_{\bb{k}}^{\rm(m)}=\bb{k}\times(\bb{B}_0\times\delta\bb{v}_{\bb{k}}^{\rm(m)})/\omega_{\bb{k}}^{\rm(m)}$ and $\delta\rho_{\bb{k}}^{\rm(m)}=\rho_0\bb{k}\cdot\delta\bb{v}_{\bb{k}}^{\rm(m)}/\omega_{\bb{k}}^{\rm(m)}$, where $\omega_{\bb{k}}^{\rm(m)}$ is the linear frequency of the mode $m$ at a given wavevector $\bb{k}$ (so, e.g., $\delta\bb{B}_{\bb{k}}^{\rm(A)}=-\frac{k_\parallel}{|k_\parallel|}B_0\,\delta\bb{v}_{\bb{k}}^{\rm(A)}/v_{\rm A}$ and $\delta\rho_{\bb{k}}^{\rm(A)}=0$).
Once the fluctuations are decomposed, we compute the total energy fraction as
\begin{equation}\label{eq:mode_EnergyFrac}
\frac{E_\mathrm{tot}^{\rm(m)}}{E_{\rm tot}} = \frac{E_{\rm mag}^{\rm(m)} + E_{\rm kin}^{\rm(m)}}{E_{\rm tot}} 
= 
\frac{\sum_{\bb{k}} \left|\delta \bb{B}_{\bb{k}}^{(m)}\right|^2 + \sum_{\bb{k}} \left|\sqrt{\rho_{\bb{k}}^{\rm(m)}}\,\delta \bb{v}_{\bb{k}}^{\rm(m)}\right|^2 }
{\sum_{\bb{k}} \left( \left|\delta \bb{B}_{\bb{k}}\right|^2 + \left|\sqrt{\rho_{\bb{k}}}\,\delta \bb{v}_{\bb{k}}\right|^2 \right)}\,,
\end{equation}
where $\rho_{\bb{k}}^{\rm(m)}=\rho_0+\delta\rho_{\bb{k}}^{\rm(m)}$ (the factor $\sqrt{4\pi}$ is absorbed in the definition of $\bb{B}$ in the PLUTO code). Note that \eqref{eq:mode_EnergyFrac} differs from the definition of \cite{Makwana2020} as in compressible MHD turbulence, the kinetic-energy density term $E_{\rm kin}^{\rm(m)}$ should explicitly account for density fluctuations.
Finally, we stress that the mode decomposition should be formally limited to small-amplitude fluctuations (e.g., $\bb{k}_\|$ is identified as parallel to $\bb{B}_0$), and therefore must not be over-interpreted (see Appendix~\ref{app:modes_Helm} for a comparison between mode and Helmholtz decompositions, and Appendix~\ref{app:omegak} for a ($\omega$, $k_z$, $k_\perp$) Fourier analysis).

The Helmholtz decomposition reveals a picture that is consistent with the behavior of density fluctuations in Figure~\ref{fig:Jrms}. In fact, the upper panel of Figure~\ref{fig:vHelm_time} shows that at $\beta=1$ a significant amount of compressible fluctuations develops over time regardless of the nature of initial fluctuations, eventually reaching values of order $\delta v_\mathrm{rms}^\mathrm{(comp)}/v_{\rm A}\sim0.1$ even when Alfvénic fluctuations are initialized.
Therefore, while the relative importance between $\delta v_{\rm rms}^{\rm(comp)}$ and $\delta v_{\rm rms}^{\rm(sol)}$ may be different for different type of injection, i.e., $\delta v_{\rm rms}^{\rm(comp)}/\delta v_{\rm rms}^{\rm(sol)}\sim 1$ for fast-magnetosonic injection and $\delta v_{\rm rms}^{\rm(comp)}/\delta v_{\rm rms}^{\rm(sol)}\sim0.3$ for Alfv\'enic injection, we stress that even for ``nominally incompressible'' low-amplitude injection, the amount of compressive fluctuations that are nonlinearly generated in fully developed turbulence is not negligible.
Clearly, when low-amplitude Alfv\'enic fluctuations are injected in a lower-compressibility plasma (i.e., $\beta=200$), both density fluctuations and compressive fluctuations are instead negligibly small, i.e., $\delta\rho/\rho_0\sim0.01$ and $\delta v_{\rm rms}^{\rm(comp)}\ll\delta v_{\rm rms}^{\rm(sol)}$ (orange lines in Figure~\ref{fig:vHelm_time}).
On the other hand, when the initial fluctuations are highly compressive as for fast-magnetosonic injection, they are rapidly damped through shock formation until they reach near-equipartition with the solenoidal component in fully developed turbulence. An analogous behavior is found when varying the plasma $\beta$ and the amplitude $\delta B/B_0$ of initial fluctuations (upper panel of Figure~\ref{fig:vHelm_time2}), showing that these two parameters are the only one controlling the final value of the ratio $\delta v_{\rm rms}^{\rm(comp)}/\delta v_{\rm rms}^{\rm(sol)}$ for a given type of injection (becoming negligible for $\beta\gg1$ and $\delta B/B_0 \lesssim 1$). Finally, as for the $\beta=1$ case, the absolute value of $\delta v_{\rm rms}^{\rm(comp)}/v_{\rm A}$ that is eventually achieved in fully developed turbulence at given $\beta$ and $\delta B/B_0$ does not significantly depend on the nature of the injected fluctuations (not shown).

The mode decomposition further tells us the relative importance of different type of fluctuations in these turbulent systems. From the lower panel of Figure~\ref{fig:vHelm_time} it is evident that, at $\beta=1$, the contribution of fast-magnetosonic fluctuations to the total energy budget in fully developed turbulence remains relevant (i.e., $\sim50\%$ of the total fluctuation energy) only when the initial injection consists exclusively of fast modes. In such case, the rest of the turbulent energy is mainly represented by Alfv\'enic fluctuations ($\sim40\%$ of $E_{\rm tot}$). For other types of injection at $\beta=1$, the fully turbulent turbulent state is always characterized by mainly Alfv\'enic fluctuations ($\sim80\%$ of the total fluctuation energy), while fast-magnetosonic fluctuations only account for $\sim10\%$ of the turbulent energy. Finally, regardless of the injection type, slow-magnetosonic fluctuations are always sub-dominant at $\beta=1$, representing only $\lesssim10\%$ of the $E_{\rm tot}$ in fully developed turbulence. 
Also at $\beta>1$, the fully developed turbulent state is mainly populated by Alfv\'enic fluctuations for any initial amplitude (orange line in Figure~\ref{fig:vHelm_time} and green, magenta and cyan lines in Figure~\ref{fig:vHelm_time2}), except when the initial perturbations consist of purely fast-mode, large-amplitude fluctuations (Figure~\ref{fig:vHelm_time2}, ocher lines). In this latter case, however, a large fraction of fast-magnetosonic fluctuations are quickly dissipated by shocks and the final turbulent state consist of a near equipartition between the three MHD modes ($\sim30\%$ of $E_{\rm tot}$ in each type of fluctuation). For all other type of injection at $\beta>1$, the contribution of fast-magnetosonic fluctuations to the total turbulent energy remains almost negligible (i.e, $\lesssim10\%$ of $E_{\rm tot}$). In the large-$\beta$ regime, slow-magnetosonic fluctuations become more relevant than at $\beta=1$ even when injection consists of purely Alfv\'enic fluctuations: in the fully turbulent state, slow mode contributes to up to $\sim30\%$ of the fluctuation energy for large-amplitude initial fluctuations (a contribution that then decreases for decreasing initial $\delta B/B_0$). However, we remind that mode decomposition is not fully reliable at large fluctuation amplitudes and that MHD modes exhibit a certain degree of degeneracy at $\beta\gg1$, and thus this analysis should not be over-interpreted (see Appendix~\ref{app:modes_Helm}).

\begin{figure*}[th!]
    \centering
    \includegraphics[width=0.49\linewidth]{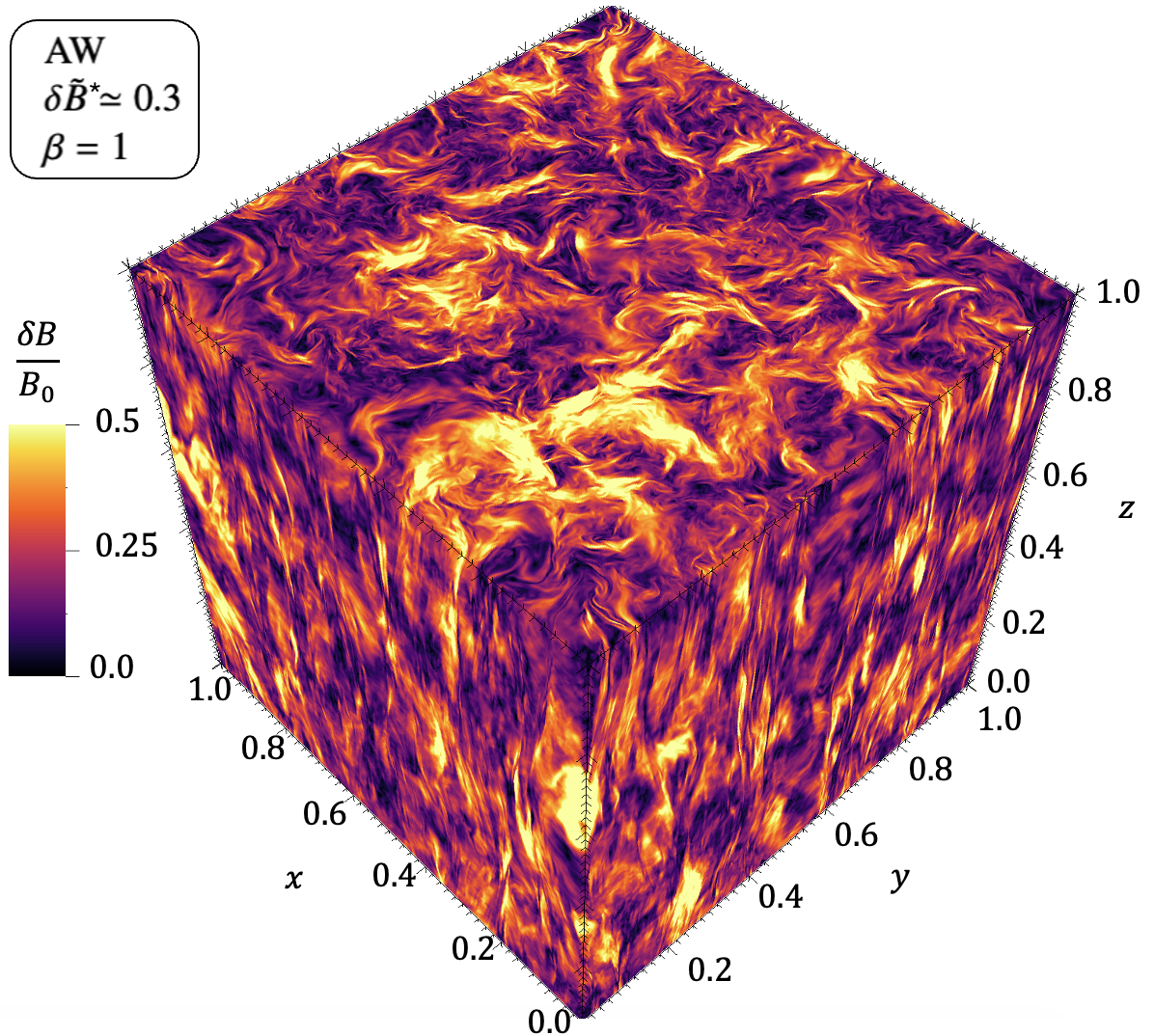}
    \includegraphics[width=0.49\linewidth]{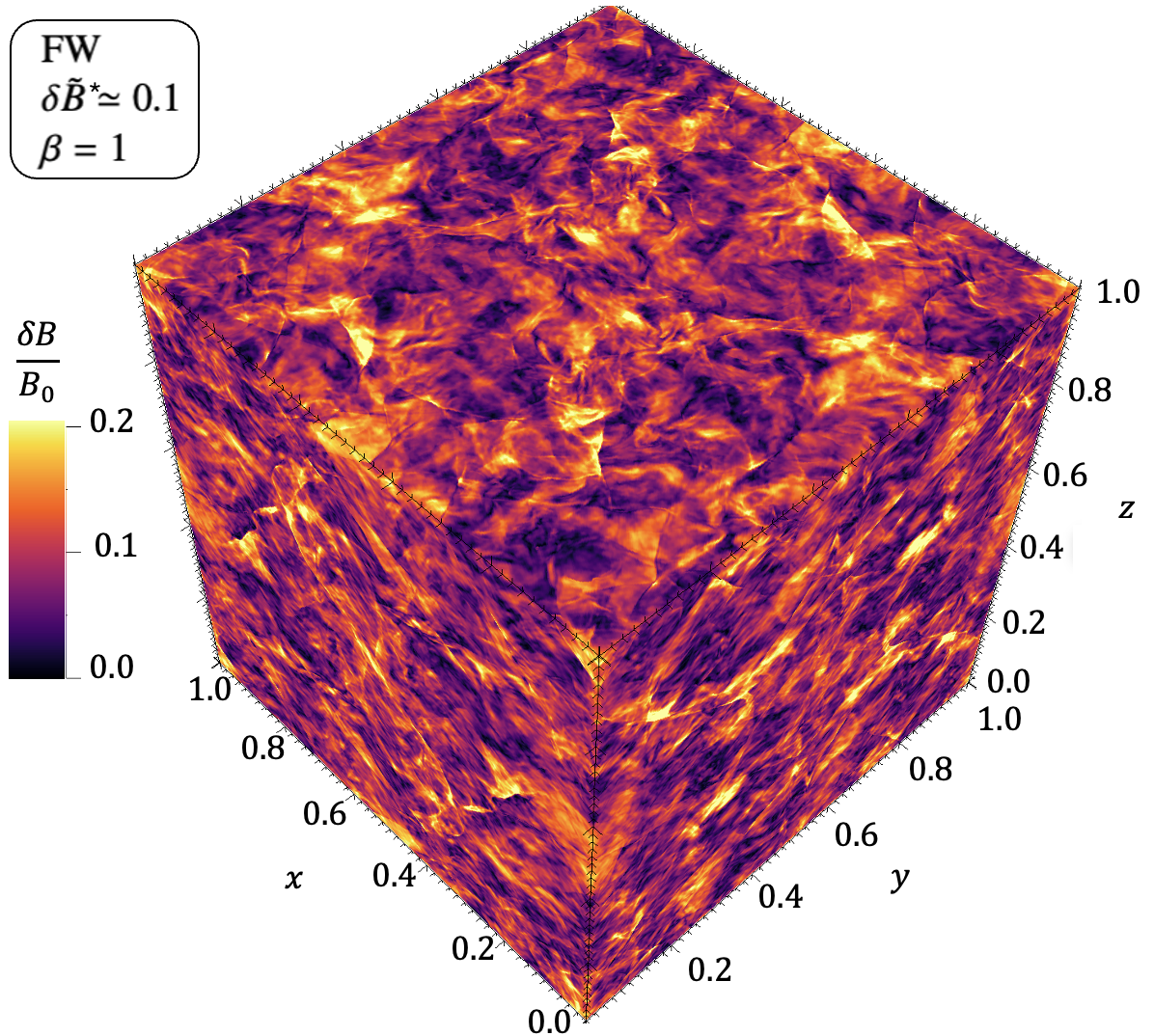}

    \includegraphics[width=0.49\linewidth]{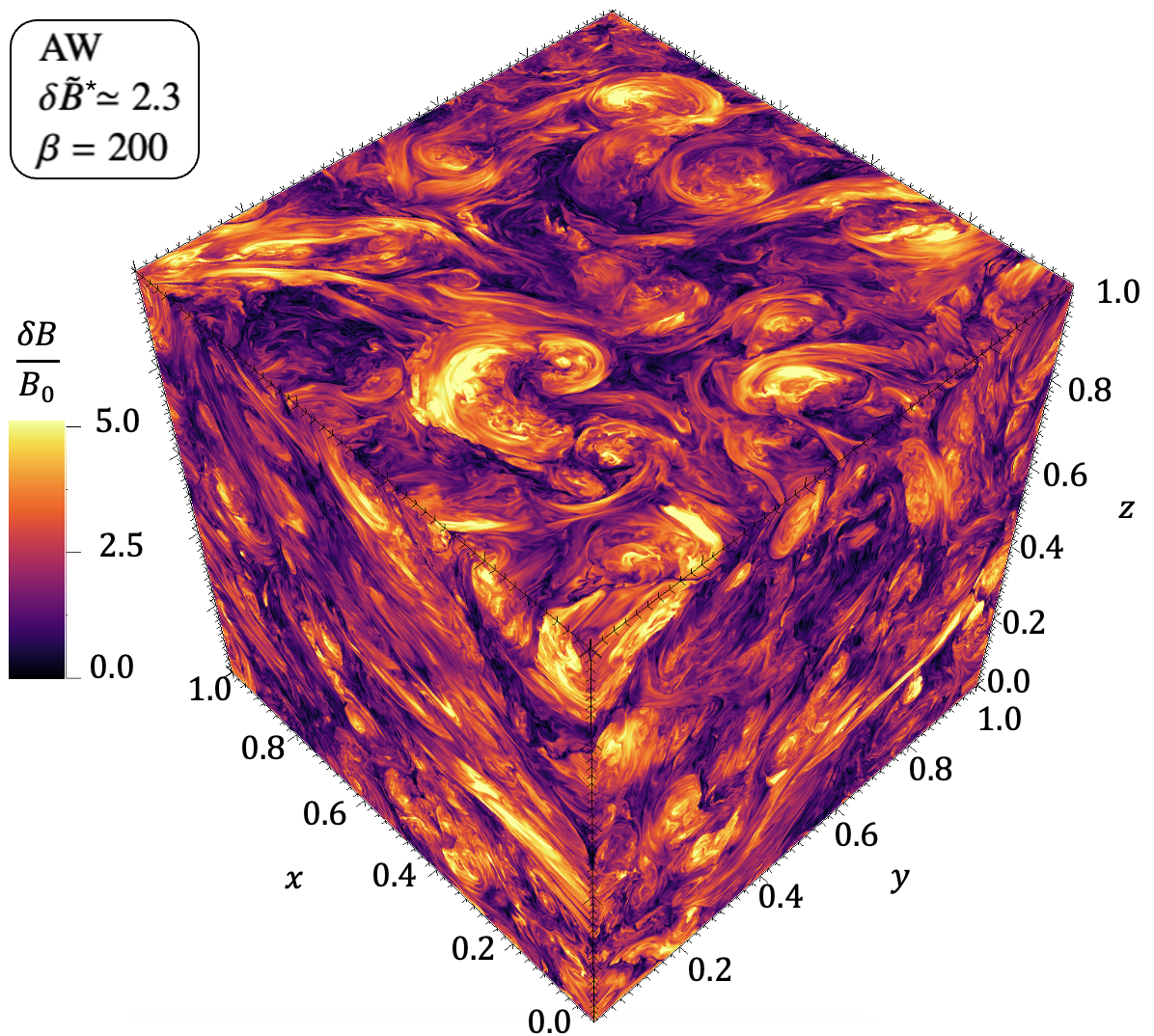}
    \includegraphics[width=0.49\linewidth]{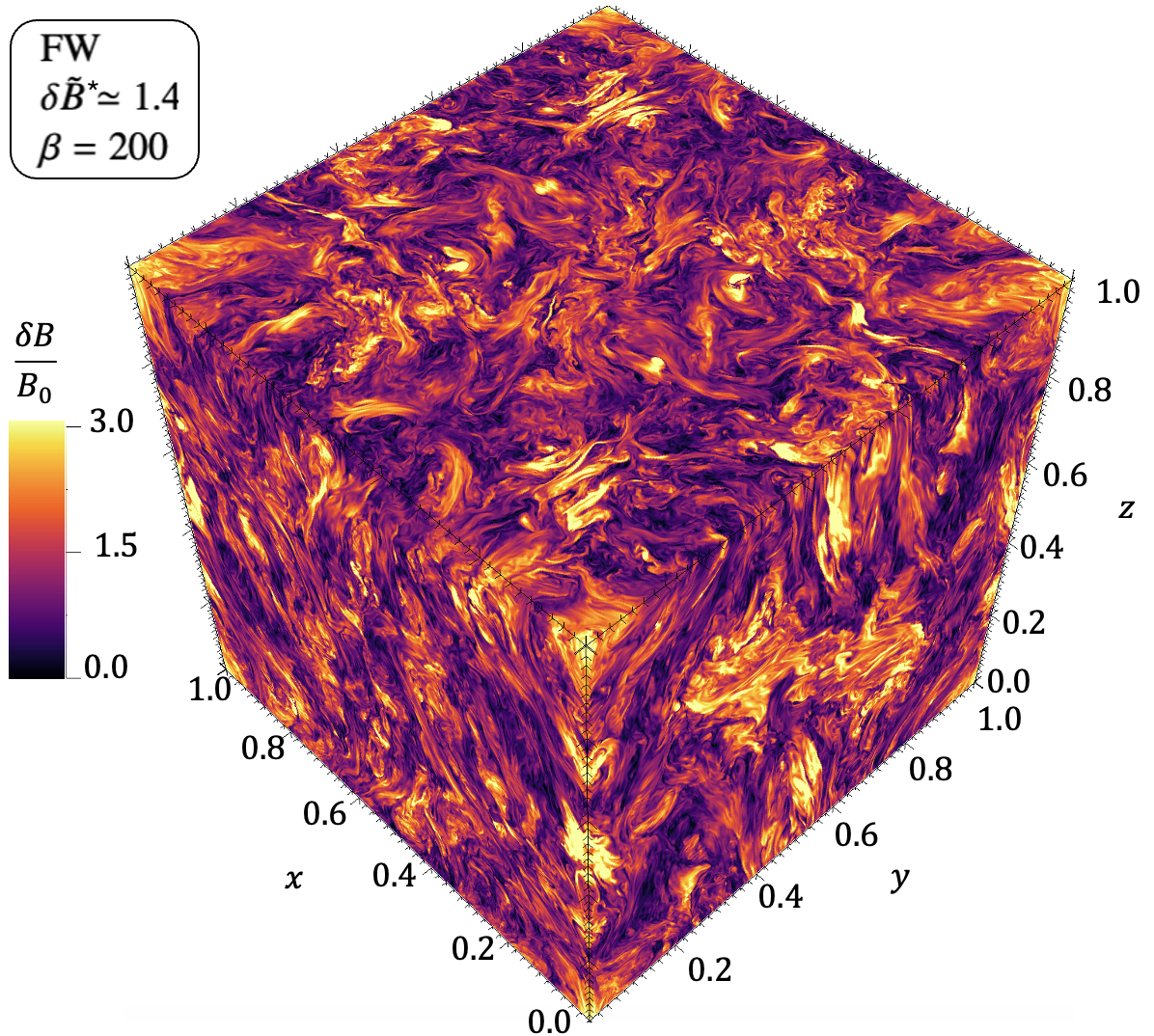}

    \caption{3D plots of magnetic-fluctuation amplitude $\delta B/B_0$ in fully developed turbulence. \textit{Top panels:} low-amplitude Alfvénic (left) and fast-magnetosonic (right) injection at $\beta =1$. \textit{Bottom panels:} Same as top panels but for large-amplitude turbulence at $\beta = 200$.}
    \label{fig:B_3D}
\end{figure*}

\subsection{Formation of turbulent structures in real space}\label{sec:structures}

The formation of magnetic structures and patterns in real space is an important aspect of turbulence~\citep[e.g.,][and references therein]{Vlahos2023,Ntormousi2024}, which in turn can strongly affect the transport of cosmic rays~\citep[e.g.,][]{LazarianXu2021,Lemoine2023,Kempski2023,Lubke2025}.

Figure \ref{fig:B_3D} shows 3D visualizations of magnetic-field fluctuations for Alfvénic (left) and fast-magnetosonic (right) injections, at $\beta = 1$ (upper panels) and $\beta = 200$ (lower panels), evaluated at $t = t^*$, when turbulence is fully developed.
At $\beta = 1$, the Alfvénic injection (upper left panel) exhibits the expected anisotropic structure: sheet-like features in the perpendicular plane that are much more elongated along the direction of the mean magnetic field $\bb{B}_0=B_0\bb{e}_z$.
In contrast, the fast-magnetosonic injection (upper right panel) is characterized by the presence of shock-like structures in both perpendicular and parallel planes, resulting in an approximately isotropic morphology, qualitatively consistent with the inferred nature of the injected modes from sub-Alfv\'enic simulations~\citep{Cho2002b} and with a weak-turbulence regime~\citep{KochurinKuznetsovPRL2024}; see Appendix~\ref{app:chi}. However, a detailed analysis of the fluctuation anisotropy, presented in Section~\ref{subsec:spec_B}, reveals some deviation from this apparent anisotropy.
At $\beta = 200$, the behavior changes significantly. In the fast-magnetosonic case (lower right panel), despite large-amplitude fluctuations being initialized, a sub-sonic regime is quickly reached due to the high plasma $\beta$ (while still being in a super-Alfv\'enic regime) and shocks are no longer present in the fully turbulent state. As a result, the system develops structures in the perpendicular plane that vaguely resemble those of the Alfvénic case at $\beta=1$.
It is also noteworthy that local peaks as large as $\delta B/B_0 \simeq 3$ are observed: these can be seen in the contours as patches and ``blobs'' of intense magnetic-field fluctuations present at intermediate and small scales (these will likely act as mirroring sites for cosmic rays). 
For the case of initial large-amplitude, Alfvénic fluctuations at $\beta = 200$ (lower left panel), large-scale vortices and filaments emerge in both parallel and perpendicular planes. Although fully developed turbulence looks qualitatively isotropic at $\delta B/B_0>1$ due to the fact that this regimes significantly reduces the influence of a background field, a certain degree of anisotropy is still present when magnetic-field fluctuations are analyzed (see Section~\ref{subsec:spec_B}). Moreover, the morphology of emerging magnetic structures are significantly different for Alfv\'enic or fast-magnetosonic injection.
\begin{figure}[t!]
    \centering
    \includegraphics[width=0.42\linewidth]{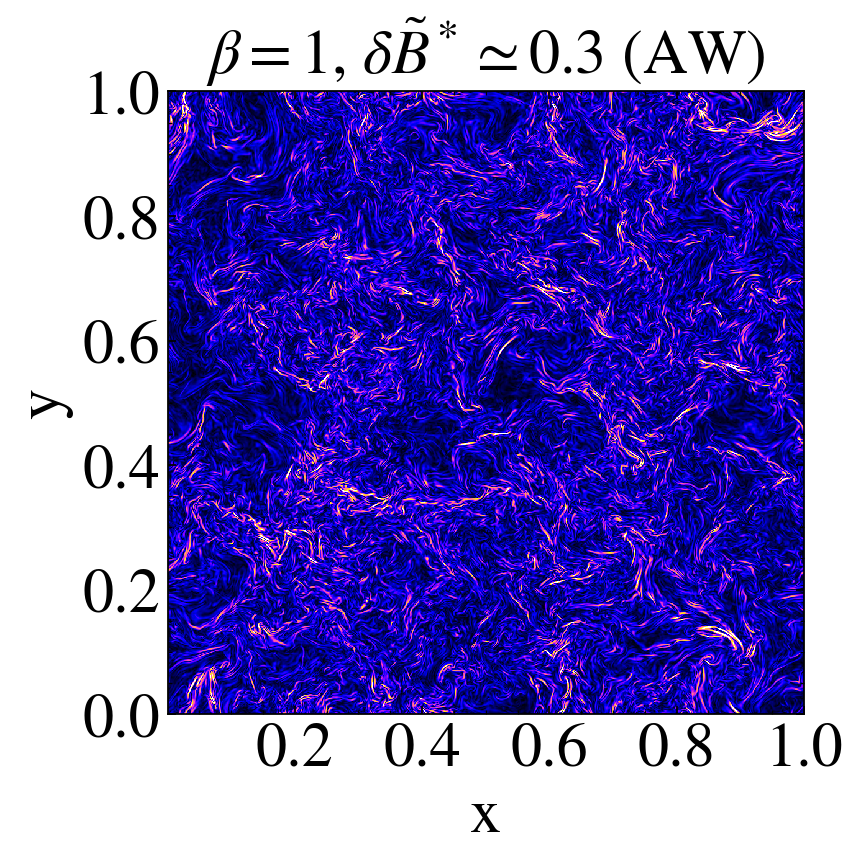}
    \includegraphics[width=0.495\linewidth]{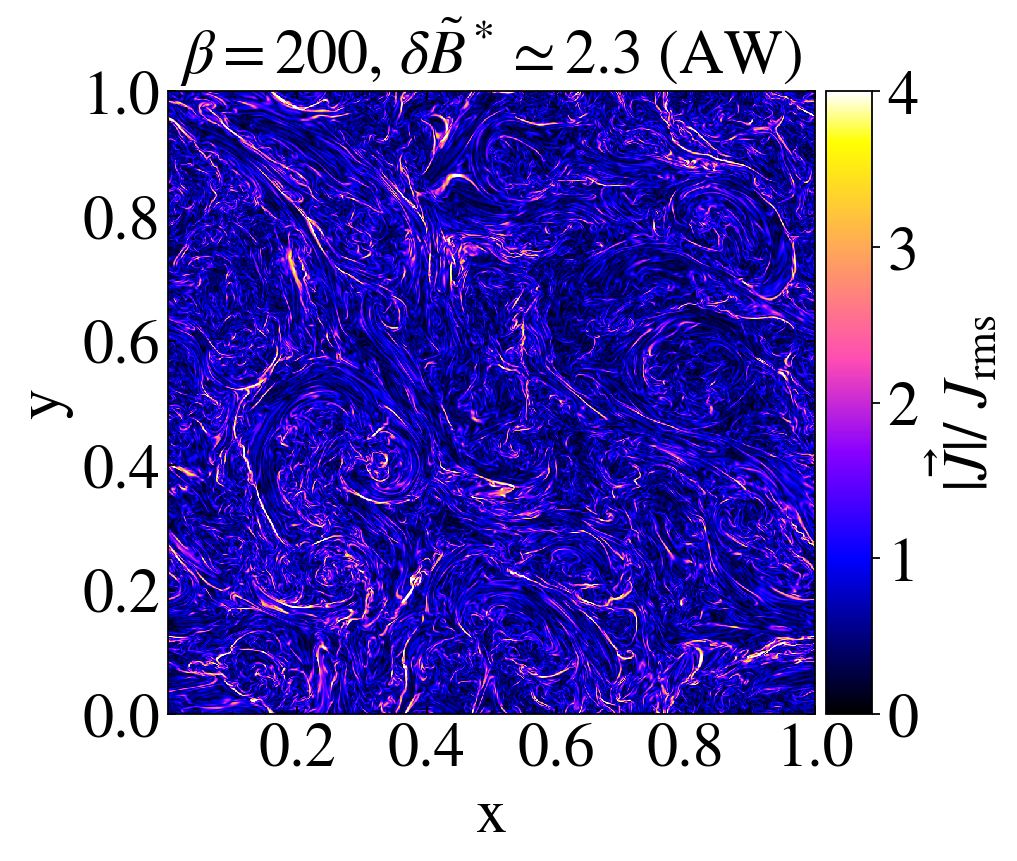}
 \\ 
    \includegraphics[width=0.42\linewidth]{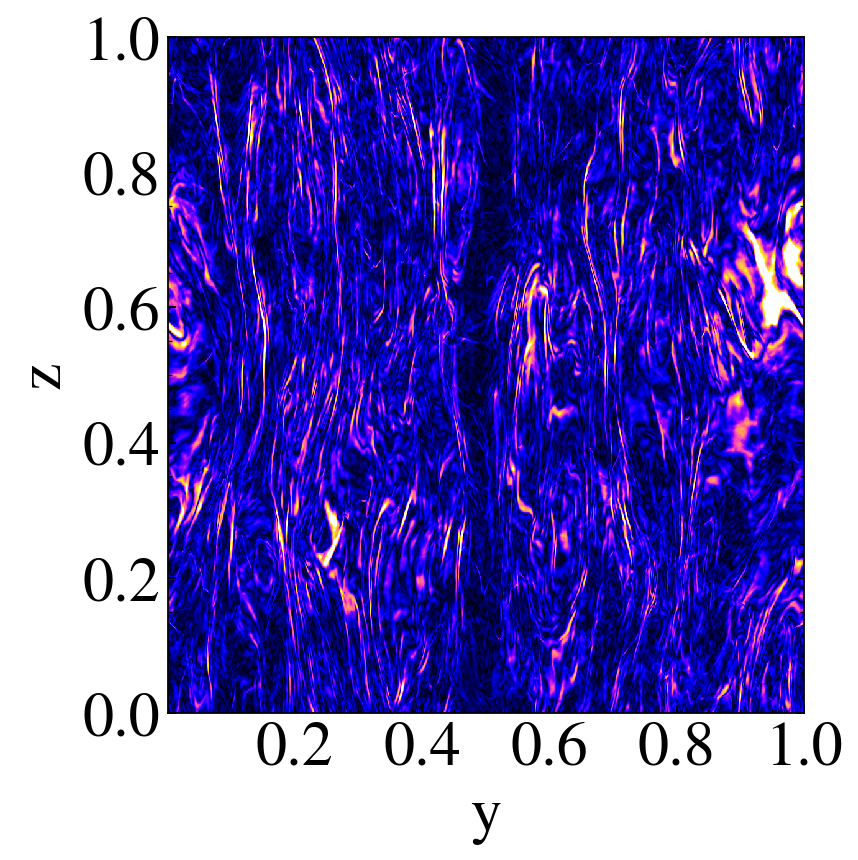}
    \includegraphics[width=0.495\linewidth]{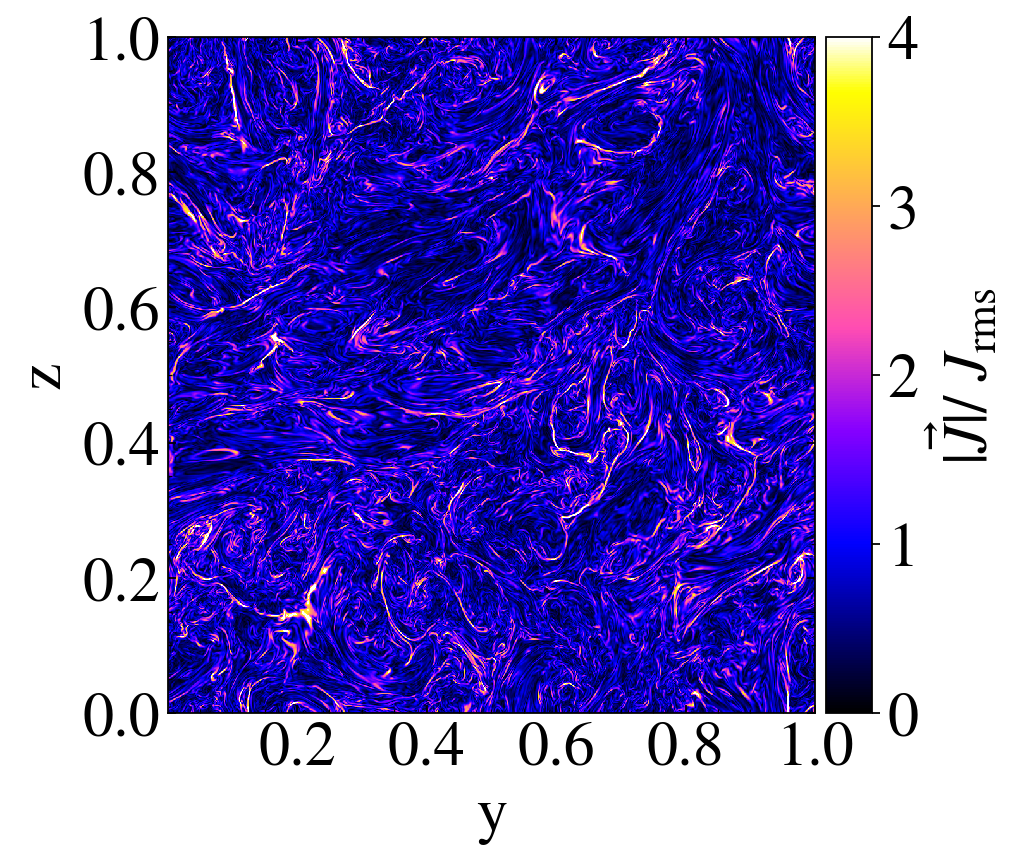}
    
    \caption{Normalized current-density magnitude $|\bb{J}|/J_\mathrm{rms}$ in fully developed turbulence for Alfv\'enic injection with $\delta\tilde{B}\simeq0.3$ at $\beta=1$ (left column) and $\delta\tilde{B}\simeq2.3$ at $\beta=200$ (right column). \textit{Top panels}: ($x$, $y$) plane at $z = 0$. \textit{Bottom panels:} ($y$, $z$) plane at $x = 0$.}
    \label{fig:current}
\end{figure}

\begin{figure}[t!]
    \centering
    \includegraphics[width=0.42\linewidth]{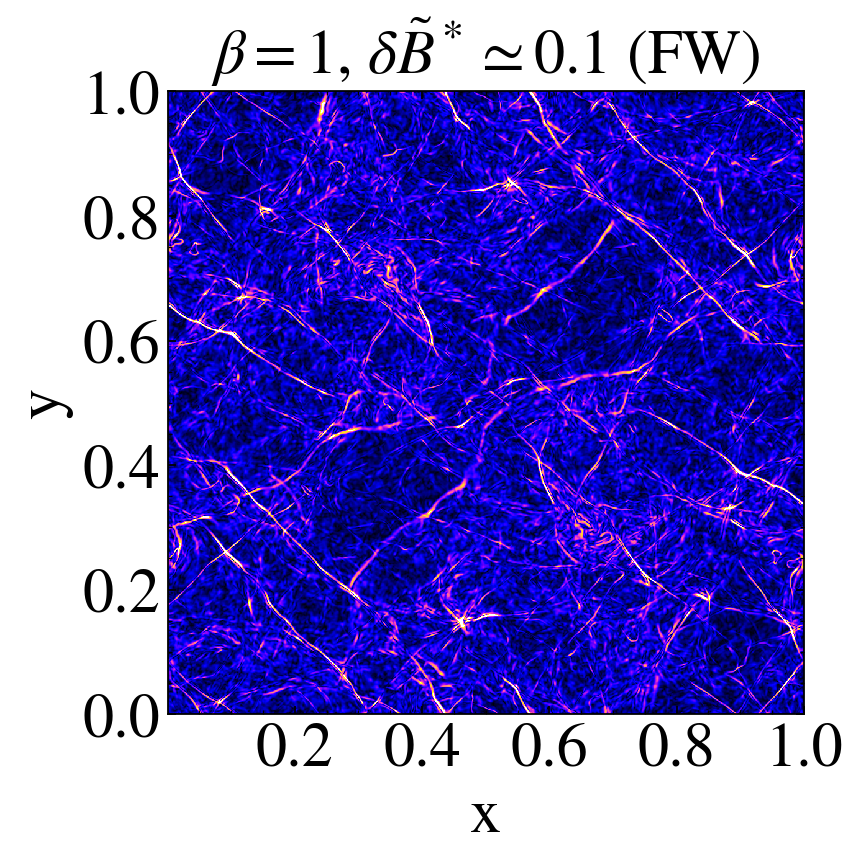}
    \includegraphics[width=0.495\linewidth]{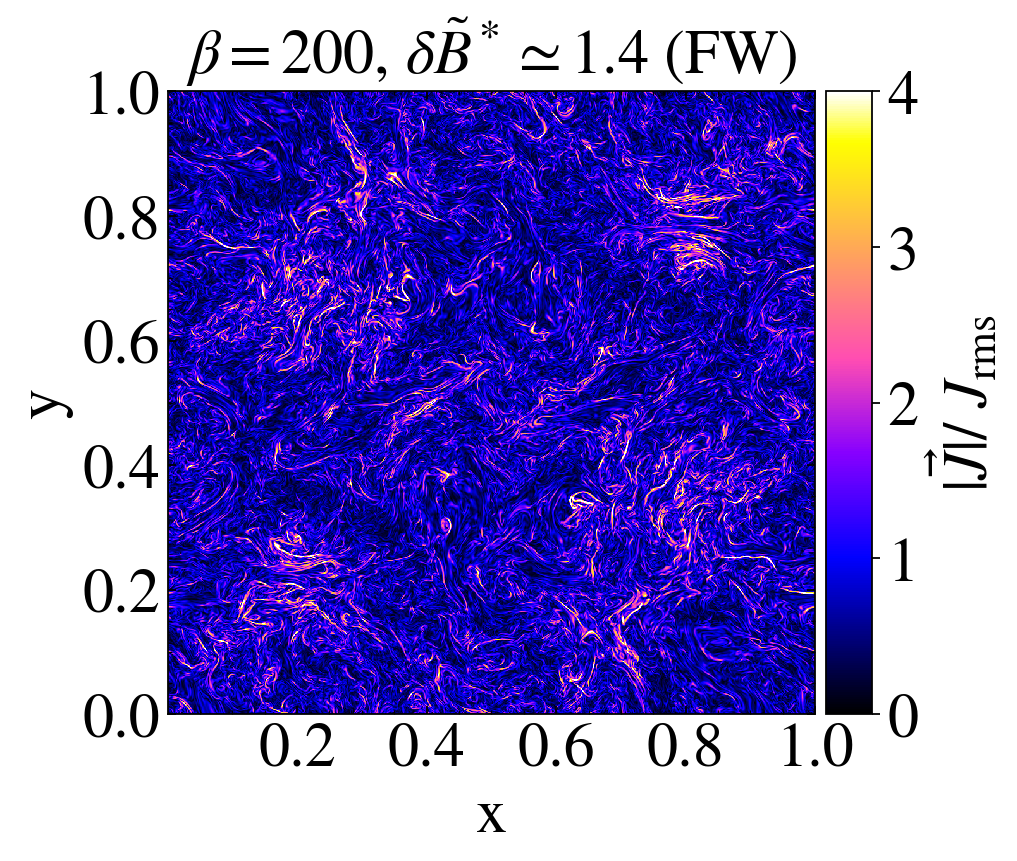}\\ 
    \includegraphics[width=0.42\linewidth]{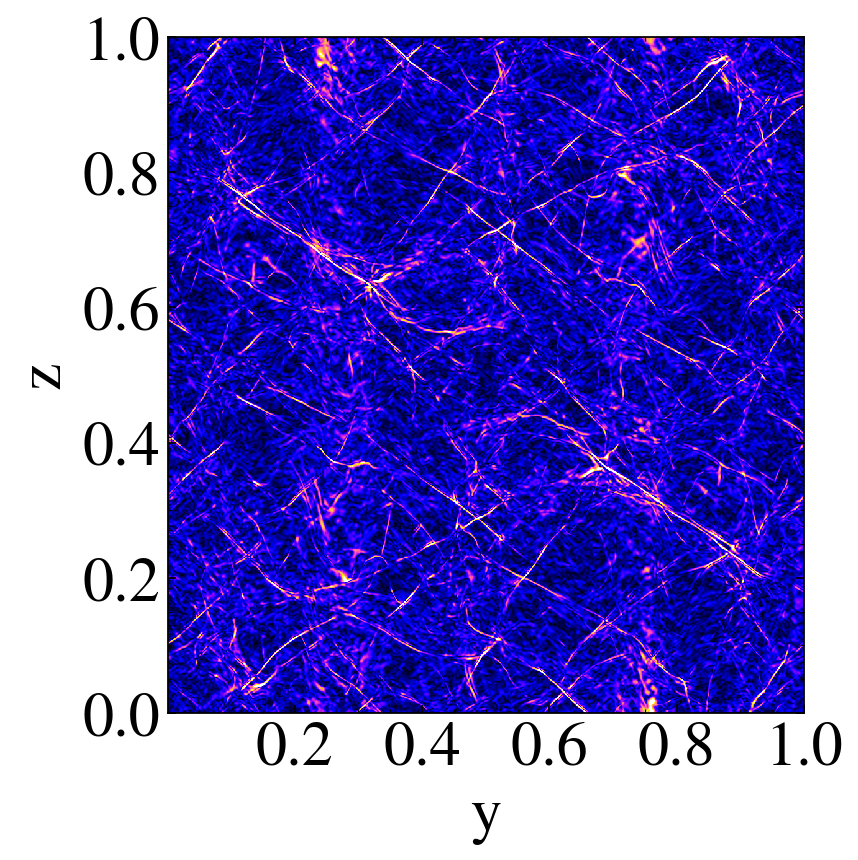}
    \includegraphics[width=0.495\linewidth]{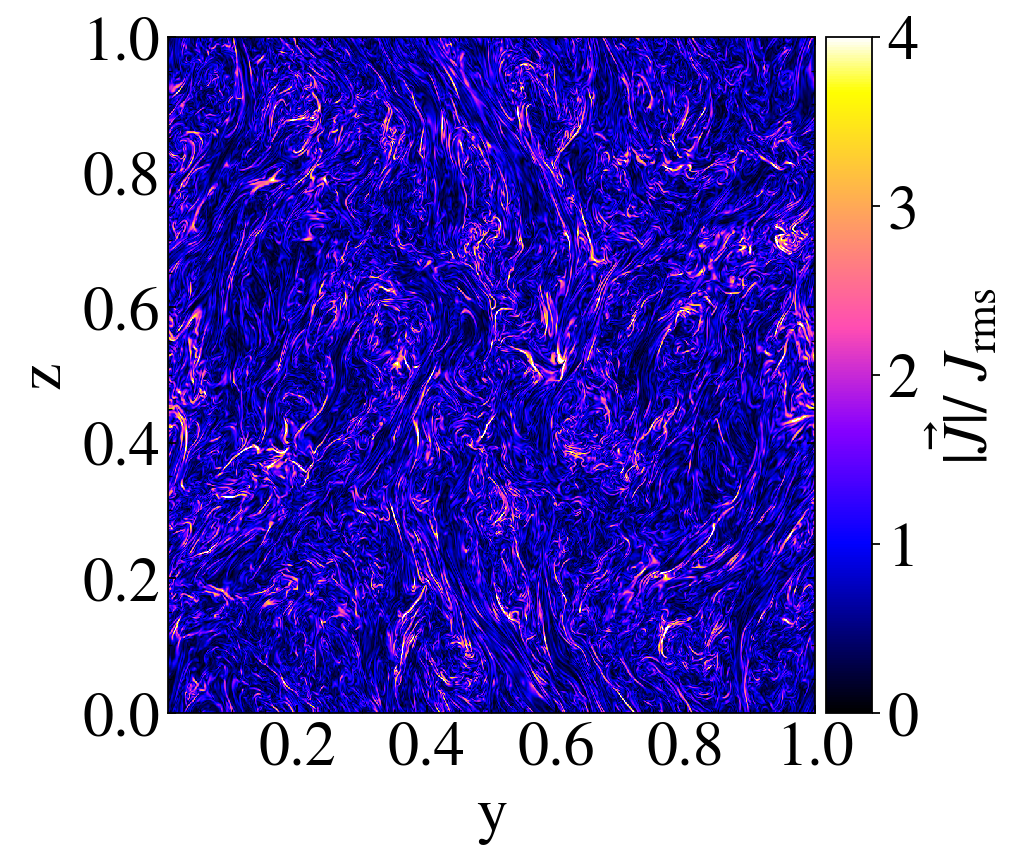}
    
    \caption{Same as Figure \ref{fig:current}, but for fast-magnetosonic injection.}
    \label{fig:current3}
\end{figure}

Some of the above-mentioned features are further illustrated in Figures \ref{fig:current} and \ref{fig:current3}, showing snapshots of the normalized current density $J/J_{\rm rms}$ at $t = t^*$ in the perpendicular ($x$, $y$) plane at $z=0$ (upper panels) and in the plane ($y$, $z$) at $x=0$ that contains $\bb{B}_0$ (lower panels) for Alfv\'enic and fast-magnetosonic injection, respectively. In the low-amplitude cases ($\delta B/B_0 \ll 1$), the system displays current-sheet structures elongated in the $\bb{B}_0$ direction for either purely Alfv\'enic (left panels of Figure \ref{fig:current}) or mixed (not shown) injection; the same holds for low-amplitude Alfv\'enic injection at $\beta=200$ (not shown).
Increasing the amplitude of the initial Alfv\'enic fluctuations to $\delta B/B_0\gg1$ naturally leads to a more isotropic turbulent state with homogeneously distributed current-sheet structures (right panels of Figure \ref{fig:current}).
In contrast, low-amplitude fast-magnetosonic  injection at $\beta=1$ (left panels of Figure \ref{fig:current3}), clearly exhibits a nearly isotropic and homogeneous distribution of current filaments associated to shock-like structures. This is consistent with turbulence that is at least partially in a weak regime (see Appendix~\ref{app:chi}) dominated by fast-magnetosonic waves, as observed in weak acoustic turbulence~\citep[e.g.,][]{KochurinKuznetsovPRL2024} and in relativistic $2$D plasma simulations~\citep{Ugarov2026}.
Increasing the amplitude of the initial fast-magnetosonic fluctuations and reducing the plasma compressibility ($\beta=200$) leads to a transition from a regime dominated by shock structures to a more homogeneous turbulent state dominated by current-sheet structures (right panels of Figure \ref{fig:current3}), similar to the large-amplitude Alfv\'enic case (right panels of Figure~\ref{fig:current}). However, the compressive nature of fast-mode fluctuations makes the spatial distribution of current structures for this case more clumpy than in the Alfv\'enic case. This transition is associated with a stronger turbulence regime (see Appendix \ref{app:chi}).

\subsection{Statistics of magnetic-field curvature and magnetic-mirror scales}\label{subsec:B_curv} 

A quantity that has recently gained attention in cosmic-ray scattering is the parallel magnetic-field curvature $K_\parallel = |(\hat{\bb{b}}\cdot\bb{\nabla}) \hat{\bb{b}}|$, where $\hat{\bb{b}}= \bb{B}/B$ \citep[e.g.,][]{Kempski2023,Lemoine2023}, whose statistics turns out to be significantly sensitive to the properties and nature of large-scale injection. 
Furthermore, magnetic-mirror structures have been recently invoked to play an important role in the transport of cosmic rays~\citep[e.g.,][]{LazarianXu2021,ZhangXu2023}. Therefore, we also present a study on the statistics of the field-aligned gradient of magnetic-field strength, $K_\mathrm{M} = |\hat{\bb{b}} \cdot \bb{\nabla} B / B| = |\hat{\bb{b}} \cdot \bb{\nabla} \ln B|$. This is a quantity related to the (inverse) parallel scale of magnetic-mirror structures (hereafter called {\it mirroring curvature}, in analogy with the actual field-line curvature $K_\parallel$) whose statistics also strongly depends on the properties of injected large-scale fluctuations.

\begin{figure*}[t!]
    \centering
    \includegraphics[width=0.495\linewidth]{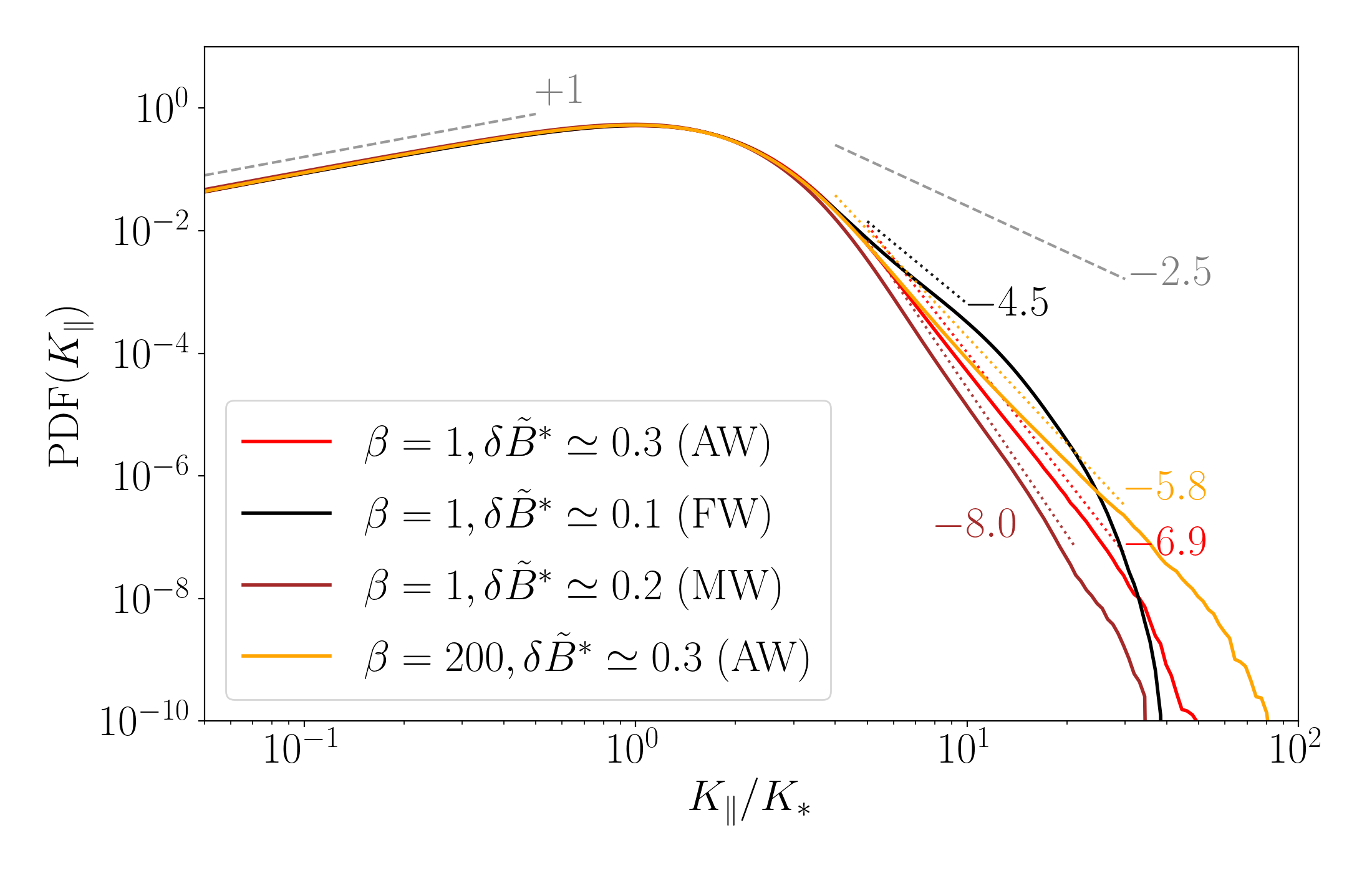}
    \includegraphics[width=0.495\linewidth]{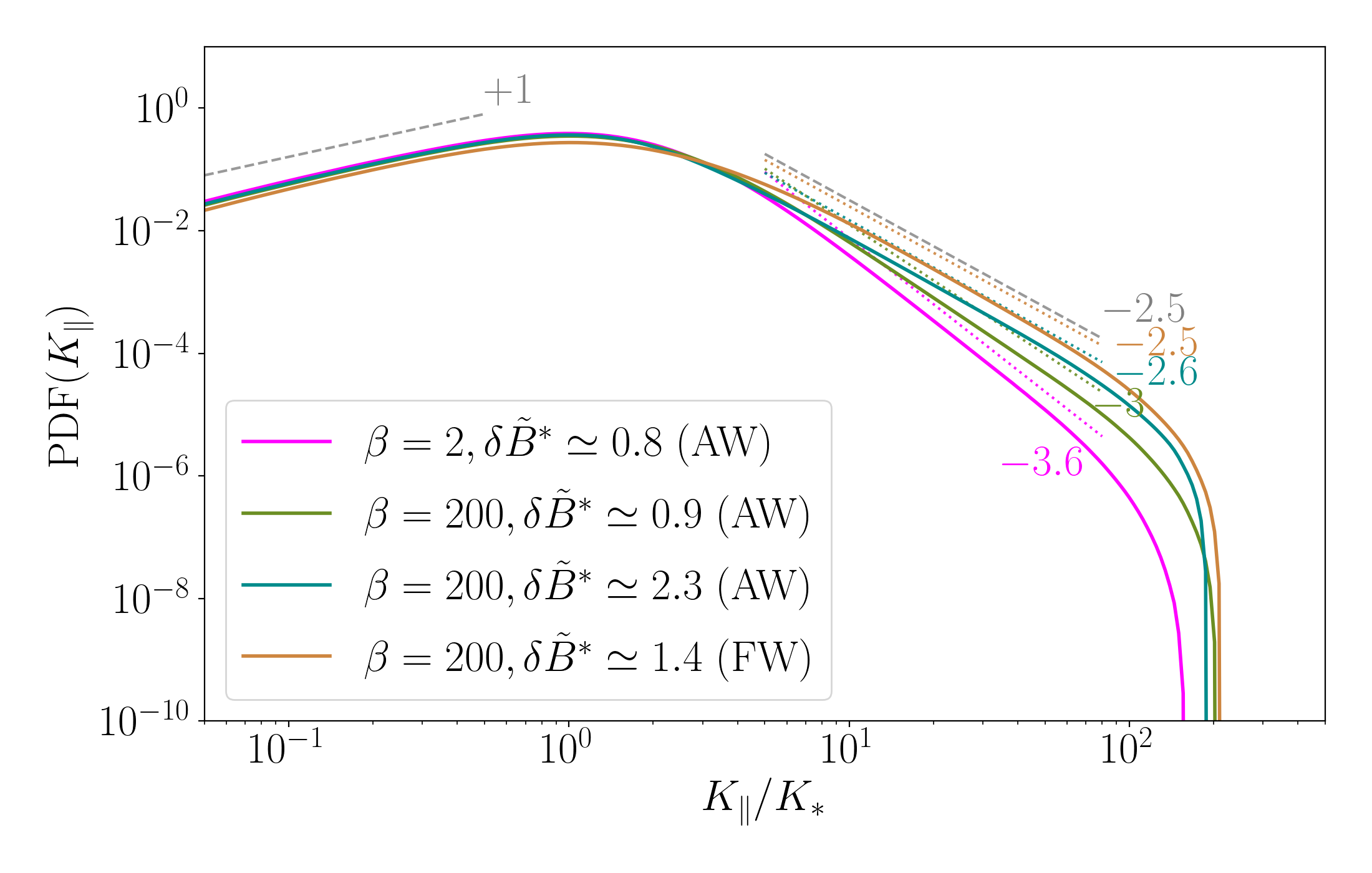}
    \caption{PDF of the magnetic-field curvature $K_\parallel$ versus $K_\parallel/K_*$ ($K_*$ is the value of $K_\parallel$ that maximizes the PDF), averaged over times $1 \lesssim t/t^* \lesssim 1.25$. {\it Left panel:} small-amplitude turbulence ($\delta B/B_0\ll1$).
    {\it Right panel:} large-amplitude turbulence ($\delta B/B_0\gtrsim1$). Power-law fits and indices are shown using dotted lines and numbers of the corresponding colors. 
    }
    \label{fig:PDF_K}
\end{figure*}
\begin{figure*}[th!]
    \centering
    \includegraphics[width=0.49\linewidth]{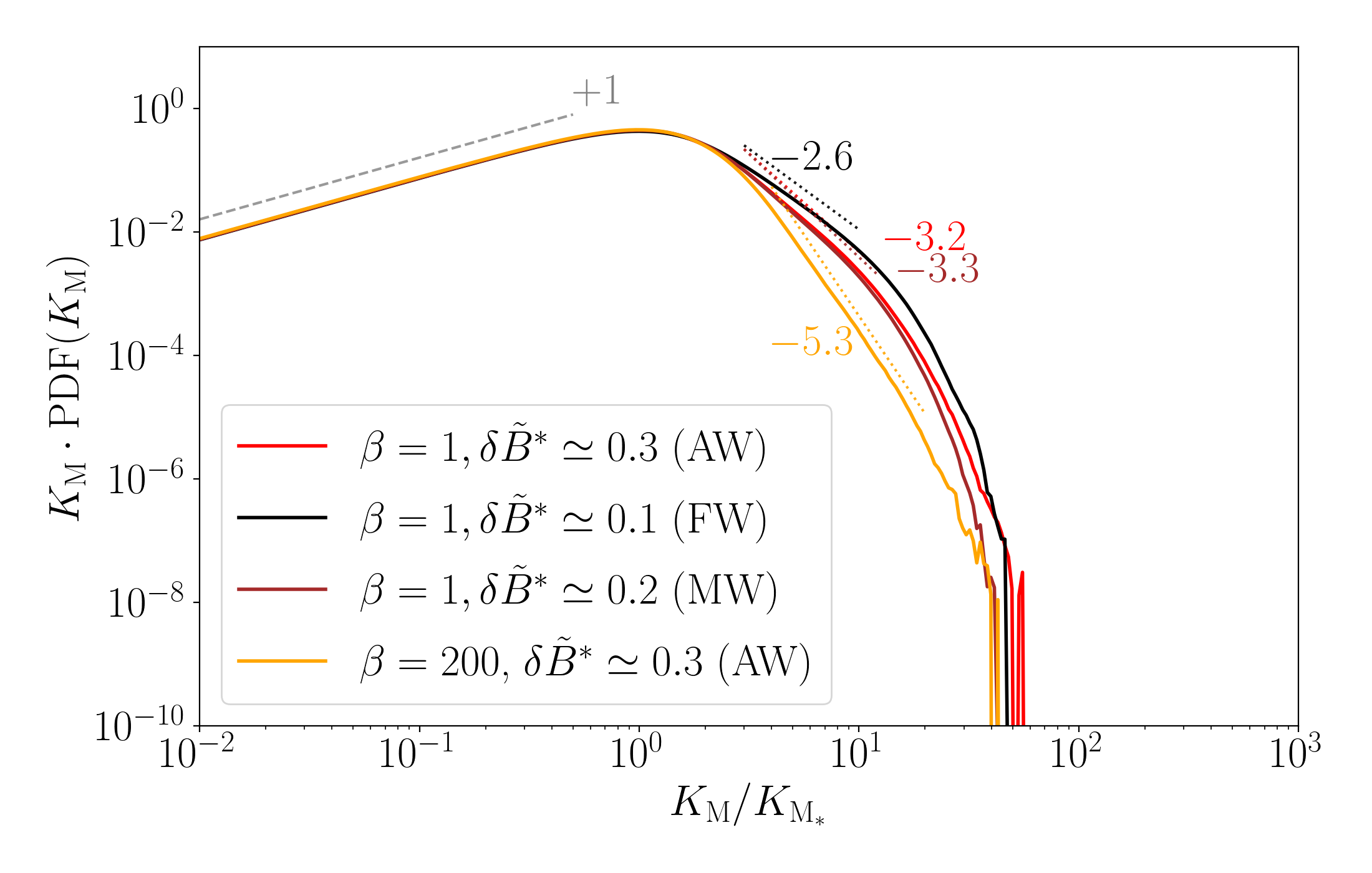}
    \includegraphics[width=0.49\linewidth]{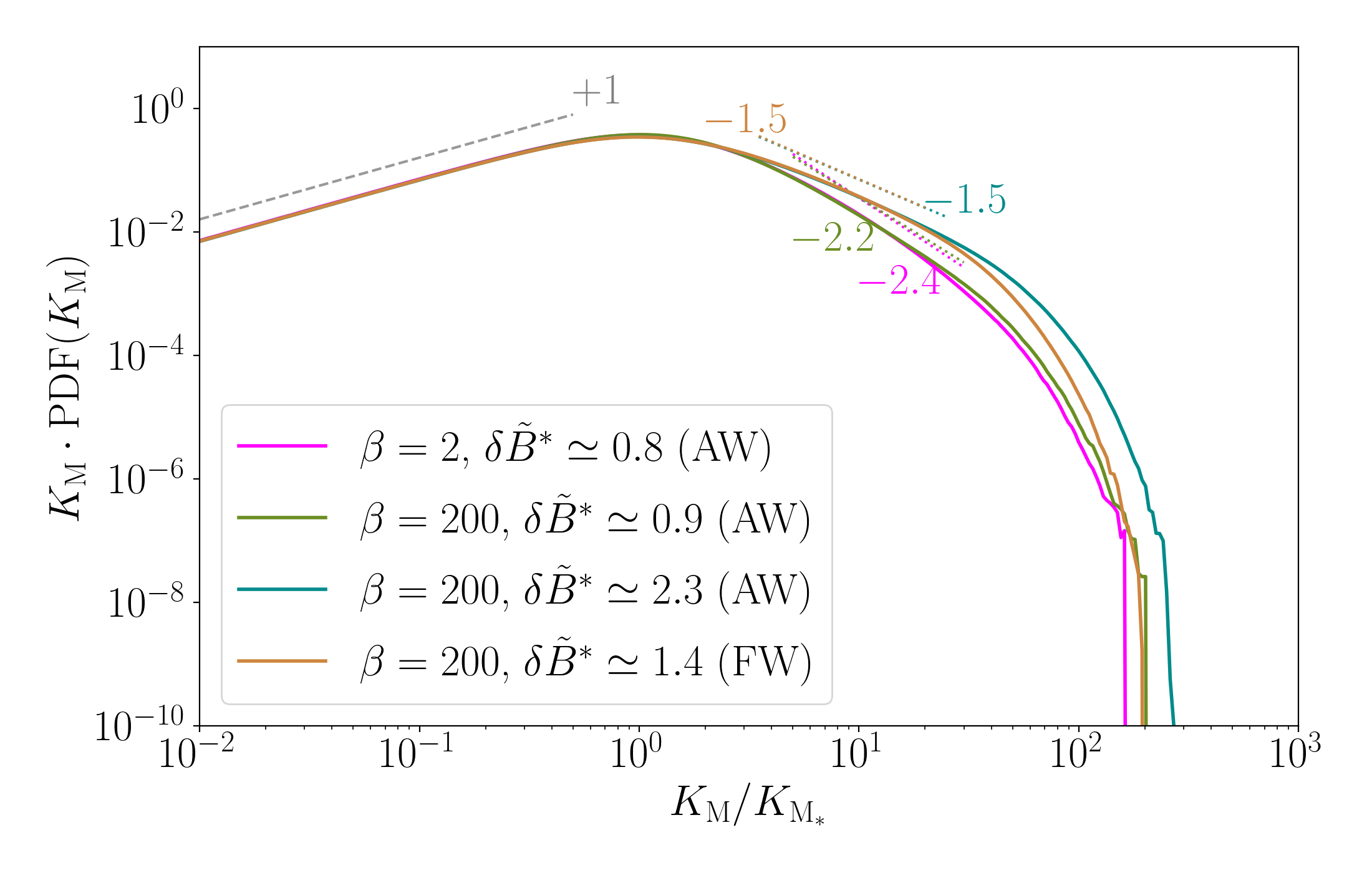}
    \caption{PDF of the field-aligned gradient of magnetic-field strength $K_\mathrm{M}$ (multiplied by $K_\mathrm{M}$ for better visualization) versus $K_\mathrm{M} / K_\mathrm{M_*}$ ($K_\mathrm{M_*}$ is the $K_\mathrm{M}$ value that maximizes $K_\mathrm{M} \cdot$ PDF$(K_\mathrm{M})$),  averaged over $1 \lesssim t/t^* \lesssim 1.25$. {\it Left panel:} small-amplitude turbulence ($\delta B/B_0\ll1$). {\it Right panel:} large-amplitude turbulence ($\delta B/B_0\gtrsim1$). Power-law fits and indices are color-coded as in Figure~\ref{fig:PDF_K}.}
    \label{fig:mirror}
\end{figure*}

Figure \ref{fig:PDF_K} shows the probability density functions (PDFs) of $K_\parallel$ averaged over the interval $1 \lesssim t/t^* \lesssim 1.25$ and normalized by the value $K_*$ at which each PDF peaks.
The left panel shows the results for small-amplitude fluctuations ($\delta B/B_0\ll1$), while the right panel reports large-amplitude injection cases ($\delta B/B_0\gtrsim 1$).
At low curvatures, $K_\|/K_*\ll1$, all cases develop a $\mathrm{PDF}(K_\|)\propto K_\parallel$ scaling, regardless of the nature and amplitude of the injected fluctuations or of the plasma $\beta$. This may be consistent with the fact that low-curvature regions are qualitatively correlated with small variations of the magnetic-field modulus~\citep[][]{Yang2019}. In fact, within these regions of strong guide field, the specific polarization of $\delta\bb{B}$ seems to be irrelevant for what concerns the curvature statistics~\citep[e.g., see also][]{Bouchet_arXiv2026}.
Curvature statistics seems to become sensitive to the properties of large-scale initial fluctuations only for large curvatures, $K_\|/K_*\gg1$.
In fact, $\mathrm{PDF}(K_\|)$ develops a steep power law for small-amplitude Alfv\'enic or mixed-wave initial fluctuations at $\beta=1$, i.e., $\mathrm{PDF}(K_\|)\propto K_\|^{\alpha}$ with $\alpha\simeq -6.9$ and $\alpha\simeq -8.0$, respectively (red and brown lines, respectively). On the other hand, when the large-scale initial fluctuations consist of exclusively small-amplitude fast-magnetosonic perturbations, the distribution of $K_\|$ exhibits a flatter power law with an index of approximately $\alpha\simeq-4.5$, before steepening again for $K_\parallel/K_* \gtrsim 10$ (black line).
This excess of large-curvature events is likely associated with the presence of shocks (see Section \ref{sec:structures}).
The high-$K_\|$ power-law tail of the PDF is also strongly affected by the plasma $\beta$ and the fluctuations' amplitude.
Low-amplitude Alfv\'enic injection yields a noticeably harder and more extended power law when passing from $\beta=1$ to $\beta=200$, i.e., decreasing the plasma compressibility (left panel of Figure \ref{fig:PDF_K}, red and orange line, respectively).
The power law $\mathrm{PDF}(K_\|)\propto K_\|^\alpha$ becomes significantly shallower when increasing $\delta B/B_0\gtrsim1$ and at $\beta>1$ (right panel of Figure \ref{fig:PDF_K}), eventually approaching the ideal power law, $\alpha\to-2.5$, that was predicted and observed in specific regimes~\citep{Yang2019,Bandyopadhyay2020,Lemoine2023,Kempski2023}.
Slightly different dependence on the plasma $\beta$ and injection type is instead found for the perpendicular curvature, $K_\perp=|\hat{\bb{b}}\times(\hat{\bb{b}}\times\bb{\nabla}\ln B)|$ (see Appendix~\ref{app:Kperp}). 

Figure~\ref{fig:mirror} shows the PDF of $K_\mathrm{M}$
multiplied by $K_\mathrm{M}$ for visualization purposes, i.e., $K_{\rm M}\cdot\mathrm{PDF}(K_{\rm M})$, averaged over the time interval $1 \lesssim t/t^* \lesssim 1.25$, and plotted versus $K_{\rm M}/K_{\rm M*}$ ($K_{\mathrm{M}*}$ is the value at which $K_\mathrm{M} \cdot \mathrm{PDF}(K_\mathrm{M})$ peaks).
As found for the PDF of the parallel curvature, also the low-$K_{\rm M}$ end of $\mathrm{PDF}(K_\mathrm{M})$, i.e., $K_{\rm M}/K_{\rm M*}\ll1$, is independent of the nature and amplitude of the injected fluctuations, as well as of the plasma compressibility, showing a flat distribution ${\rm PDF}(K_{\rm M})\propto K_{\rm M}^0$ (so that $K_{\rm M}\cdot{\rm PDF}(K_{\rm M})\propto K_{\rm M}$, in the Figure). 
Significant differences in the PDF of the mirroring curvature emerge only in its high-$K_{\rm M}$ tail, depending on turbulence amplitude $\delta B/B_0$ and on plasma compressibility (controlled by $\beta$).
Even for small-amplitude turbulence ($\delta B/B_0\ll1$; left panel), if compressible effects are important (i.e., $\beta=1$; red, brown, and black lines) the PDF of mirroring curvature develops a non-negligible power-law tail at high-$K_{\rm M}$, namely ${\rm PDF}(K_{\rm M})\propto K_{\rm M}^\alpha$ with $\alpha\approx-4.3$ and $\alpha\approx-3.6$ (where fast-magnetosonic injection, in black, exhibits the harder power law, despite having an rms fluctuation amplitude that is lower than the other two cases). This high-$K_{\rm M}$ power-law tail drastically steepens when low-amplitude Alfv\'enic fluctuations are injected into a plasma with lower compressibility (i.e., $\beta=200$; orange line).
At larger turbulence amplitudes ($\delta B/B_0\gtrsim 1$; right panel), the power-law tail at high-$K_{\rm M}$ becomes significantly harder, i.e., ${\rm PDF}(K_{\rm M})\propto K_{\rm M}^\alpha$ with $\alpha\gtrsim-3.4$, eventually approaching $\alpha\to-2.5$ for both Alfv\'enic and fast-magnetosonic injection with $\delta B/B_0\gg1$. However, the plasma beta seems to play a less relevant role than the fluctuations' amplitude, once large-enough amplitudes are reached: in fact, for Alfv\'enic injection with $\delta B/B_0\sim1$, the power law does not change significantly between the $\beta=2$ and $\beta=200$ cases (magenta and olive green lines in the right panel of Figure~\ref{fig:mirror}, respectively).
The mirroring curvature seems to be sensitive also to the nature of injected fluctuations: at high-$K_{\rm M}$, fast-magnetosonic injection develops the same power law as the Alfv\'enic case ($\alpha\approx -2.5$) despite a lower turbulent level than the latter (although the power-law tail extends to larger $K_{\rm M}$ values in the Alfv\'enic case).

\begin{figure*}[t]
    \centering
    \includegraphics[width=0.49\linewidth]{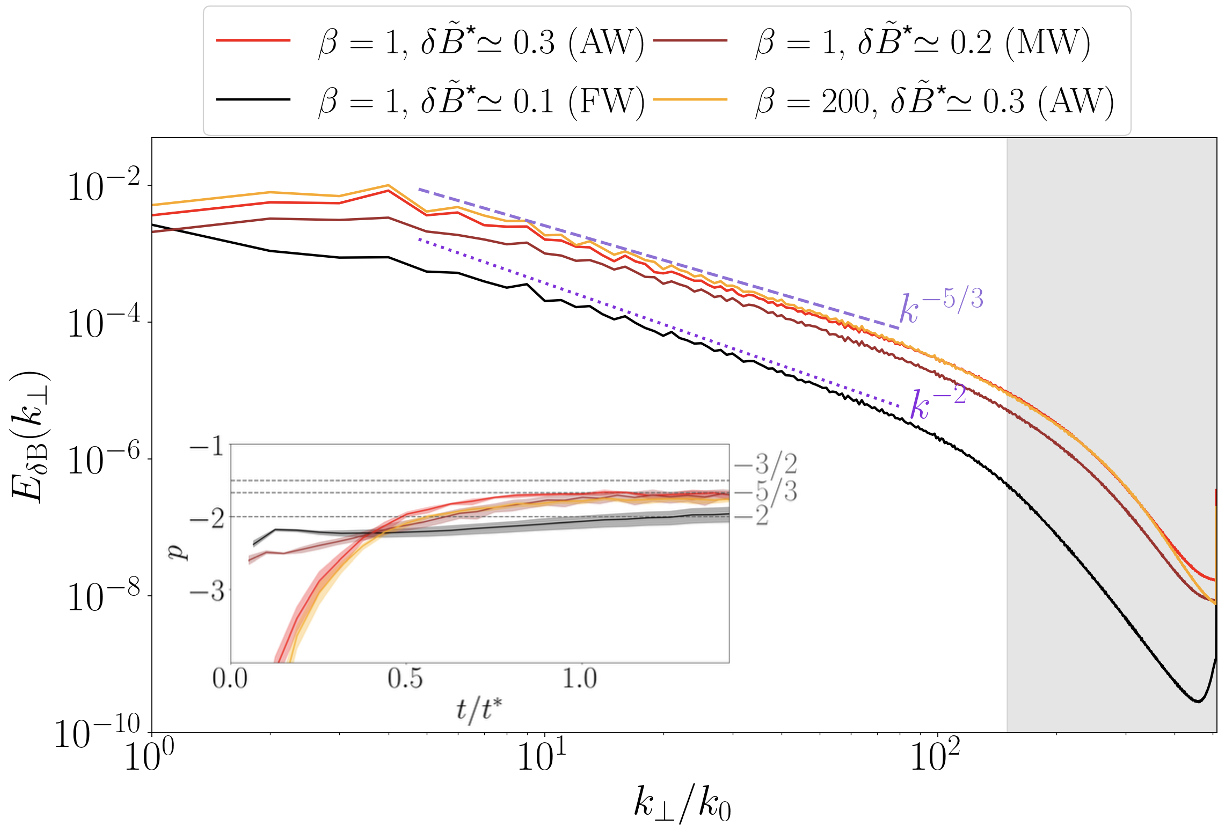}
    \includegraphics[width=0.49\linewidth]{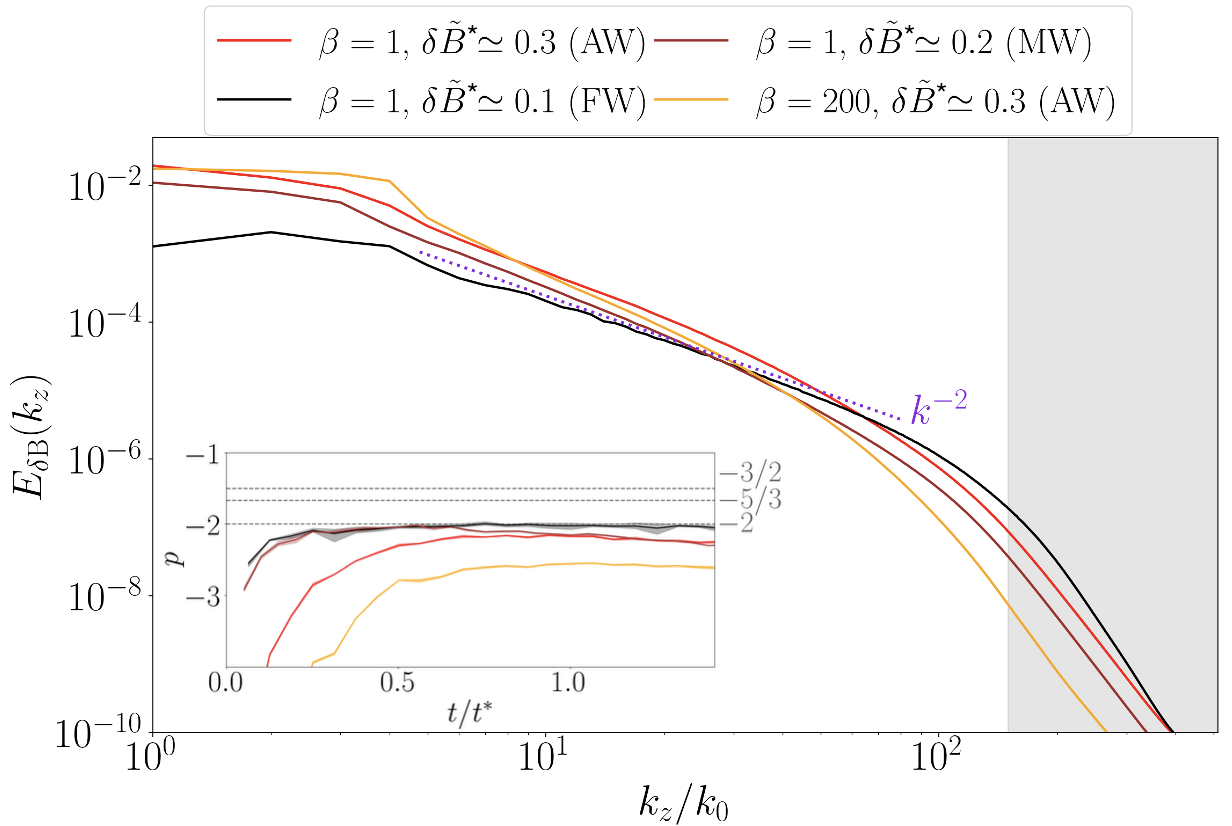}
    
    \caption{Magnetic-field spectra as a function of $k_\perp$ (left) and $k_z$ (right), averaged over times $1 \lesssim t/t^* \lesssim 1.25$, for low-amplitude turbulence ($\delta B/B_0\ll1$). Power laws are provided for reference. The gray shaded area indicates the dissipative range. The insets show the time evolution of the inertial-range spectral index $p$ ($E_\mathrm{\delta B} (k) \propto k^p$) for each case.}
    \label{fig:spectra}
\end{figure*}

\begin{figure*}[t]
    \centering
    \includegraphics[width=0.49\linewidth]{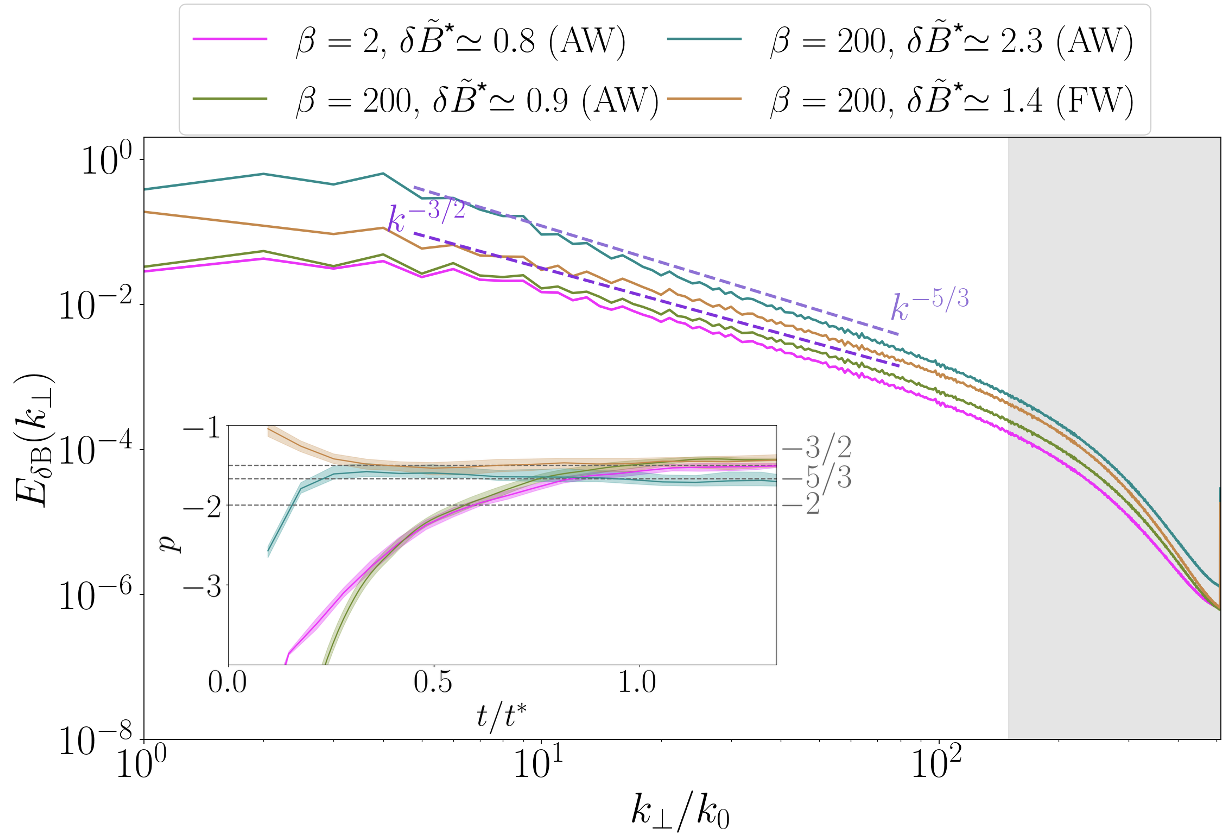}
    \includegraphics[width=0.495\linewidth]{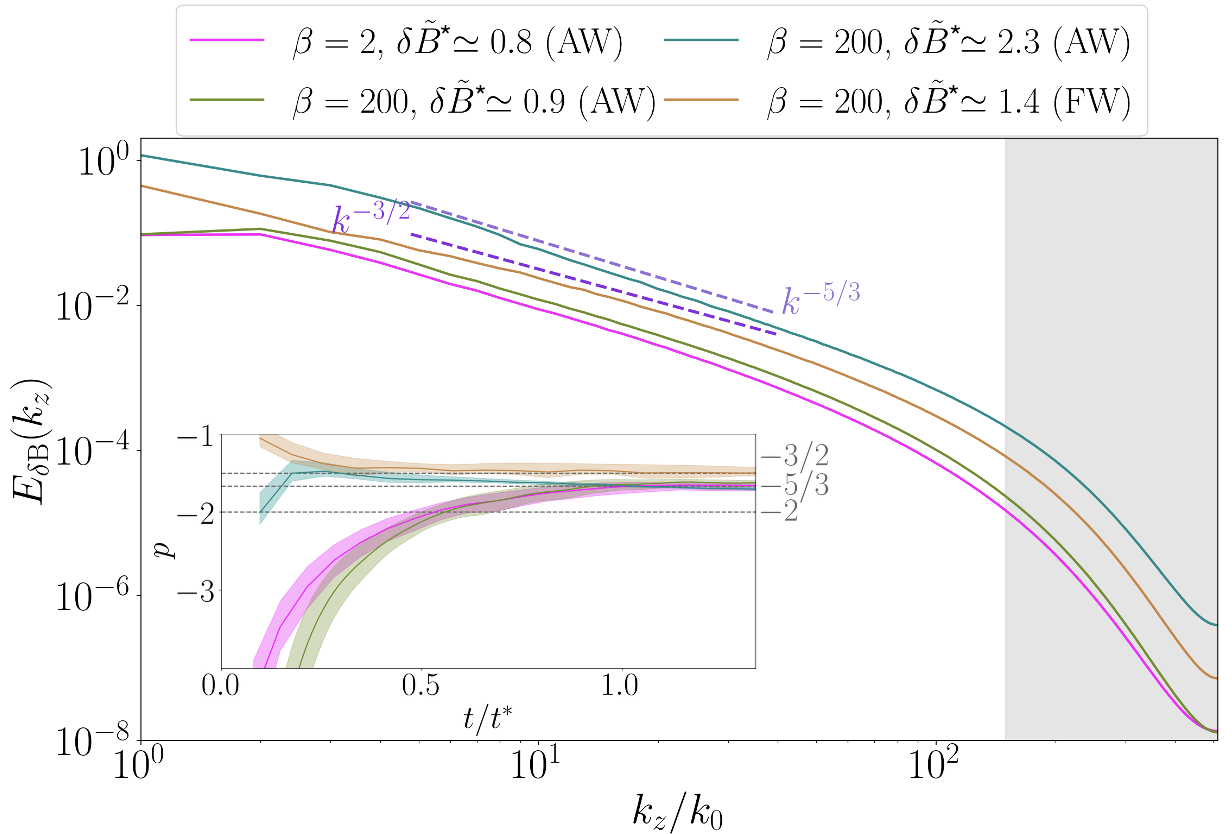}
    \caption{Same as Figure \ref{fig:spectra}, but for large-amplitude turbulence ($\delta B/B_0\gtrsim1$).}
    \label{fig:spectra_high}
\end{figure*}

\section{Spectral properties of magnetic-field fluctuations}
\label{sec:results_spectra}

In standard quasi-linear theory (QLT) of cosmic-ray diffusion, the scattering efficiency is mainly determined by the power spectrum of magnetic fluctuations and its intrinsic anisotropy~\citep[e.g.,][and references therein]{Chandran2000,Shalchi2009,Blasi2013}. Therefore, we explore how these features depend on the nature and properties of large-scale injection.

\subsection{Power spectra}
\label{subsec:spec_B}

Figure \ref{fig:spectra} shows the magnetic field fluctuation spectra as functions of $k_\perp$ (left) and $k_z$ (right) for the different injection types, averaged over the interval $1 \lesssim t / t^* \lesssim 1.25$.
For low-amplitude injection at $\beta=1$, initializing Alfv\'enic (red) or mixed (brown) fluctuations results in the same anisotropic spectrum of fully developed turbulence, with a power law roughly consistent with $ k_\perp^{-5/3}$ perpendicular to $\bb{B}_0$ and a parallel spectrum that is close to $k_z^{-2}$ (but only in the range $k_z/k_0\lesssim15$, while it is noticeably steeper at smaller scales). This is qualitatively consistent with the intrinsic anisotropy of standard Alfv\'enic turbulence by \citet{Goldreich1995} (see also Section~\ref{subsec:anis_B}).
On the other hand, fast-magnetosonic injection (black) exhibits a more isotropic $\propto k^{-2}$ spectrum (i.e., steeper than the other two cases in the perpendicular direction, but noticeably more extended towards smaller scales along $\bb{B}_0$). 
Such spectrum is likely associated to the presence of shock-like structures (see Section~\ref{sec:structures}), which are a signature of a weaker turbulent regime (see Appendix~\ref{app:chi}) and survive in fully developed turbulence only when the initial fluctuations consist exclusively of fast-magnetosonic perturbations, while they quickly dissipate for mixed-wave injection.
The above analysis is supported by the spectra of fluctuations decomposed into the three MHD modes (see Appendix \ref{app:modes_Helm}) and is in agreement with recent numerical simulations~\citep{Makwana2020} reporting a $k_z^{-2}$ scaling for fast modes and a steeper parallel spectrum at small scales for the Alfvénic and slow modes. 
Concurrently, if the same small-amplitude Alfv\'enic fluctuations that were initialized at $\beta=1$ (red line) are now injected in a less compressible plasma, i.e., $\beta=200$ (orange line), the perpendicular spectrum is still $\propto k_\perp^{-5/3}$, but the parallel spectrum is noticeably steeper than $-2$ (i.e., roughly $\propto k_z^{-3}$). This could be due to an enhanced numerical dissipation in the strongly subsonic regime $M_{\rm s}\ll1$ \citep{Watt2025}.
However, this point requires further investigation that is beyond the scope of the present work.

\begin{figure*}[t!]
    \centering
    \includegraphics[width=0.49\linewidth]{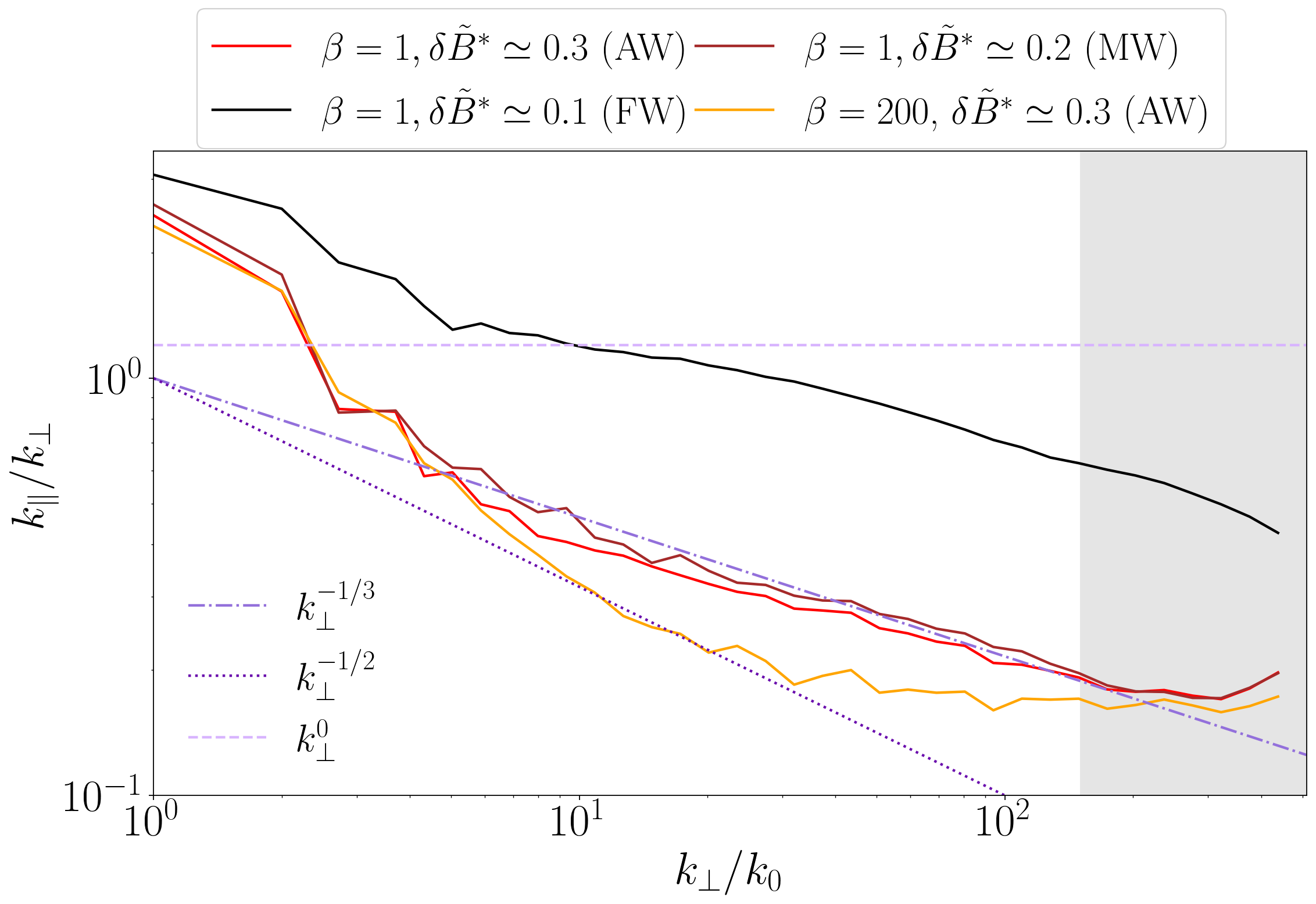}
    \includegraphics[width=0.49\linewidth]{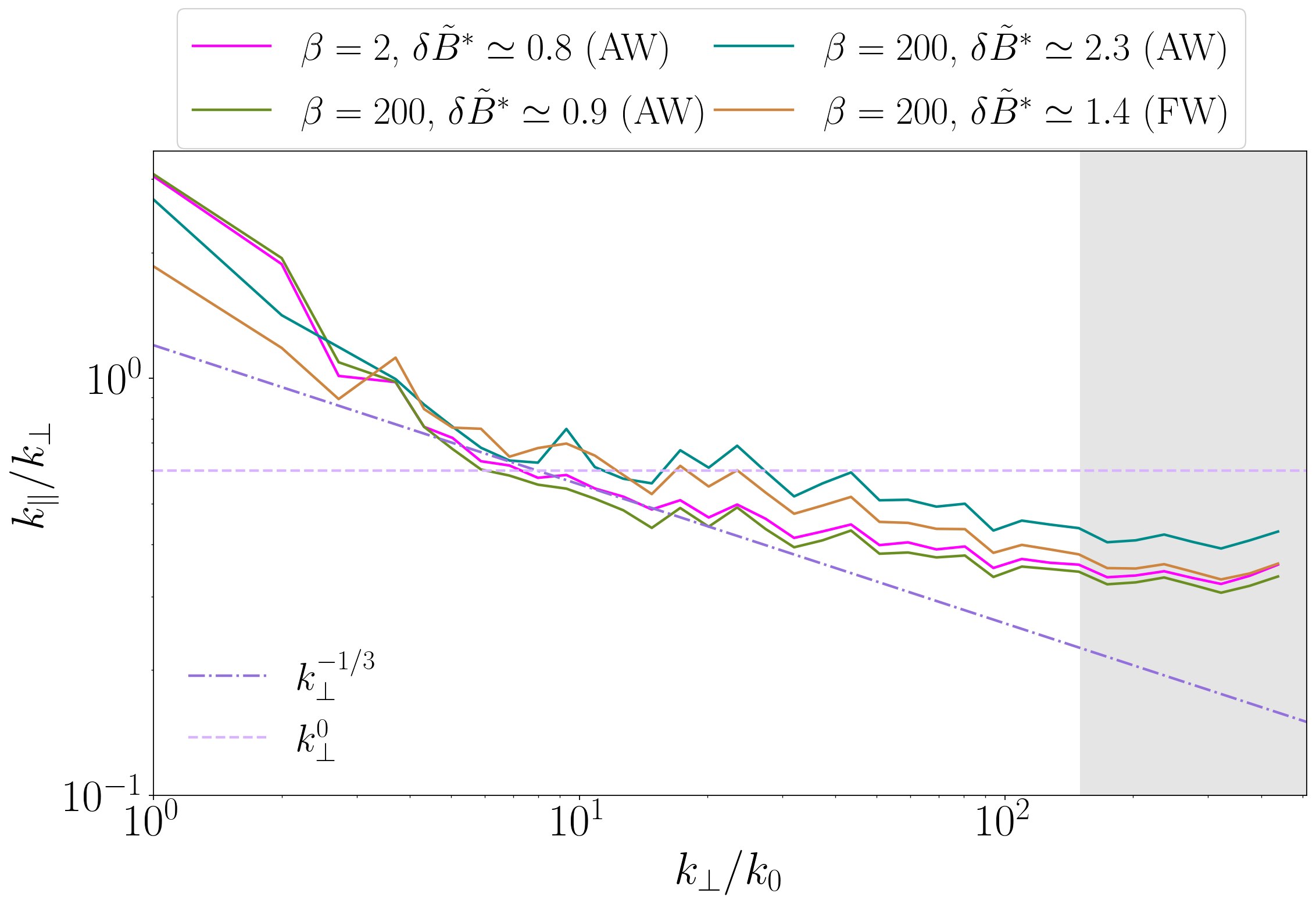}
    \caption{Anisotropy of magnetic-field fluctuations $k_\parallel/k_\perp$ versus $k_\perp$ at $t\approx t^*$, along with reference power laws. The gray area indicates the dissipative range. \textit{Left panel:} small-amplitude turbulence ($\delta B/B_0\ll1$). \textit{Right panel:} large-amplitude turbulence ($\delta B/B_0\gtrsim1$).}
    \label{fig:anis_B}
\end{figure*}

Figure~\ref{fig:spectra_high} reports the case of larger-amplitude initial fluctuations (i.e., $\delta B/B_0\gtrsim1$). In this regime, the spectra obviously become more isotropic. For very large amplitude Alfv\'enic injection (dark cyan), fully developed turbulence is roughly consistent with a Kolmogorov-like $k^{-5/3}$ spectrum.
On the other hand, very large amplitude fast-magnetosonic injection (ocher), yields an isotropic Iroshnikov-Kraichnan-like spectrum (i.e., $\propto k^{-3/2}$).
Interestingly, Alfv\'enic injection with $\delta B/B_0\sim1$ at both $\beta=2$ and $\beta=200$ (magenta and olive green, respectively) still develop a small (but noticeable) degree of anisotropy in the spectra (see also Section~\ref{subsec:anis_B}): both cases exhibit a spectrum that is very close to $-3/2$ in the perpendicular direction and a slightly steeper parallel spectrum, consistent with a $-5/3$ power law. 
While an isotropic $k^{-3/2}$ spectrum would be consistent with the Iroshnikov-Kraichnan theory of Alfv\'enic turbulence~\citep{Iroshnikov1963,Kraichnan1965}, these spectra are not exactly isotropic. On the other hand, an anisotropic spectrum with a $k_\perp^{-3/2}$ scaling would emerge in Alfv\'enic turbulence due to dynamic alignment~\citep{Boldyrev2006}, but in such a scenario it should be accompanied by a steep $-2$ spectrum along $\bb{B}_0$ (which is not the case here).
However, as for the $\beta=200$ case in Figure~\ref{fig:spectra}, this (slight) steepening at $\beta>1$ could be due to enhanced numerical dissipation in the subsonic regime~\citep{Watt2025}.

The nature of fluctuations' spectra for some cases is further discussed with mode versus Helmholtz decompositions as well as with frequency--wavenumber analysis in Appendices \ref{app:modes_Helm} and \ref{app:omegak}, respectively.
Furthermore, an analysis of the scale-dependent nonlinear parameter $\chi = k_\perp \delta v_\perp(k_\perp)[k_\parallel(k_\perp) v_A]^{-1}$ is reported in Appendix \ref{app:chi}.
The possible role of dynamic alignment will be the focus of a dedicated future work.

\subsection{Spectral anisotropy}\label{subsec:anis_B}

To quantify the scale-dependent anisotropy 
$k_\parallel(k_\perp)$ of turbulent fluctuations, we follow the approach of \cite{Cho2002a}. Given a vector field $\bb{F}$ (e.g., $\bb{F}=\bb{v}$, $\bb{B}$), its anisotropy at scale $\ell_\perp\sim k_\perp^{-1}$ with respect to the mean-magnetic-field direction at such scale is computed as follows: we define a large-scale magnetic field $\bb{B}_{\rm L}$ as the low-pass filter of $\bb{B}$ with $|\bb{k}_\perp'|\leq k_\perp/2$, and a small-scale fluctuating field $\bb{f}_{\ell_\perp}$ at scale $\ell_\perp$ as the high-pass filter of $\delta\bb{F}=\bb{F}-\langle\bb{F}\rangle$ with $|\bb{k}_\perp'|\geq k_\perp$.
For each shell $k_\perp$, the anisotropy of $\delta\bb{F}$ with respect to $\bb{B}_L$ is then computed in Fourier space as
\begin{equation}    
    k_\parallel(k_\perp) \approx 
    \sqrt{\frac{\sum_{k_\perp}|\{\bb{B}_{L}\cdot\nabla\bb{f}_{\ell_\perp}\}_{\bb{k}}|^2}{|\bb{B}_L|^2\,\sum_{k_\perp}|\{\bb{f}_{\ell_\perp}\}_{\bb{k}}|^2}\,}
    , 
\label{eq:anis}
\end{equation}
where $\sum_{k_\perp}$ represents the ring summation over the wavevectors ($k_x$, $k_y$) belonging to that $k_\perp$ shell.

Figure~\ref{fig:anis_B} shows the wavevector anisotropy, $k_\parallel / k_\perp$ versus $k_\perp$, of magnetic-field fluctuations $\delta B$ for different cases of low-amplitude ($\delta B/B_0\ll1$; left panel) and large-amplitude ($\delta B/B_0\gtrsim1$; right panel) turbulence at $t\approx t^*$. 
At $\beta=1$, when low-amplitude fluctuations of Alfvénic (red) or mixed (brown) type are injected, the anisotropy rapidly increases across injection scales ($k_\perp/k_0\lesssim5$) and $k_\|/k_\perp$ eventually follows the $k_\perp^{-1/3}$ scaling in the inertial range (corresponding $k_\parallel \propto k_\perp^{2/3}$), while for low-amplitude fast-magnetosonic injection (black) the anisotropy increases much slowly in the first part of the inertial range (i.e., $5\lesssim k_\perp/k_0\lesssim25$), remaining close to the $k_\|/k_\perp\sim 1$ level and then it increases for $k_\perp/k_0\gtrsim30$ (roughly as $k_\|\propto k_\perp^{2/3}$).  
When low-amplitude Alfv\'enic fluctuations are injected in a lower-compressibility $\beta=200$ plasma (orange), the parallel cascade is suppressed (see Figure~\ref{fig:spectra}), and this is reflected in the anisotropy: $k_\|/k_\perp$ initially exhibits a very steep scaling up to $k_\perp/k_0 \lesssim 10$, almost consistent with $k_\|\approx\mathrm{const.}$, which would be typical of weak turbulence~\citep[however, the perpendicular spectrum in Figure~\ref{fig:spectra} does not agree with the $k_\perp^{-2}$ scaling of weak Alfv\'enic turbulence, e.g.,][]{NgBhattacharjee1997,Galtier2000}. At intermediate scales ($10\lesssim k_\perp/k_0\lesssim40$), the anisotropy scales nearly as $k_\|\propto k_\perp^{1/2}$, before reaching a constant-anisotropy state $k_\|\propto k_\perp$ at smaller scales. A $k_\|\propto k_\perp^{1/2}$ scaling that transitions to constant anisotropy is suggestive of a scenario where dynamic alignment~\citep{Boldyrev2006} gets eventually interrupted at small scales~\citep[see discussion in][and references therein]{Beresnyak2014,Schekochihin2022}. The change in small-scale anisotropy might be due to magnetic reconnection, which can reduce the anisotropy of Alfv\'enic turbulence~\citep[e.g.,][]{LoureiroBoldyrevPRL2017,MalletMNRAS2017,Cerri2022}. However, an in-depth investigation of these properties is out of the scope of the present study and will be presented in a future dedicated work.

At larger amplitudes, $\delta B/B_0\gtrsim1$ (right panel of Figure~\ref{fig:anis_B}), anisotropy below the injection scales (i.e., $k_\perp/k_0>4$) is clearly weaker and essentially insensitive to the nature of the injected fluctuations or to the plasma $\beta$ (i.e., the behavior of $k_\|/k_\perp$ is determined only by the level $\delta B/B_0$ of turbulent fluctuations). For turbulence with $\delta B/B_0>1$ (dark cyan and ocher lines), the anisotropy in the range $5 \lesssim k_\perp/k_0 \lesssim 30$ follows roughly a $k_\|\propto k_\perp$ scaling and then slightly increases at small scales (nearly as $k_\|\propto k_\perp^{2/3}$ in the range $30\lesssim k_\perp/k_0\lesssim90$). On the other hand, when $\delta B/B_0\sim 1$ (magenta and olive green) the effect of $\bb{B}_0$ is not completely negligible and the first part of the inertial range develops an anisotropy roughly consistent with $k_\|\propto k_\perp^{2/3}$, before smoothly achieving a constant anisotropy at dissipation scales.

\begin{figure}[t!]
    \centering
    \includegraphics[width=1.0\linewidth]{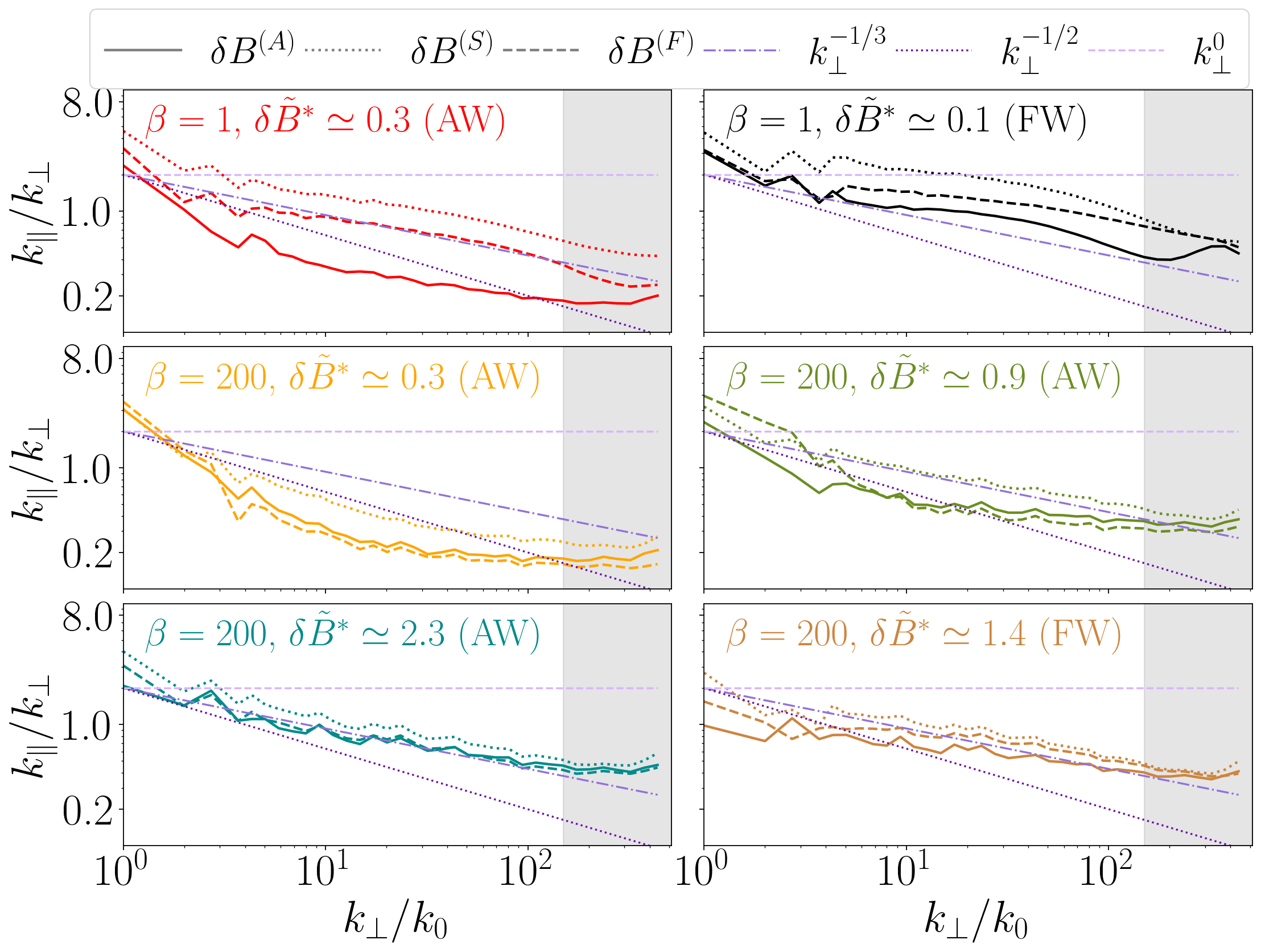}
        \caption{Scale-dependent anisotropy $k_\|/k_\perp$ of mode-decomposed magnetic fluctuations for different simulations at $t\approx t^*$.}
    \label{fig:anis_Bmodes}
\end{figure}

Figure \ref{fig:anis_Bmodes} shows the scale-dependent anisotropy, $k_\parallel/k_\perp$ as a function of $k_\perp$, of the mode-decomposed magnetic fluctuations for different simulations at $t\approx t^*$, exhibiting interesting differences between the various type of fluctuations.
For low-amplitude Alfv\'enic injection at $\beta=1$ (red lines), the anisotropy of (nonlinearly generated) fast- and slow-magnetosonic fluctuations ($\delta B^{\rm(F)}$ and $\delta B^{\rm(S)}$, respectively) seems essentially decoupled from the one of Alfv\'enic fluctuations ($\delta B^{\rm(A)}$): both magnetosonic modes exhibit a $k_\|\propto k_\perp^{2/3}$ scaling almost through the entire $k_\perp$ range, while $\delta B^{\rm(A)}$ initially develops a stronger anisotropy scaling (roughly consistent with $k_\|\propto k_\perp^{1/2}$), becoming $k_\|\propto k_\perp^{2/3}$ only at $k_\perp/k_0\gtrsim15$. On the other hand, fast-magnetosonic injection in the same regime (black lines) show a stronger coupling between the three modes, i.e., all type of fluctuations develop a similar anisotropy scaling (nearly constant up to $k_\perp/k_0\lesssim 30$). At the same time, when low-amplitude Alfv\'enic fluctuations are injected in a lower-compressibility plasma (i.e., $\beta=200$; orange lines), the slow-magnetosonic mode remains decoupled from the Alfv\'enic component like in the $\beta=1$ case, but the fast-mode exhibits a tighter coupling with the Alfv\'enic counterpart (however, we recall that fast-magnetosonic fluctuations are energetically negligible in this regime, i.e., $E_{\rm tot}^{\rm(F)}/E_{\rm tot}\lesssim1\%$; see Figure~\ref{fig:vHelm_time}). A similar behavior is observed also for Alfv\'enic injection with $\delta B/B_0\sim1$ at $\beta=200$ (olive green lines): the anisotropy of $\delta B^{\rm(S)}$ scales as $k_\|\propto k_\perp^{2/3}$ throughout the $k_\perp$ range and is essentially decoupled from the Alfv\'enic component, while significant coupling between the anisotropy of $\delta B^{\rm(F)}$ and $\delta B^{\rm(A)}$ is observed (and both exhibit a $k_\|\propto k_\perp^{2/3}$ scaling up to $k_\perp/k_0\lesssim 30$, followed by nearly constant anisotropy $k_\|\propto k_\perp$ at $k_\perp/k_0\gtrsim 30$). As soon as $\delta B/B_0>1$, all the three MHD modes are tightly coupled---also due to the intrinsic limitation of the mode decomposition (see Appendix~\ref{app:modes_Helm}). However, it is interesting to notice that, while the anisotropy of the mode-decomposed fluctuations for fast-magnetosonic injection (ocher lines) essentially agrees with the one of the total magnetic fluctuations (ocher line in the right panel of Figure~\ref{fig:anis_B}), the case of Alfv\'enic injection (dark cyan lines) shows an anisotropy scaling consistent with $k_\|\propto k_\perp^{2/3}$ through the whole $k_\perp$ range for all $\delta B^{\rm(m)}$ (m=A,F,S), whereas $\delta B$ clearly developed a range of constant anisotropy at intermediate scales (dark cyan line in the right panel of Figure~\ref{fig:anis_B}). This is likely due to the fact that all the modes are energetically important and that local (in $k_\perp$) differences between $k_\|/k_\perp$ values for the different components combine in a non-trivial way, so that looking at the scaling of the mode-decomposed fluctuations leaves out some information about the whole nonlinear system (see discussion in Appendix~\ref{app:modes_Helm}).

\section{Conclusions}
\label{sec:conclusions}

We have presented a study on how properties of sub-sonic MHD turbulence, that are expected to influence CR transport, depend on some key physical parameters. In particular, we explored the effects of (i) the nature of large-scale injection (considering both nearly incompressible Alfvénic and strongly compressible fast-magnetosonic initial fluctuations), (ii) the fluctuation amplitude (considering both $\delta B/B_0 \ll 1$ and $\delta B/B_0 \gtrsim 1$), and (iii) the plasma compressibility (controlled by the $\beta$ parameter, varied from $\beta\sim1$ to $\beta =200$).\\
The main results and implications can be summarized as follow:
\begin{itemize}
    \item The level of density fluctuations in fully developed turbulence is surprisingly insensitive to the nature of large-scale injection (i.e., Alfv\'enic or fast-magnetosonic), but is determined only by the turbulent sonic Mach number: increasing $M_{\rm s}$ leads to larger $\delta\rho/\rho_0$, whose magnitude is approximately the same for Alfv\'enic and fast-magnetosonic injection. Indeed, when strongly compressive fast-magnetosonic fluctuations are initialized, the system rapidly forms shocks that dissipate the compressible component of the velocity until it reaches a turbulent decay phase that is the same as in the case of nearly incompressible Alfvénic injection.\\ Although this result seems obvious {\it a posteriori}, it is non-trivial, since some other turbulent properties such as the relative importance of compressive and solenoidal velocity fluctuations (i.e., $\delta v_{\rm comp}/\delta v_{\rm sol}$),  the statistics of field-line curvature and mirroring sites indeed critically depend on the nature of injected fluctuations. This also means that both the applicability of incompressible theories to Alfv\'enic turbulence and the relevance of structures (e.g., sharp field-line bends and magnetic mirrors) for CR transport is strongly sensitive to the $M_{\rm s}$ regime (i.e., on a combination of fluctuations' level and plasma $\beta$) and cannot be overlooked.
    
    \item Turbulence arising from small-amplitude ($\delta B/B_0\ll1$) Alfv\'enic and mixed-wave injection is strongly anisotropic, with a $\propto k_\perp^{-5/3}$ spectrum and a parallel spectrum roughly $\propto k_z^{-2}$ (which becomes noticeably steeper than $-2$ for Alfv\'enic injection at $\beta\gg1$, i.e., in the strongly sub-sonic regime). On the other hand, small-amplitude fast-magnetosonic injection leads to a homogeneous turbulent state populated by (weak) shocks and characterized by a quasi-isotropic $\propto k^{-2}$ spectrum. At large amplitudes (i.e., $\delta B/B_0>1$), turbulence tends to develop a nearly isotropic Kolmogorov ($\propto k^{-5/3}$) or Iroshnikov-Kraichnan ($\propto k^{-3/2}$) spectrum for Alfv\'enic or fast-magnetosonic injection, respectively. At the same time, Alfv\'enic turbulence with $\delta B/B_0\sim 1$ still develops a certain degree of anisotropy, characterized by spectra that are possibly consistent with $\propto k_\perp^{-3/2}$ and $\propto k_z^{-5/3}$ power laws.\\ 
    The spectrum of small-amplitude Alfv\'enic turbulence is thus only partially in agreement with the theory of \citet{Goldreich1995}, and compressive fluctuations may play a relevant role in sustaining parallel nonlinearities. At the same time, small-amplitude fast-magnetosonic turbulence is clearly more isotropic than the Alfv\'enic case, but far from the $k^{-3/2}$ spectrum predicted by weak-turbulence theories~\citep{Zakharov1970,Galtier2023} and found in early low-resolution simulations by \citet{Cho2002b,Cho2003} (which, however, investigated regimes very different from the ones presented here, i.e., the low-$\beta$, super-sonic and high-$\beta$, sub-Alfv\'enic regimes). Indeed, \citet{KochurinKuznetsovPRL2024} found that the $k^{-3/2}$ scaling of acoustic turbulence is recovered only in a particular weak regime, while in general such turbulence turns into an ensemble of random shocks described by the $k^{-2}$ Kadomtsev-Petviashvili spectrum~\citep{KadomtsevPetviashvili1973}; such spectrum is also consistent with recent fully kinetic 2D simulations of fast-magnetosonic turbulence in the weak regime~\citep{Ugarov2026}. On the other hand, at large amplitudes the power spectrum is more consistent with theories of homogeneous and isotropic turbulence~\citep{Kolmogorov1941a,Iroshnikov1963,Kraichnan1965}, although hints about the presence of a small degree of anisotropy definitely requires further investigations. These results raise a concern about QLT modeling of CR diffusion coefficients, since the underlying turbulence entering their expression is usually assumed to have either a Goldreich-Sridhar or Iroshnikov-Kraichnan scaling for the Alfv\'enic or fast-magnetosonic spectrum, respectively~\citep[e.g.,][and references therein]{EvoliYan2014,Fornieri2021}.

    \item The statistics of magnetic-field curvature $K_\|=|(\bb{b}\cdot\bb{\nabla})\bb{b}|$ and of mirroring curvature $K_{\rm M}=|\bb{b}\cdot\bb{\nabla}\ln B|$ are sensitive to the fluctuations' amplitude and plasma compressibility. Low-amplitude turbulence and/or high plasma compressibility (i.e., $\delta B/B_0\ll1$ and/or $\beta\sim1$) translates into very steep power-law tails of the PDF of field-line curvature at high $K_\|$; the ideal harder power-law scaling $\mathrm{PDF}(K_\|)\propto K_\|^{-2.5}$ is recovered only in the limit of large-amplitude fluctuations in a low-compressibility plasma (i.e., $\delta B/B_0>1$ and $\beta\gg1$). The PDF of mirroring curvature exhibits a similar behavior to the distribution of field-line curvature and a hard power-law scaling $\mathrm{PDF}(K_{\rm M})\propto K_{\rm M}^{-2.5}$ is achieved in the $\delta B/B_0>1$ limit. For both quantities, in the $\delta B/B_0\ll1$ regime fast-magnetosonic injection seems to produce PDFs with harder power-law tails than Alfv\'enic injection.\\
    The fact that the predicted $-2.5$ power-law scaling for the magnetic-field curvature~\citep{Yang2019} emerges only in the limit of large-amplitude, nearly incompressible turbulence, severely challenges the applicability of CR-transport models based on curvature scattering, since these models assume $\mathrm{PDF}(K_\|)\propto K_\|^{-2.5}$ to obtain the right dependence of the CR mean-free path on their energy~\citep{Lemoine2023,Kempski2023}. Nevertheless, the emergence of mirroring structures in compressible turbulence and the behavior of their statistics may prove significant in complementing curvature scattering for CR confinement in environments characterized by higher plasma compressibility and/or by a non-negligible guide field $B_0$.

\end{itemize}

Our results highlight that many features of sub-sonic compressible MHD turbulence that can directly affect cosmic-ray transport strongly depend on large-scale conditions. The effect of these features is not retained in the presently available CR-transport models and will be investigated in Paper II.
We stress that this work is limited to the $M_{\rm s}\lesssim1$ regime and that more efforts are needed to further explore the parameter space with high-resolution simulations.


\begin{acknowledgements}
      E.P. and S.S.C. are supported by the French government, through the National Research Agency (ANR) grant ``MiCRO” with the reference number ANR-23-CE31-0016. This work was performed using HPC resources from GENCI-TGCC, France (Grants 2026 - AD010416202R2, 2025/2026 - A0180416202, and 2024/2025 - AD010415960). The authors would like to warmly thank Yoann Genolini, Alexandre Marcowith, Philipp Mertsch, Hui Li and Andrea Mignone for many useful discussions, as well as the referee for valuable comments. 

\end{acknowledgements}

%

\bibliographystyle{aa}
\bibliography{refs}







   
  



\appendix
\twocolumn
\nolinenumbers





\section{Arc-polarized Alfvén waves}\label{app:arc}

The arc-polarized Alfv\'en wave is derived under the following conditions: 
\begin{equation}\label{eq:arcpol_AWs_conditions}
    \begin{cases}
      \bb{k} \cdot \delta \bb{B}=0\\
      |\bb{B}| = \text{const}\\
      \delta\bb{v} = \pm  \frac{\delta \bb{B}}{B_0}v_{\rm A}
    \end{cases}
\end{equation}
where $v_{\rm A}=B_0/\sqrt{4\pi\rho_0}$ is the Alfv\'en speed.
The first condition in \eqref{eq:arcpol_AWs_conditions} states that magnetic-field fluctuations lie in a plane perpendicular to $\bb{k}$, so that the solenoidal nature $\bb{\nabla}\cdot\delta\bb{B}=0$ of magnetic-field fluctuations is ensured. The second condition in \eqref{eq:arcpol_AWs_conditions} imposes constant magnetic-field modulus, $|\bb{B}|=|\bb{B}_0+\delta\bb{B}|=\text{const.}$, while the third and last condition in \eqref{eq:arcpol_AWs_conditions} ensures the Alfv\'enic nature of the wave for any amplitude of the fluctuations~\citep{Barnes1974}. 
In the following, we consider a generic wavevector 
\begin{equation}
    \bb{k} = k\,(\cos \theta\,\hat{\bb{e}}_z + \sin\theta\,\sin\alpha\, \hat{\bb{e}}_x + \sin\theta\,\cos\alpha\,\hat{\bb{e}}_y)\,,
\end{equation}    
where $k=|\bb{k}|$, $\theta$ is the angle between the wave vector and $\bb{B}_0$ (assumed to be along $z$), and $\alpha$ is the angle between the projection of $\bb{k}$ onto the plane perpendicular to $\bb{B}_0$ (i.e., $\bb{k}_\perp$) and the $y$ axis. 
We now define a basis ($\hat{\bb{e}}_{\perp, 1}, \hat{\bb{e}}_{\perp, 2}$) in the plane perpendicular to $\bb{k}$, where the magnetic-field fluctuations are confined. 
Imposing that $\hat{\bb{e}}_{\perp, 1}$ lies in the plane spanned by $\bb{k}$ and $\bb{B}_0$, then this unit vector must satisfy the following conditions:
\begin{equation}\label{eq:e1_conditions}
    \begin{cases}
      \hat{\bb{e}}_{\perp, 1} \cdot \bb{k} =0\\
      \hat{\bb{e}}_{\perp, 1} = a_1 \bb{k} + a_2 \bb{B}_0\\
      |\hat{\bb{e}}_{\perp, 1}| = 1
    \end{cases}
\end{equation}
Solving the above system \eqref{eq:e1_conditions} yields
\begin{equation}
    \hat{\bb{e}}_{\perp,1}=(\sin\theta)^{-1}\hat{\bb{e}}_z-(\tan\theta)^{-1}\hat{\bb{k}}\,,
\end{equation}
where $\hat{\bb{k}}=\bb{k}/k$, 
which, substituting the expression for $\bb{k}$, rewrites as
\begin{equation}
\label{eq:e1}
    \hat{\bb{e}}_{\perp, 1} = - \cos\theta\,\sin\alpha\,\hat{\bb{e}}_x - \cos\theta\,\cos\alpha\,\hat{\bb{e}}_y + \sin\theta\, \hat{\bb{e}}_z .
\end{equation}
For the second basis vector, $\hat{\bb{e}}_{\perp, 2}$, we require that it satisfies the following conditions:
\begin{equation}\label{eq:e2_conditions}
    \begin{cases}
      \hat{\bb{e}}_{\perp, 2} = a_3(\bb{k}\times\hat{\bb{e}}_{\perp, 1})\\
      |\hat{\bb{e}}_{\perp, 2}| = 1
    \end{cases}
\end{equation}
i.e., requiring that $\hat{\bb{e}}_{\perp, 2}$ is a unit vector perpendicular to $\hat{\bb{e}}_{\perp,1}$ and such that the triad ($\hat{\bb{e}}_{\perp,1}$, $\hat{\bb{e}}_{\perp,2}$, $\hat{\bb{k}}$) forms a right-handed orthonormal basis. Solving the above system \eqref{eq:e2_conditions} leads to 
\begin{equation}
\label{eq:e2}
    \hat{\bb{e}}_{\perp, 2} = \cos\alpha\, \hat{\bb{e}}_x - \sin\alpha\,\hat{\bb{e}}_y .
\end{equation}
The total magnetic field is then written as 
\begin{equation}
    \bb{B} = B_0\, \hat{\bb{e}}_z +  \delta B_1\, \hat{\bb{e}}_{\perp, 1} + \delta B_2\, \hat{\bb{e}}_{\perp, 2}\,.
\end{equation}
Using \eqref{eq:e1}--\eqref{eq:e2}, its squared modulus writes as
\begin{equation}
\label{eq:B}
    |\bb{B}|^2 = B_0^2 \cos^2 \theta + \delta B_2^2 + (B_0 \sin \theta + \delta B_1)^2\,.
\end{equation}
The component $\delta B_1$ is calculated by imposing the constraints
\begin{equation}\label{eq:constraints_dB_1}
    \begin{cases}
       \langle  |\bb{B}|^2 \rangle = B_0^2 + \langle \delta B_1^2 + \delta B_2^2 \rangle = B_0^2 + \delta B_{\rm rms}^2 = \text{const} \\
      \langle \delta B_1 \rangle = 0
    \end{cases}
\end{equation}
The first constraint in \eqref{eq:constraints_dB_1} can be used to rewrite \eqref{eq:B} as 
$(B_0 \sin \theta + \delta B_1)^2 + B_0^2 \cos^2 \theta + \delta B_2^2 = B_0^2 + \delta B_\mathrm{rms}^2$
and obtain
\begin{equation}
\label{eq:deltaB1}
    \delta B_1 + B_0 \sin \theta = \pm \sqrt{B_0^2 \sin^2 \theta + \delta B_\mathrm{rms}^2 - \delta B_2^2}.
\end{equation}
By rewriting $\delta B_2 = \eta\, B_0\, \cos \phi$, with $\phi=\bb{k}\cdot\bb{x}+\varphi$ and $\varphi$ a random phase, and using \eqref{eq:deltaB1}, the second constraint in \eqref{eq:constraints_dB_1} becomes
\begin{equation}
\int_0^{\frac{\pi}{2}} B_0 \sin\theta\, {\rm d}\phi =
B_0\sqrt{\sin^2\theta + b_\mathrm{rms}^2} \int_0^{\frac{\pi}{2}} \sqrt{1 - \frac{\eta^2 \cos^2 \phi}{\sin^2 \theta + 
b_\mathrm{rms}^2}}\, {\rm d}\phi 
\end{equation}
where we have defined $b_{\rm rms}=\delta B_{\rm rms}/B_0$ for short.
Finally, defining $\gamma = \eta\,(\sin^2 \theta + b_\mathrm{rms}^2)^{-1/2}$, the root-mean-square amplitude $\delta B_\mathrm{rms}$ is obtained by numerically solving the elliptic integral 
\begin{equation}
   E(\gamma) = \int_0^{\frac{\pi}{2}} \sqrt{1 - \gamma^2 \cos ^2 \phi}\, {\rm d} \phi = \frac{\pi}{2} \frac{\sin \theta}{\sqrt{\sin^2 \theta + b_\mathrm{rms}^2}} .
   \label{eq:nonlin_eq}
\end{equation}
Once $\delta B_\mathrm{rms}=b_{\rm rms}\,B_0$ is determined, $\delta B_1$ is obtained from \eqref{eq:deltaB1}, thus fully determining the fluctuations $\delta \bb{B} = \delta B_1 \hat{\bb{e}}_{\perp, 1} + \delta B_2 \hat{\bb{e}}_{\perp, 2}$ and $\delta\bb{v}=\pm\,(\delta\bb{B}/B_0)\,v_{\rm A}$ of an arc-polarized Alfv\'en wave. 
As reported in \cite{DelZanna2001}, the critical angle in Equation \ref{eq:nonlin_eq} is defined by $\sin \theta_c = 2 \eta / \pi$ and represents the minimum propagation angle for which \eqref{eq:nonlin_eq} admits a solution. In our simulations (see Table \ref{tab:ic}), the corresponding critical angles range from $\theta_c \sim 0.5^\circ$ for $\delta B_\mathrm{rms,0}/B_0 \simeq 0.33$ ($\eta = 0.0145$) to $\theta_c \sim 8^\circ$ for $\delta B_\mathrm{rms,0}/B_0 \simeq 5$ ($\eta = 0.22$).

\section{Statistics of perpendicular magnetic-field curvature}
\label{app:Kperp}

\begin{figure*}[t!]
    \centering
    \includegraphics[width=0.49\linewidth]{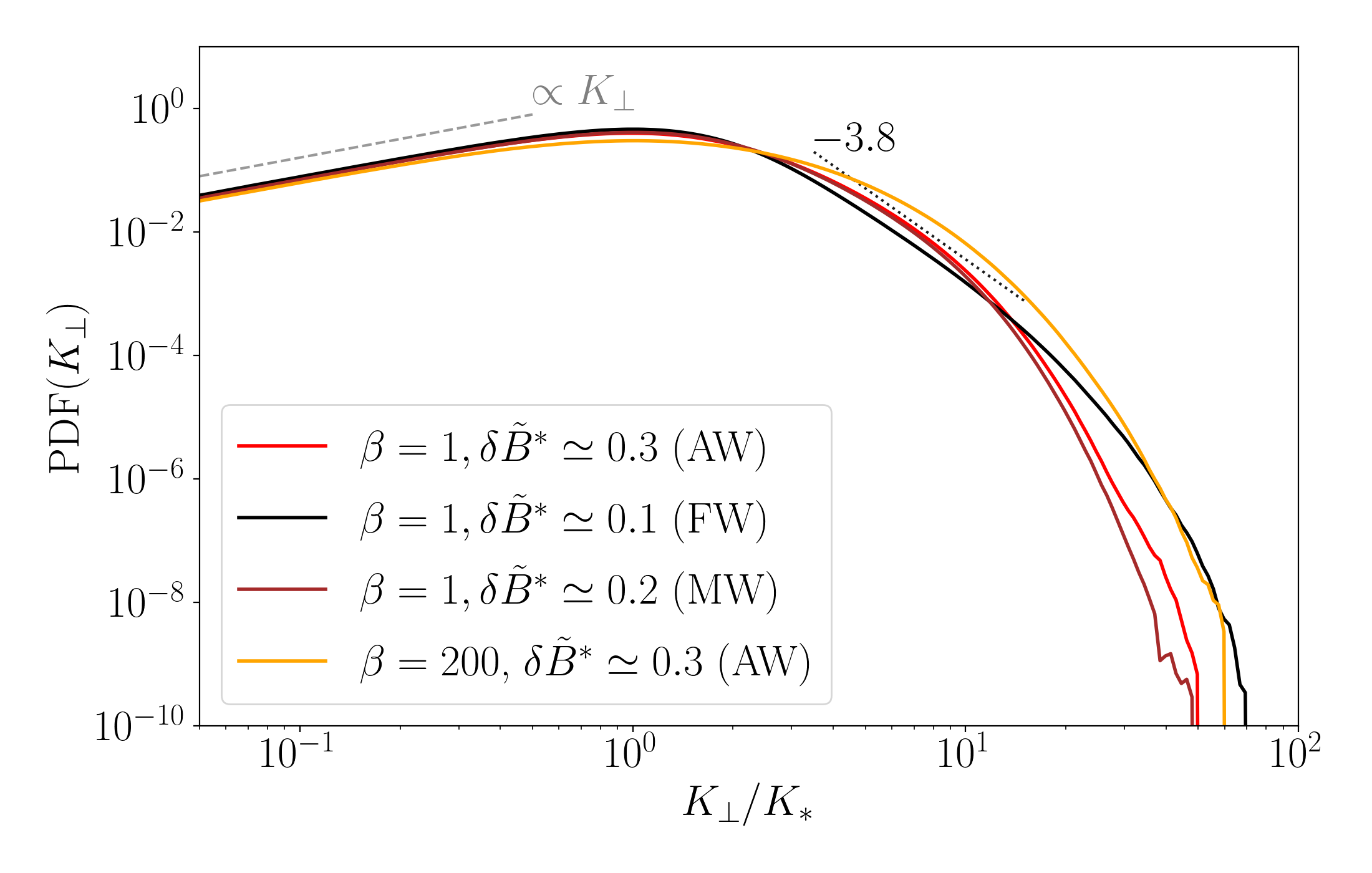}
    \includegraphics[width=0.49\linewidth]{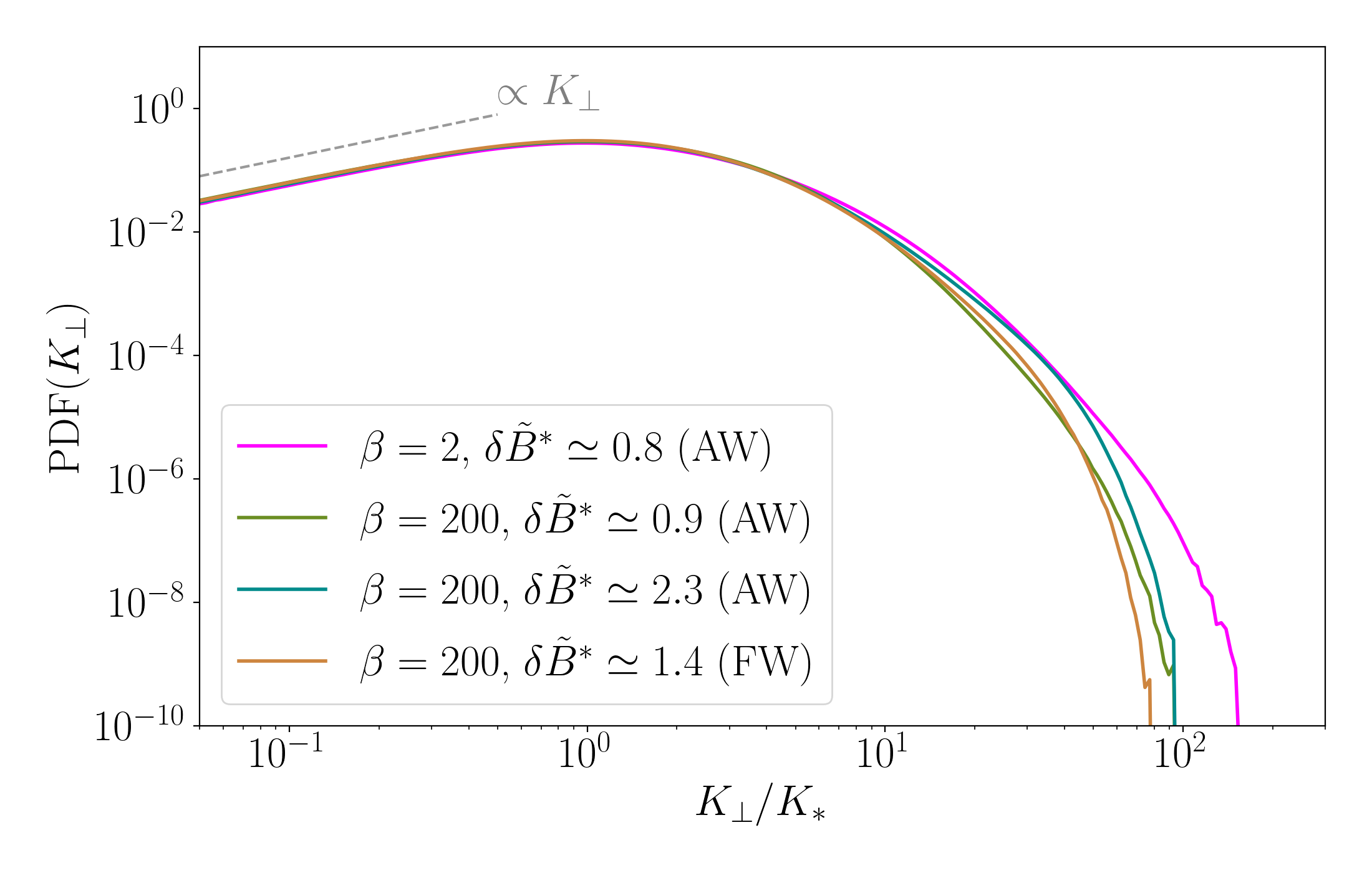}
    \caption{PDF of perpendicular magnetic-field curvature $K_\perp$ versus $K_\perp/K_*$ ($K_*$ is the value that maximizes the PDF), averaged over $1 \lesssim t/t^* \lesssim 1.25$. {\it Left:} $\beta=1$ runs with small-amplitude injection of different nature. A power law $\propto K_\perp$ at lower perpendicular curvature is shown for reference.
    {\it Right:} $\beta>1$ runs with different fluctuation amplitudes and injection type.
    }
    \label{fig:PDF_Kperp}
\end{figure*}

In addition to the statistics of $K_\parallel=|(\hat{\bb{b}}\cdot\bb{\nabla})\hat{\bb{b}}|$	(see Figure \ref{fig:PDF_K} in Section~\ref{subsec:B_curv}), we also investigated the statistics of the perpendicular curvature, $K_\perp=\big|\bb{b}\times(\bb{b}\times\bb{\nabla}\ln B)\big|$~\citep{Kempski2023}.
Figure \ref{fig:PDF_Kperp} shows the probability density functions (PDFs) of $K_\perp$ averaged over the interval $1 \lesssim t/t^* \lesssim 1.25$ and plotted versus $K_\perp/K_*$, where $K_*$ is the value $K_*$ at which each PDF peaks. 

In the low-amplitude, $\beta=1$ regime (left panel), the fast-magnetosonic injection (black) exhibits a PDF reaching systematically larger curvature values compared to the Alfvénic (red) and mixed (brown) cases, which instead follow nearly identical trends. Moreover, the fast-magnetosonic case seems to develop a power law at large $K_\perp$ with index $\approx -3.8$ (thus harder than its $K_\|$ counterpart, in the left panel of Figure \ref{fig:PDF_K}), whereas the $\text{PDF}(K_\perp)$ falls off nearly as an exponential for the other two cases. This highlights the importance of the injection type in controlling the statistics of $K_\perp$, qualitatively consistent with what is observed for $K_\|$. In addition, perpendicular curvature tends to exceed parallel curvature in this regime.
For $\beta > 1$ (right panel), the behavior of the perpendicular curvature differs from that of the parallel curvature: the PDF of curvatures reaches the largest $K_\perp$ values in the $\beta=2$ case with large-amplitude ($\delta B_\mathrm{rms}/ B_0 \sim 1$) Alfv\'enic injection (magenta), which even exceeds the larger-amplitude fast-magnetosonic (ocher) and Alfv\'enic (darkcyan) cases. 
However, the higher-$\beta$ cases with $\delta B/B_0\gtrsim 1$ seem to favor the emergence of a power-law distribution at large $K_\perp$, while lower $\beta$ and/or lower-amplitude regimes of Alfv\'enic injection tend to display a nearly exponential cutoff. This result highlights the strong influence of the plasma $\beta$, injection type, and turbulence amplitude on the statistics of $K_\perp$. 

\section{Mode-decomposed vs Helmholtz-decomposed fluctuations}\label{app:modes_Helm}

In order to investigate the nature of turbulence, magnetic and velocity fluctuations can be decomposed into their Alfv\'en, slow, and fast components (see Section~\ref{subsec:time_evo}). However, this decomposition relies on the properties of linear MHD modes, and is thus strictly valid only for small-amplitude fluctuations ($\delta B/B_0\ll1$) and not necessarily at all scales (i.e., when the scale-dependent mean field $\langle\bb{B}\rangle_\ell$ significantly differs from $\bb{B}_0$). The compressive nature of the fluctuations can thus be further investigated with an assumption-free Helmholtz decomposition (which, however, can only be applied to velocity fluctuations).

\subsection{Fluctuations' spectra}\label{subapp:modes-Helm_spectra}

\begin{figure*}[t!]
    \centering
    \includegraphics[width=0.48\linewidth]{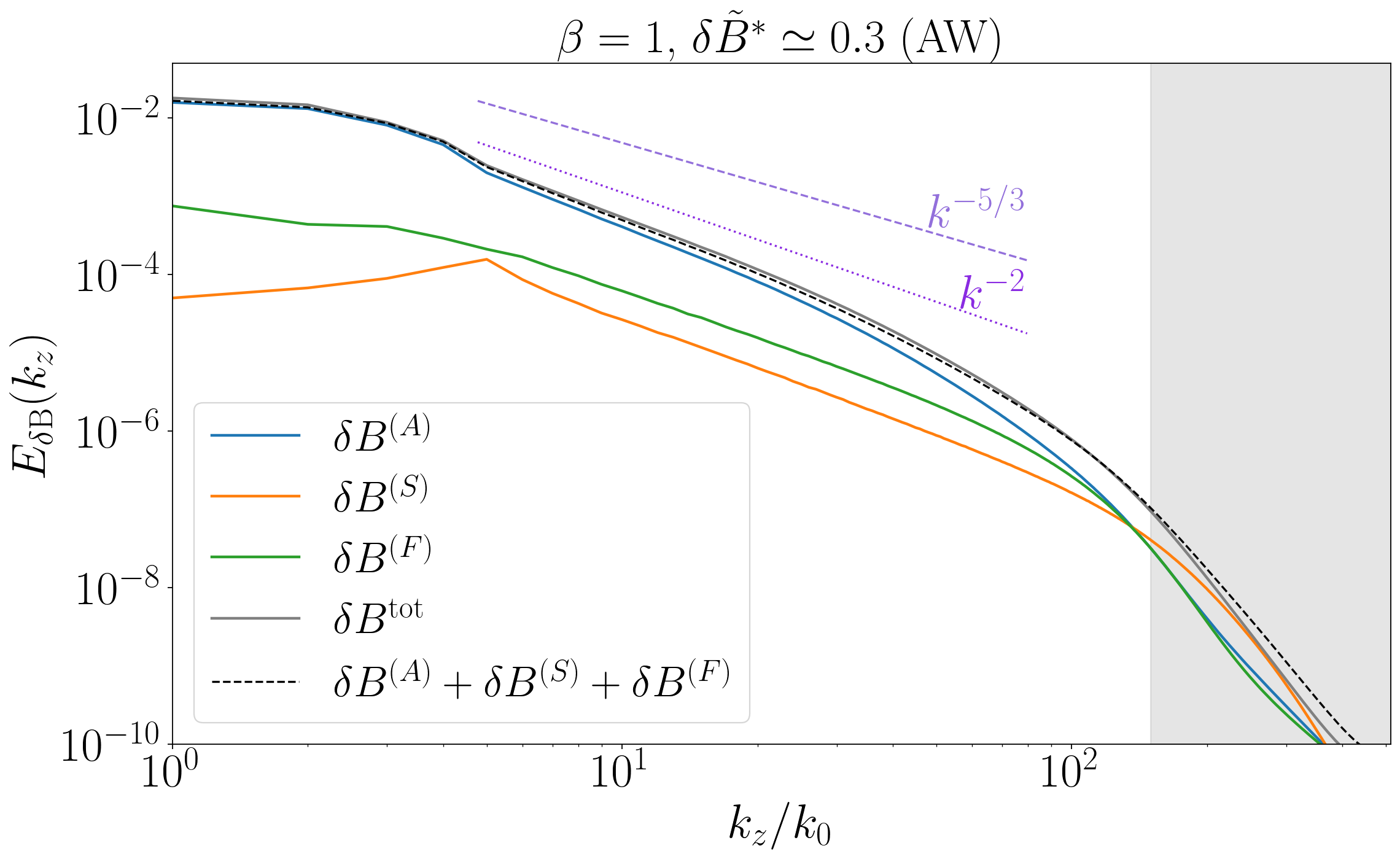}
    \includegraphics[width=0.48\linewidth]{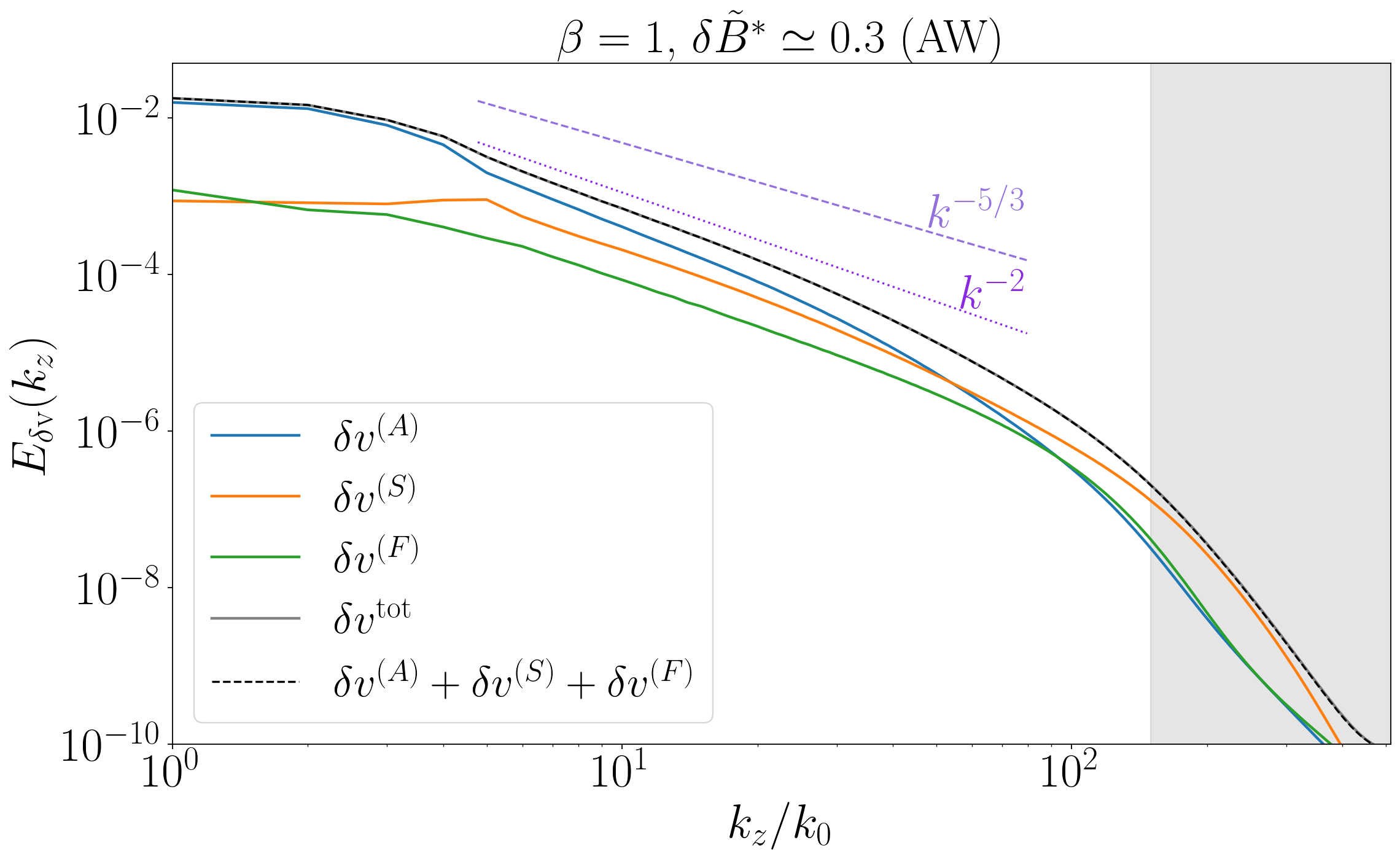}
    \includegraphics[width=0.48\linewidth]{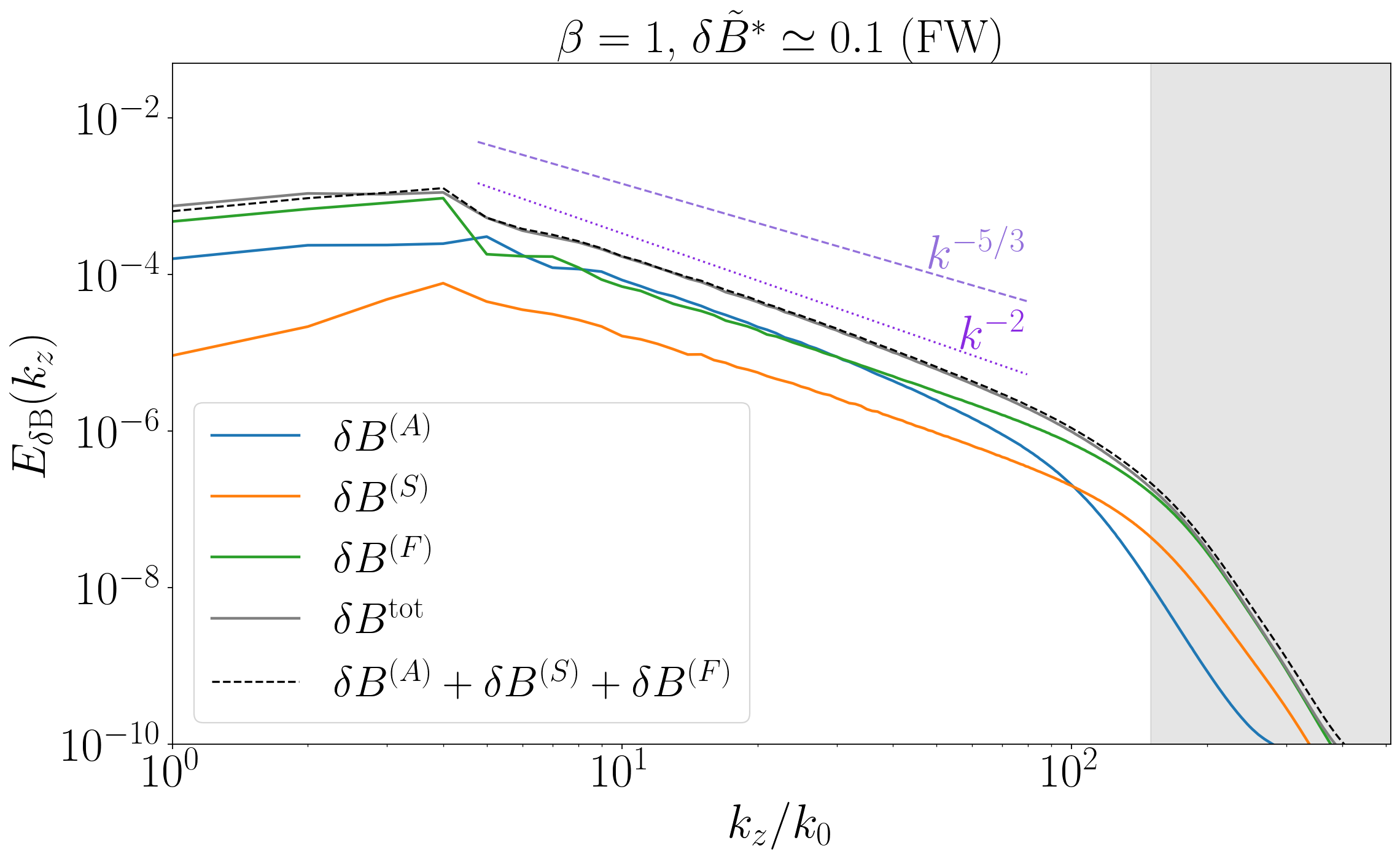}
    \includegraphics[width=0.48\linewidth]{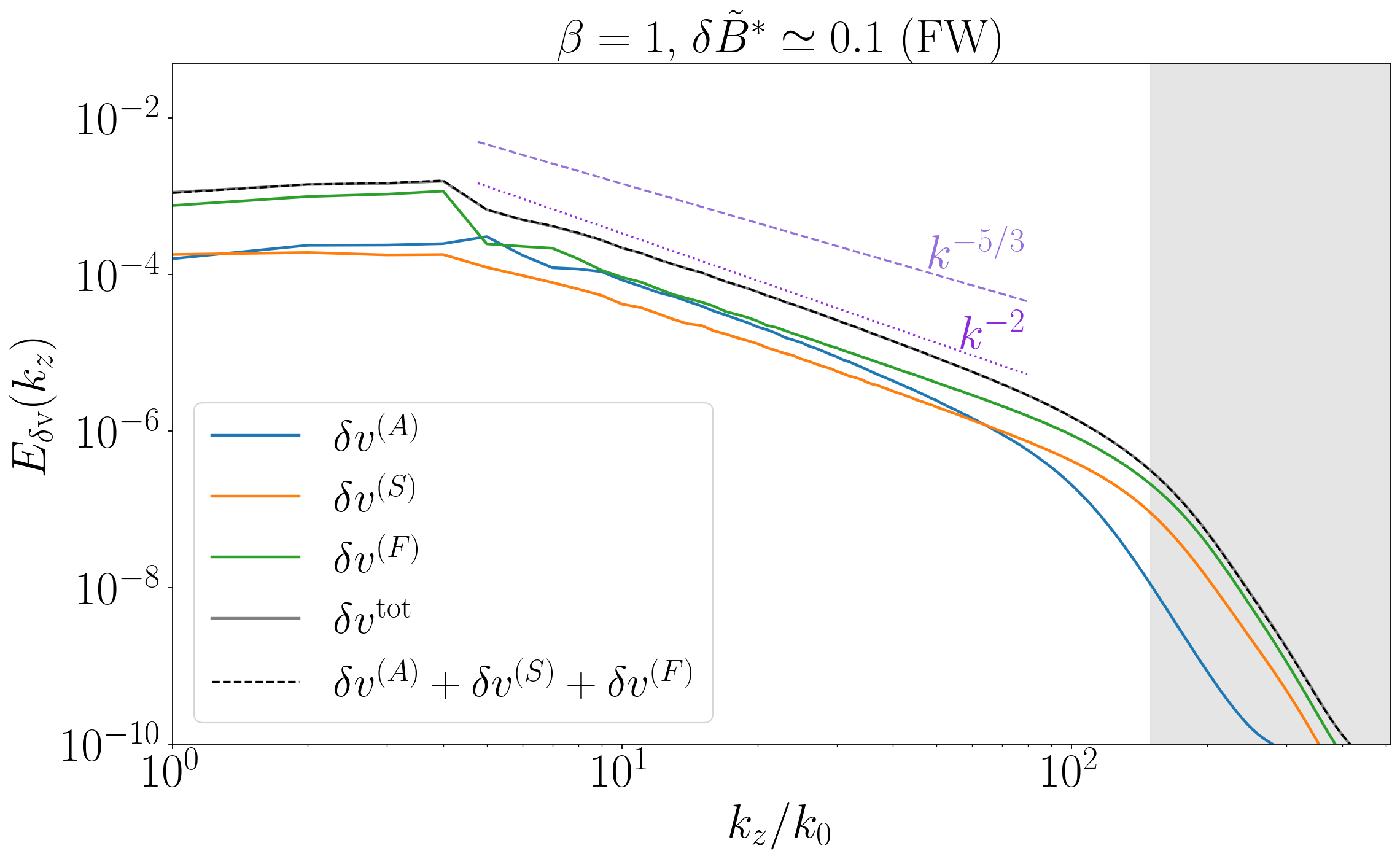}
    \caption{Parallel spectra of magnetic (left panels) and velocity (right panels) fluctuations decomposed into Alfvénic (blue), slow (orange), and fast (green) components for low-amplitude, $\beta=1$ runs. Spectra are averaged over $1 \lesssim t / t^* \lesssim 1.25$. {\it Top panels:} Alfv\'enic injection. {\it Bottom panels:} fast-magnetosonic injection. The gray shaded area denotes the dissipation range.}
    \label{fig:spectra_modes}
\end{figure*}

\begin{figure*}[t!]
    \centering
    \includegraphics[width=0.48\linewidth]{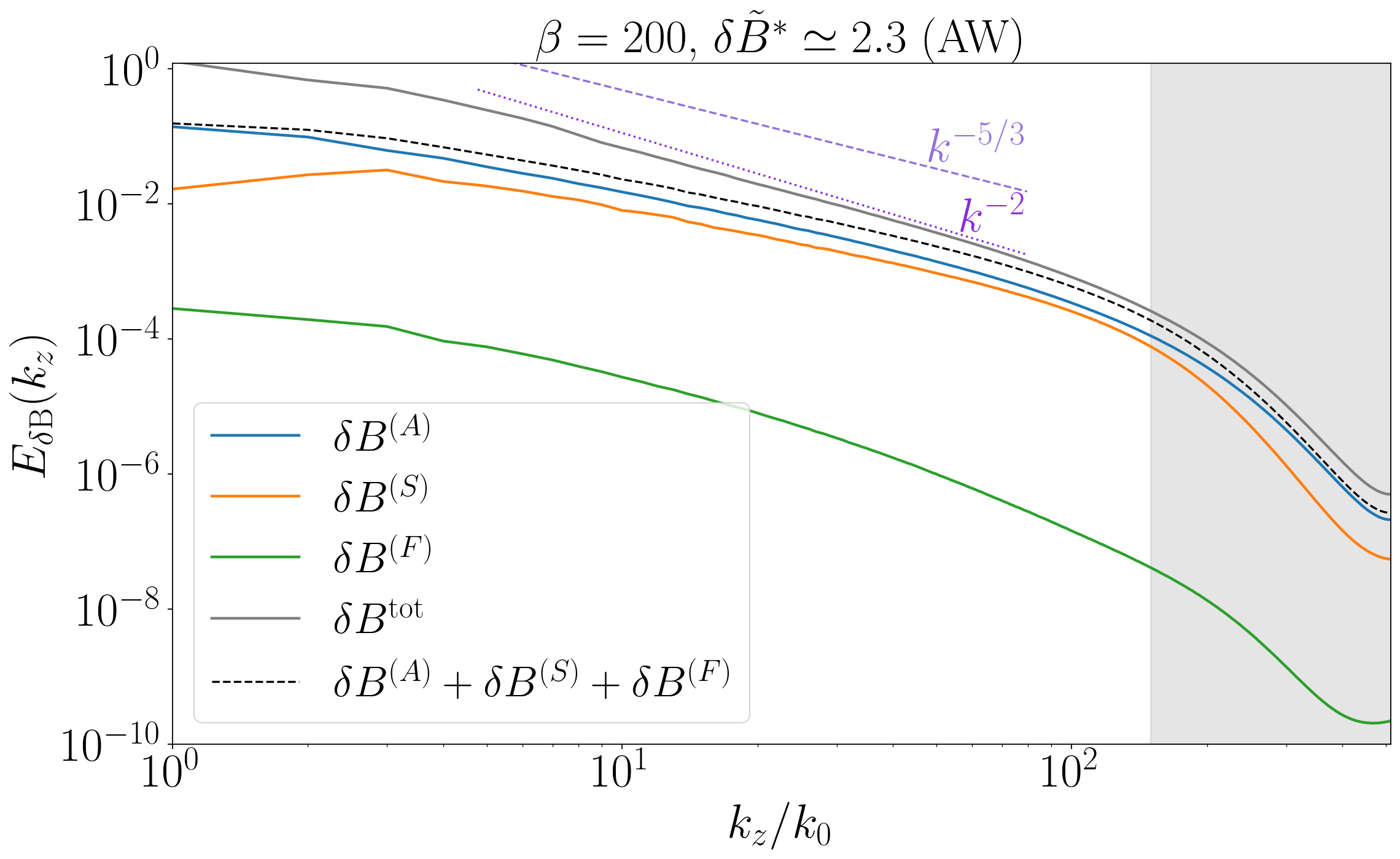}
    \includegraphics[width=0.48\linewidth]{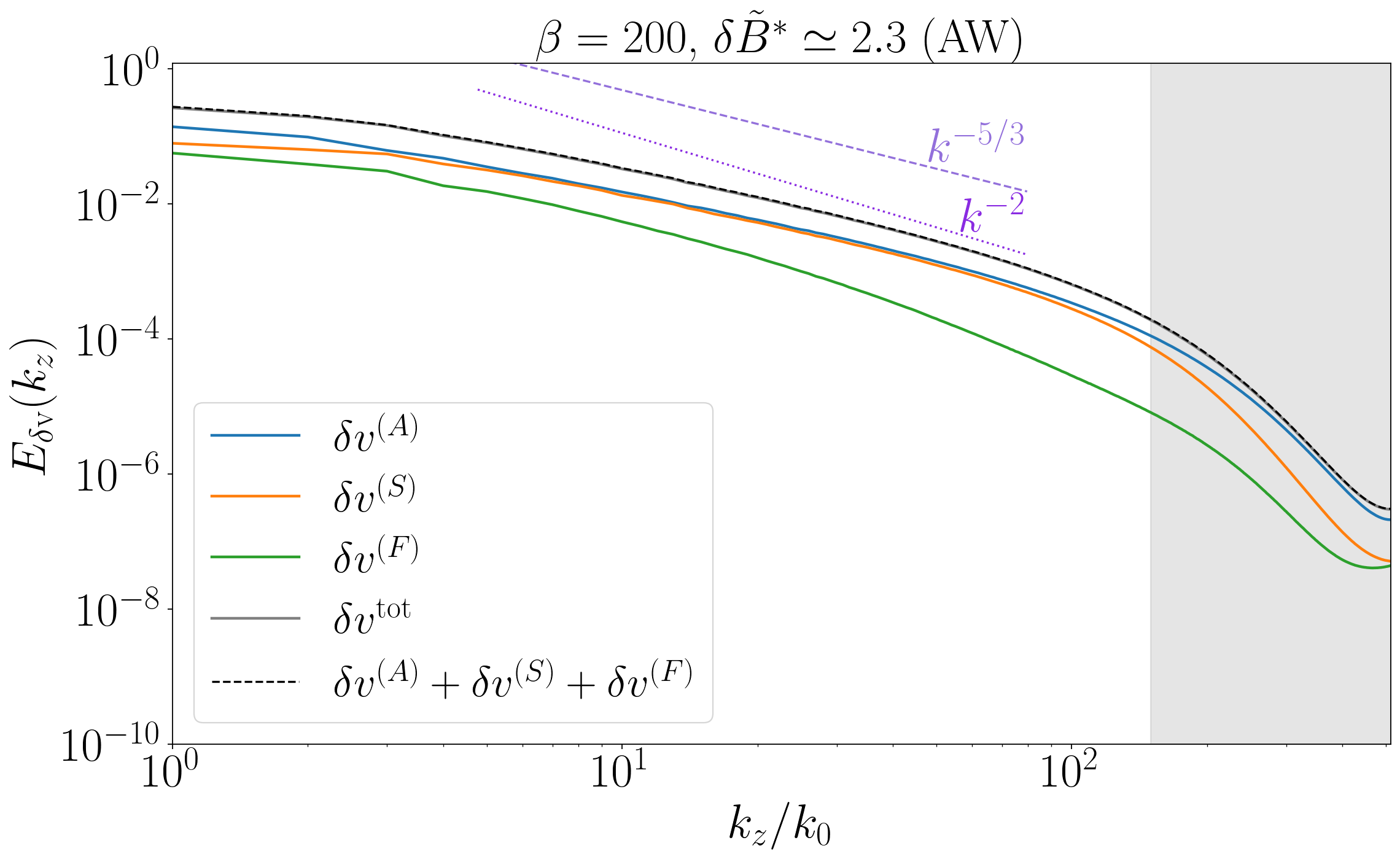}
    \includegraphics[width=0.48\linewidth]{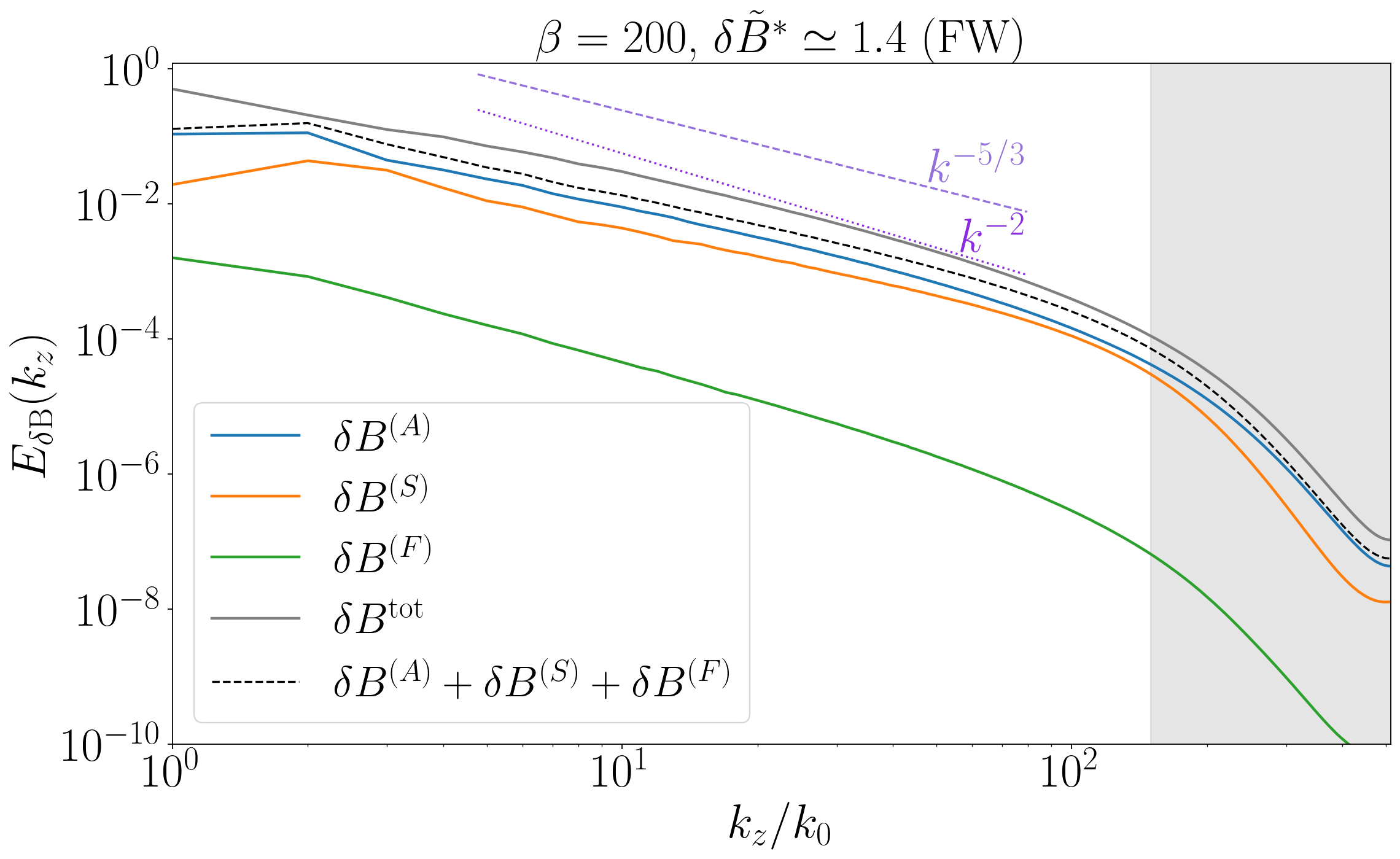}
    \includegraphics[width=0.48\linewidth]{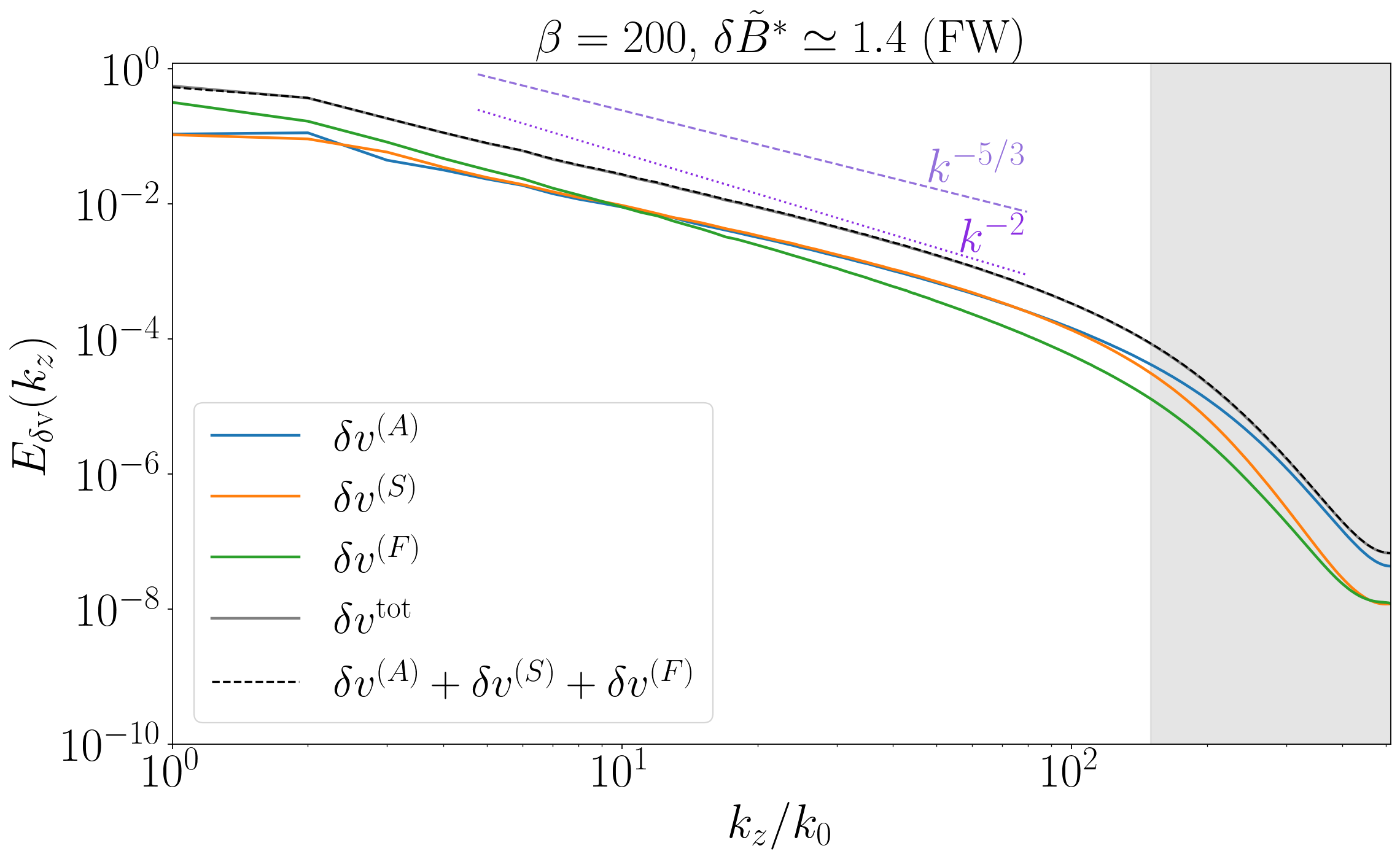}
    \caption{Same as Figure~\ref{fig:spectra_modes}, but for large-amplitude magnetic-field fluctuations with $\beta = 200$.}
    \label{fig:spectra_modes2}
\end{figure*}

Figure \ref{fig:spectra_modes} shows the time-averaged parallel spectra of the mode-decomposed magnetic (left panels) and velocity (right panels) fluctuations for low-amplitude Alfv\'enic (top panels) and fast-magnetosonic (bottom panels) injection at $\beta=1$. 
Here, we focus on the $k_z$ dependence since the parallel spectrum is particularly relevant for cosmic-ray transport in the quasi-linear framework~\citep[e.g.,][and references therein]{Shalchi2009,Blasi2013}, however a qualitatively similar behavior in terms of relative importance between the modes is found for the spectra in $k_\perp$ (not shown). 
For Alfvénic injection, the cascade of magnetic fluctuations (upper left panel) is clearly dominated by the Alfvén mode (blue) throughout the inertial range, with only minor contributions from the magnetosonic modes. In contrast, for fast-mode injection (lower left panel), both Alfvén (blue) and fast-magnetosonic (green) modes contribute at large scales, but the fast mode progressively dominates toward smaller scales (still within the inertial range). The slow mode (orange) remains subdominant in all cases. 
However, a qualitatively different picture emerges when looking at the spectra of velocity fluctuations for Alfvénic injection (upper right panel), where the slow-mode contribution is more important than the contribution from the fast mode, and even exceeds the Alfvénic contribution at smaller scales. For fast-mode injection (lower right panel), velocity and magnetic spectra are instead qualitatively consistent with each other.
The dominance of the fast-magnetosonic mode at high $k_z$ provides a natural explanation for the harder parallel spectrum and for the enhanced fluctuation power at small parallel scales for fast-magnetosonic injection in Figure \ref{fig:spectra}. In this sense, fast-magnetosonic fluctuations are likely to play a relevant role in CR parallel scattering efficiency at small scales for this regime.

Figure~\ref{fig:spectra_modes2} shows the $\beta=200$ runs with $\delta B/B_0>1$ for Alfvénic (upper panels) and fast-magnetosonic (lower panels) injection. In both cases, the spectra are dominated by both Alfvénic (blue) and slow-magnetosonic (orange) fluctuations. This is consistent with the known tendency of slow modes to become pseudo-Alfvénic at high $\beta$~\citep{Goldreich1995, Lithwick2001}. At the same time, the fast-mode contribution (green) to the magnetic spectrum appears to be strongly suppressed in this high-$\beta$ regime (left panels); this is expected for the quasi-incompressible limit at large $\beta$. 
However, the fast-mode contribution to the velocity spectrum (right panels) is always relevant at larger scales, especially for fast-magnetosonic injection (lower right panel).
This discrepancy between the magnetic and velocity amplitudes for the fast mode at large $\beta$ is due to the linear relation underlying this decomposition, i.e., $\delta B_{\rm F}/B_0\sim k\,\delta v_{\rm F}/\omega_{\rm F}\propto\beta^{-1/2}\, \delta v_{F}/v_{\rm A}$ for $\beta\gg1$ (see Eq.\eqref{eq:eigenvect_F} and paragraph below). In addition, we find that the sum of the velocity spectra associated with individual modes closely matches the total velocity spectrum, while the corresponding magnetic spectra do not fully recover the total magnetic-field spectrum.

\begin{figure*}[]
    \centering
    \includegraphics[width=0.49\linewidth]{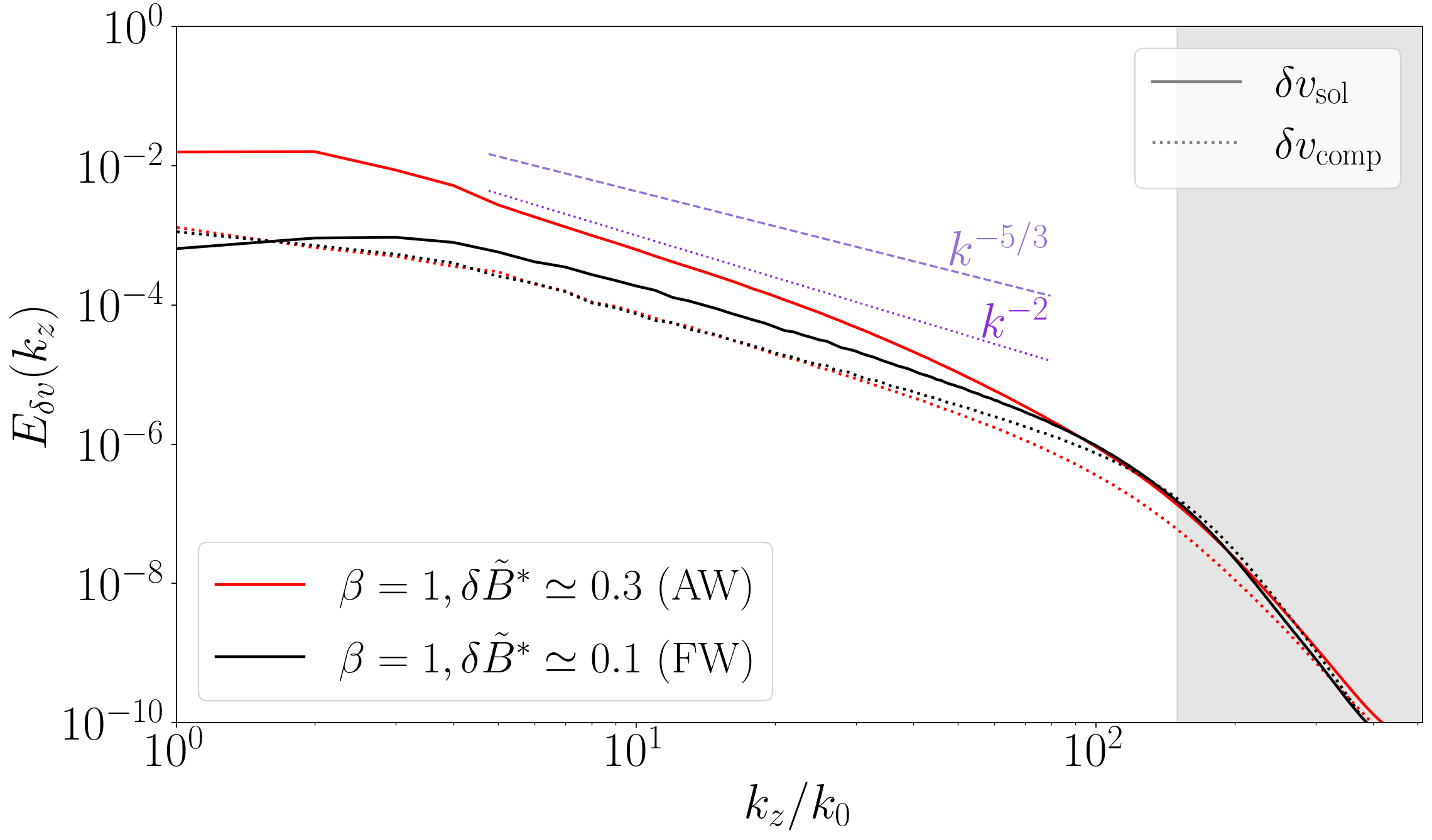}
    \includegraphics[width=0.49\linewidth]{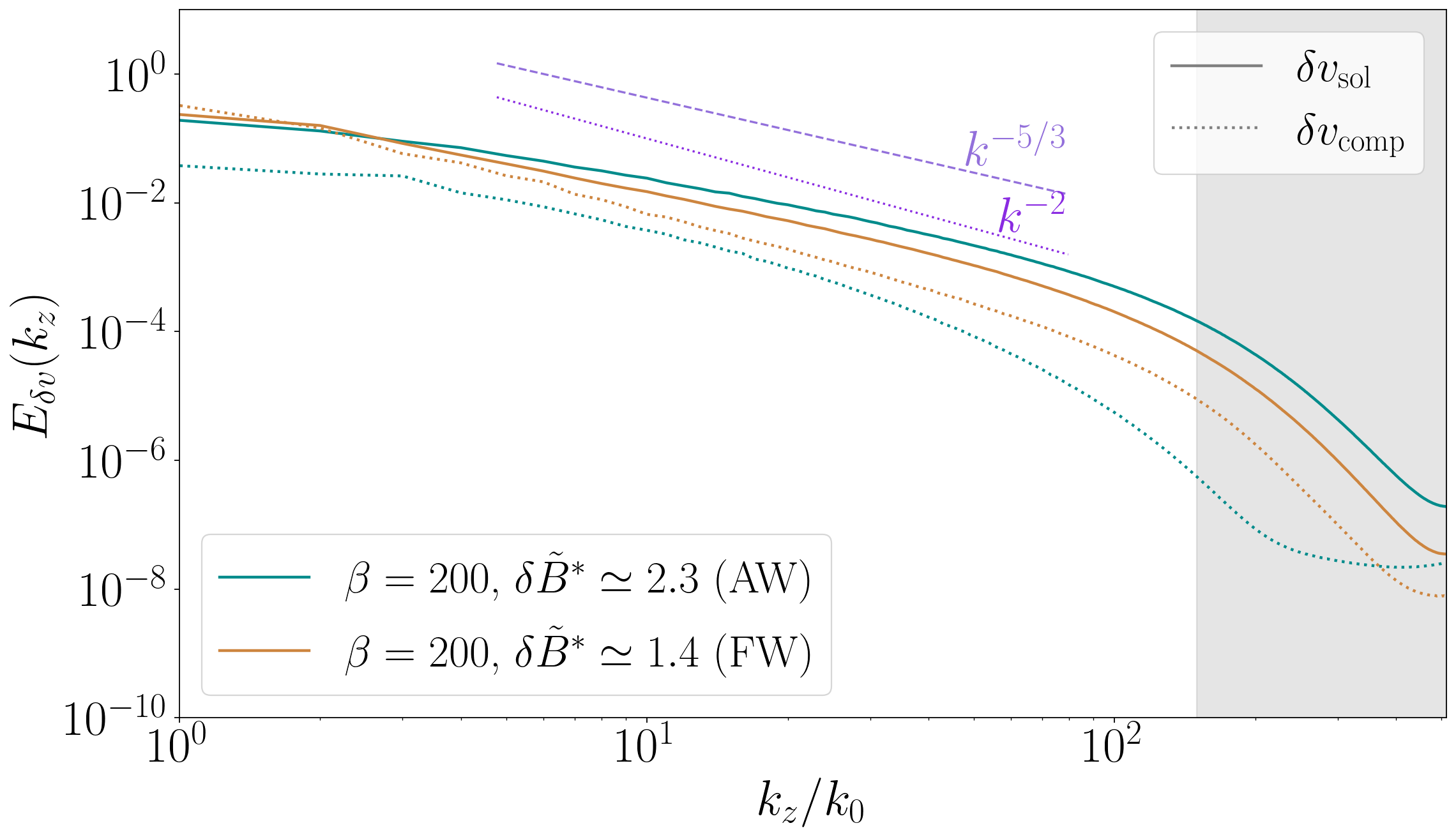}
    \caption{Solenoidal (solid lines) and compressible (dotted lines) velocity fluctuation spectra as a function of $k_z$, averaged over the time interval $1 \lesssim t/t^* \lesssim 1.25$. Power laws $k^{-5/3}$ and $k^{-2}$ are shown for reference. The gray shaded area represents the dissipation range. \textit{Left:} $\beta =1$ runs with low-amplitude Alfvénic (red) and fast-magnetosonic (black) injection. \textit{Right:} $\beta = 200$ regime of Alfv\'enic (cyan) and fast-magnetosonic (ocher) injection with $\delta B/B_0 > 1$.}
    \label{fig:spectra_vHelm}
\end{figure*}

Figure \ref{fig:spectra_vHelm} shows the parallel spectra of the solenoidal (solid lines) and compressible (dotted lines) components of the velocity for Alfv\'enic and fast-magnetosonic injection with low amplitudes at $\beta = 1$ (left panel) and with large amplitudes at $\beta=200$ (right panel).
At low amplitudes and $\beta=1$, the two injection types mainly differs for the spectrum of $\delta\bb{v}_{\rm sol}$ (solid lines), which develops a $\propto k_z^{-2}$ power law for fast-mode injection (black) while being noticeably steeper for Alfv\'enic injection (red). As expected for the small-amplitude regime, throughout the whole wavenumber range one finds $\delta v_{\rm comp}\sim\delta v_{\rm sol}$ for fast-magnetosonic injection and $\delta v_{\rm comp}\ll\delta v_{\rm sol}$ for Alfv\'enic injection. Both cases show a similar behavior for the spectrum of $\delta\bb{v}_{\rm comp}$ (dotted lines), with the case of fast-magnetosonic injection producing a slightly more extended cascade than in the case of Alfv\'enic injection.
When increasing the $\beta$ and the amplitude level (right panel), it is now the case of Alfv\'enic injection (cyan) that exhibits a spectrum of $\delta\bb{v}_{\rm sol}$ that is shallower ($\propto k_z^{-3/2}$) than the case of fast-magnetosonic injection ($\propto k_z^{-5/3}$; ocher). 
The cascade of $\delta\bb{v}_{\rm comp}$ (dotted lines) is significantly more suppressed for Alfv\'enic injection than for fast-mode injection---the former case does not even develop a clear power law, while the latter case develops a $\propto k_z^{-2}$ spectrum. Surprisingly, despite the large-amplitude regime, compressive fluctuations become less relevant towards smaller scales for both injection types.

Overall, what can be inferred from the mode-decomposed velocity spectra is not fully consistent with an analysis based on the Helmholtz decomposition (right panels of Figures~\ref{fig:spectra_modes}--\ref{fig:spectra_modes2} and Figure~\ref{fig:spectra_vHelm}). On the other hand, the mode-decomposed magnetic spectra (left panels of Figures~\ref{fig:spectra_modes}--\ref{fig:spectra_modes2}) seem to be qualitatively consistent (at best) with what can be inferred from the spectra of Helmholtz-decomposed velocity fluctuations only for low-amplitude turbulence ($\delta B/B_0\ll1$). For large-amplitude turbulence ($\delta B/B_0>1$), the two type of analysis yeld apparently inconsistent results.
In general, one must bear in mind that, while Alfv\'enic modes are characterized only by solenoidal velocity fluctuations (i.e., $\delta\bb{v}_{\rm comp}^{\rm(A)}=0$) and the magnetosonic modes are the only ones that contribute to compressive velocity fluctuations ($\delta\bb{v}_{\rm comp}=\delta\bb{v}_{\rm comp}^{\rm(F)}+\delta\bb{v}_{\rm comp}^{\rm(S)}$), magnetosonic modes can also contribute to the solenoidal component of the velocity (i.e., $\delta\bb{v}_{\rm sol}^{\rm(F)},\delta\bb{v}_{\rm sol}^{\rm(S)}\neq0$). Therefore, identifying $\delta\bb{v}_{\rm sol}$ with $\delta\bb{v}^{\rm(A)}$ would overestimate the contribution of Alfv\'enic fluctuations---analogously, identifying $\delta\bb{v}_{\rm comp}$ with $\delta\bb{v}^{\rm(F)}+\delta\bb{v}^{\rm(S)}$ would underestimate the contribution of magnetosonic fluctuations. Furthermore, distinguishing between the fast- and the slow-mode contribution to $\delta\bb{v}$ is beyond the grasp of Helmholtz decomposition. Finally, the relative contribution of these three modes to magnetic fluctuations can be significantly different from their relative contribution to velocity fluctuations: this is another feature that cannot be estimated with the Helmholtz decomposition or, in general, without the help of relations based on linear theory, as in the mode-decomposed analysis.

\section{Scale-dependent nonlinear parameter}\label{app:chi}

\begin{figure*}[t!]
    \centering
    \includegraphics[width=0.49\linewidth]{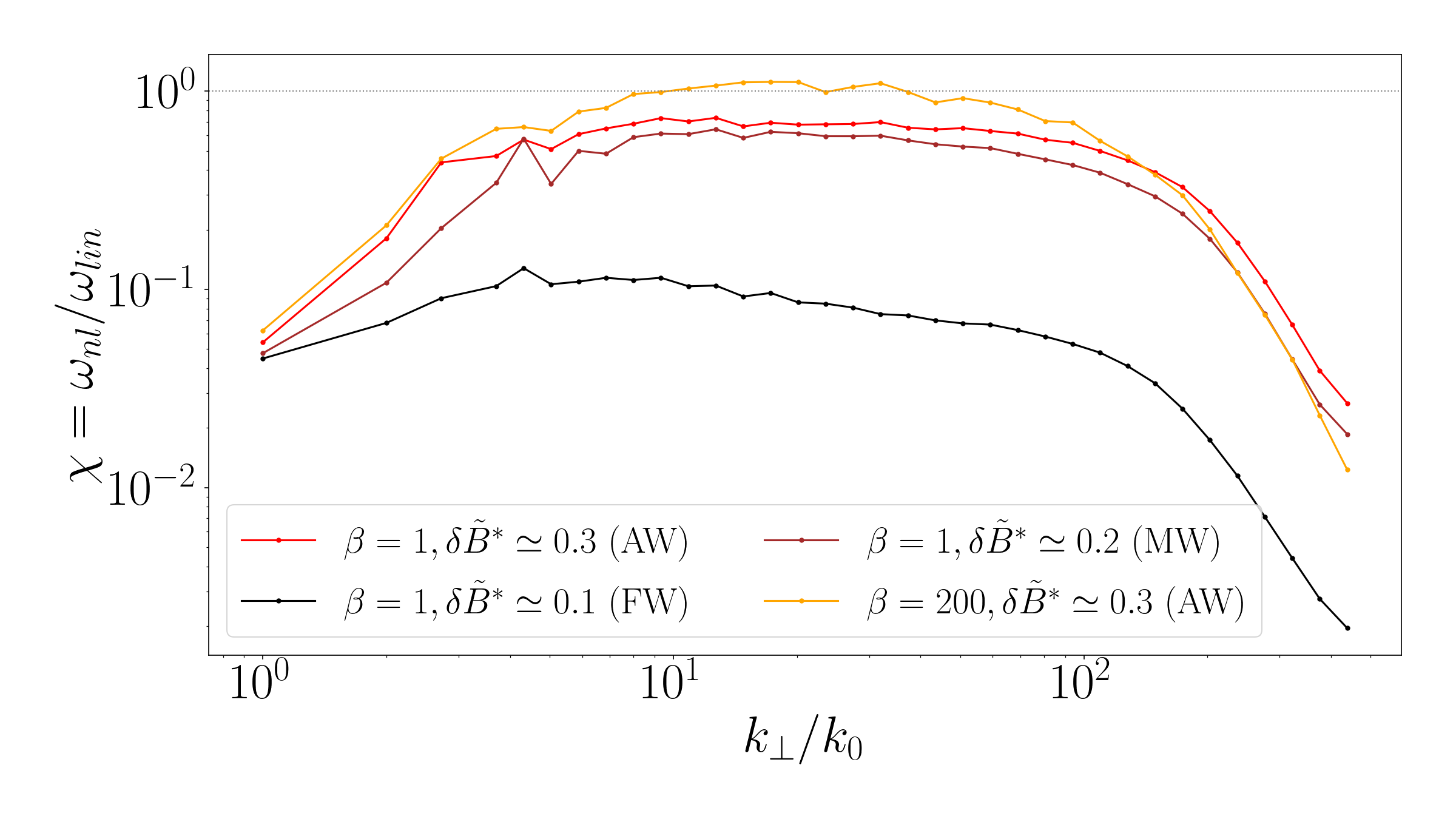}
    \includegraphics[width=0.49\linewidth]{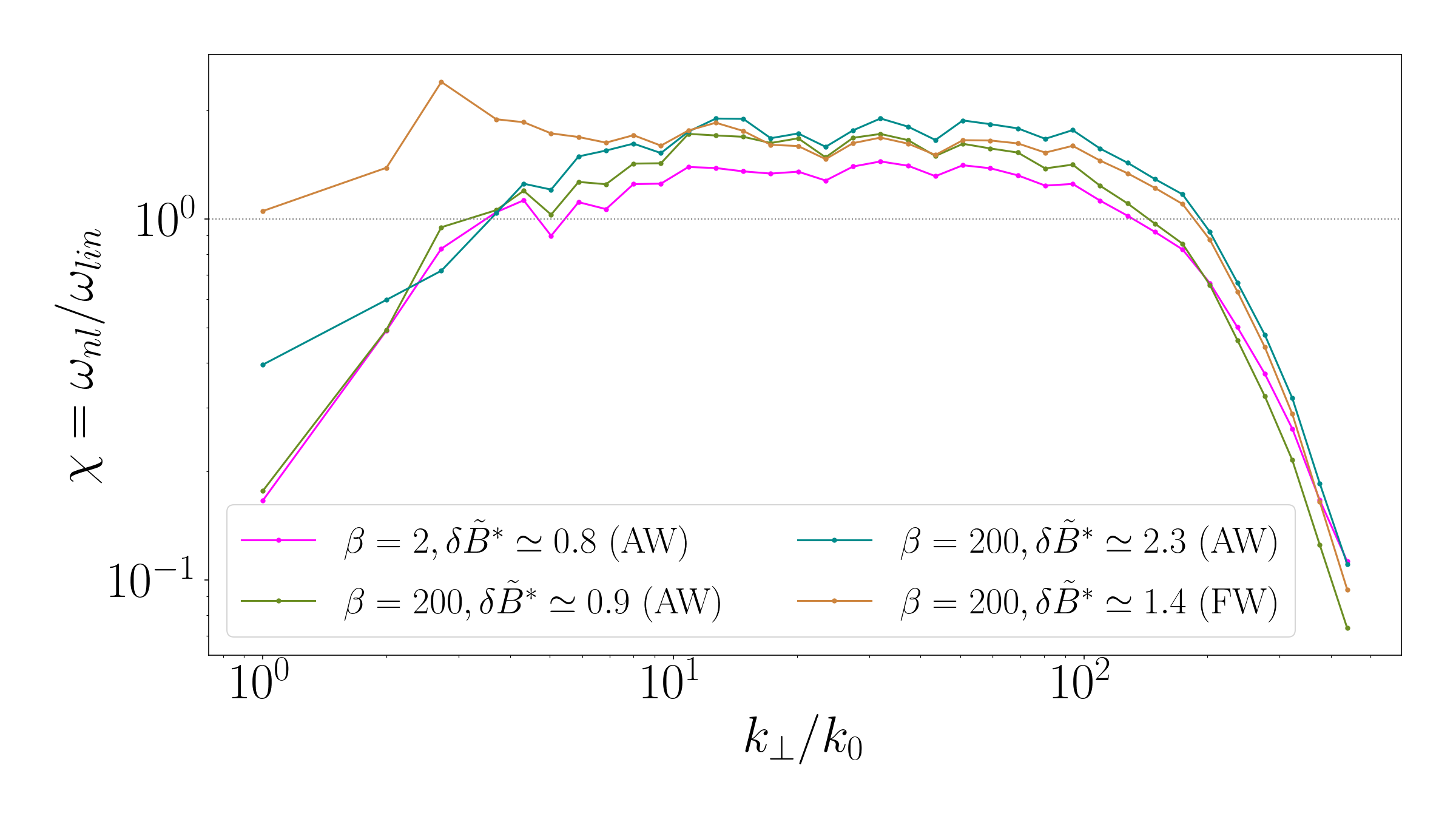} 
    \caption{Scale-dependent nonlinear parameter $\chi(k_\perp)$, averaged over the interval $1 \lesssim t/t^* \lesssim 1.25$. The horizontal dashed line marks the critical balance condition $\chi = 1$. \textit{Left panel}: small-amplitude runs. \textit{Right panel}: large-amplitude runs.}
    \label{fig:chi_kperp}
\end{figure*}

We estimate the strength of turbulence via the scale-dependent nonlinear parameter,
\begin{equation}
    \label{eq:chi}
    \chi = \frac{\omega_\mathrm{nl}}{\omega_\mathrm{lin}} \simeq \frac{k_\perp\, \delta v_\perp(k_\perp)}{k_\parallel(k_\perp)\, v_A},
\end{equation}
where $k_\|(k_\perp)$ is obtained with the method of \citet{Cho2002a} from $\delta\bb{v}$ fluctuations (see Equation \eqref{eq:anis} in Section \ref{subsec:anis_B}) and $\delta v_\perp(k_\perp)=\sqrt{k_\perp E_{\delta v_\perp}(k_\perp)}$. 
The average value of $\chi$ is found to be close to $\delta \tilde{B}^*$ and/or $M_A^*$, within a factor of $\sim 2$ at most.
We have verified that using the full nonlinear parameter $\chi= (k\,\delta v)/(k_\|\,v_{\rm A})$, instead of its perpendicular part in \eqref{eq:chi}, will mainly affect the large-scale modes and thus the following results would not qualitatively change.

Figure \ref{fig:chi_kperp} shows the values of $\chi$ as a function of $k_\perp$, averaged over the time interval $1 \lesssim t/t^* \lesssim 1.25$, for the small-amplitude (left panel) and large-amplitude (right panel) runs.
For the low-amplitude cases, the Alfvénic (red) and mixed (brown) injections develop an inertial range close to ($\chi\approx0.7$) the critical-balance condition $\chi = 1$ (horizontal dashed line), with the Alfvénic $\beta=200$ case (orange) being predominantly consistent with $\chi \sim 1$. In contrast, the fast-magnetosonic injection (black) remains in a weak-turbulence regime, with $\chi \lesssim 0.1$ at all scales. When the fluctuation amplitude is increased, $\chi$ slightly exceeds unity (namely, $1\lesssim \chi\lesssim 2$) throughout the inertial range in all simulations, indicating a transition toward a strong nonlinear regime regardless of the injection nature. This also explains the difference between the large-amplitude and small-amplitude fast-magnetosonic injection cases, i.e., why turbulence is dominated by several (weak) shocks in the latter, while these structures are no longer the main feature in the large-amplitude case (right panels of Figure \ref{fig:B_3D}).

\section{3D Spectrum in ($k_z$, $k_\perp$, $\omega$) space}
\label{app:omegak}

\begin{figure*}[th!]
    \centering
    \includegraphics[width=0.49\linewidth]{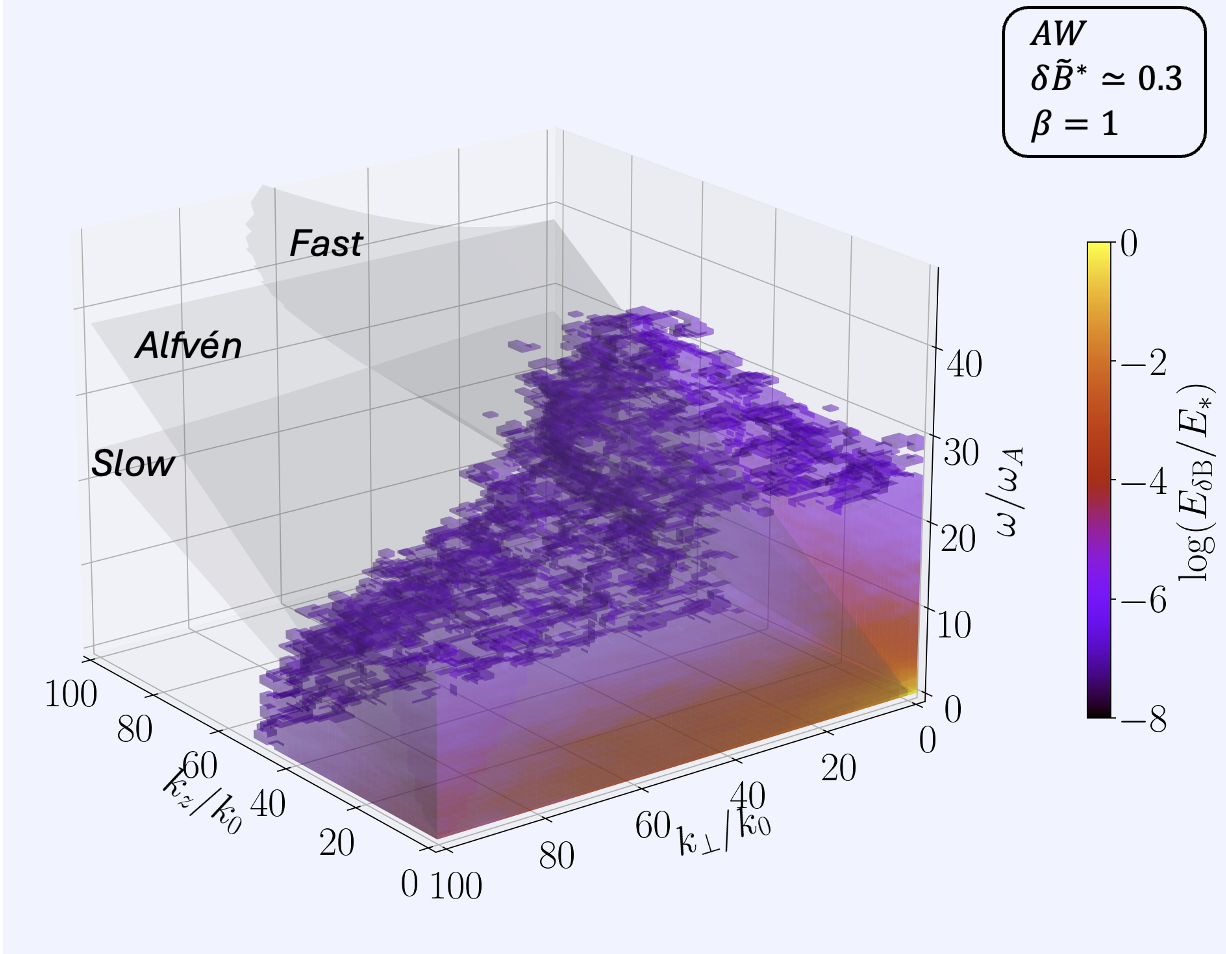}
    \includegraphics[width=0.49\linewidth]{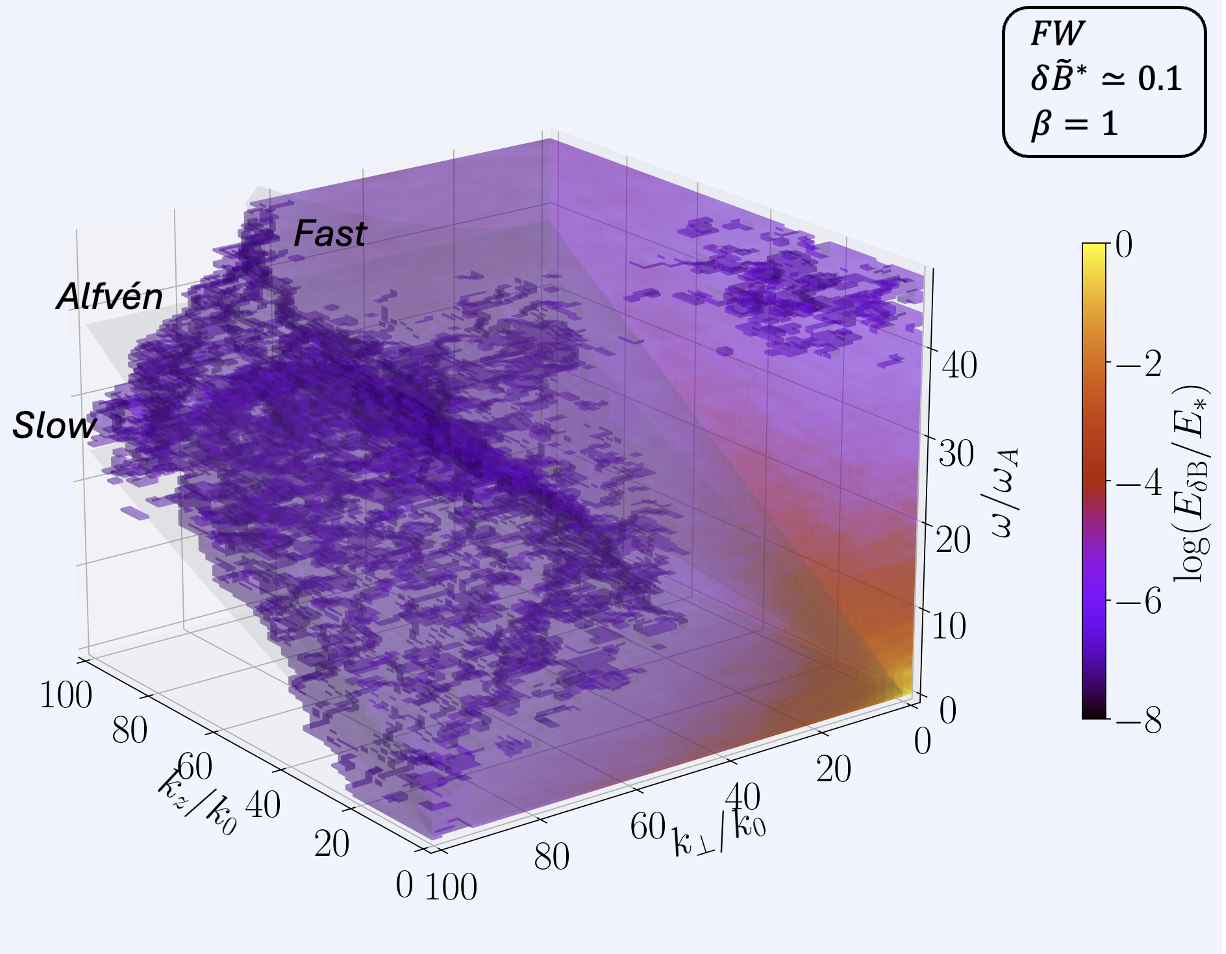}
    \includegraphics[width=0.49\linewidth]{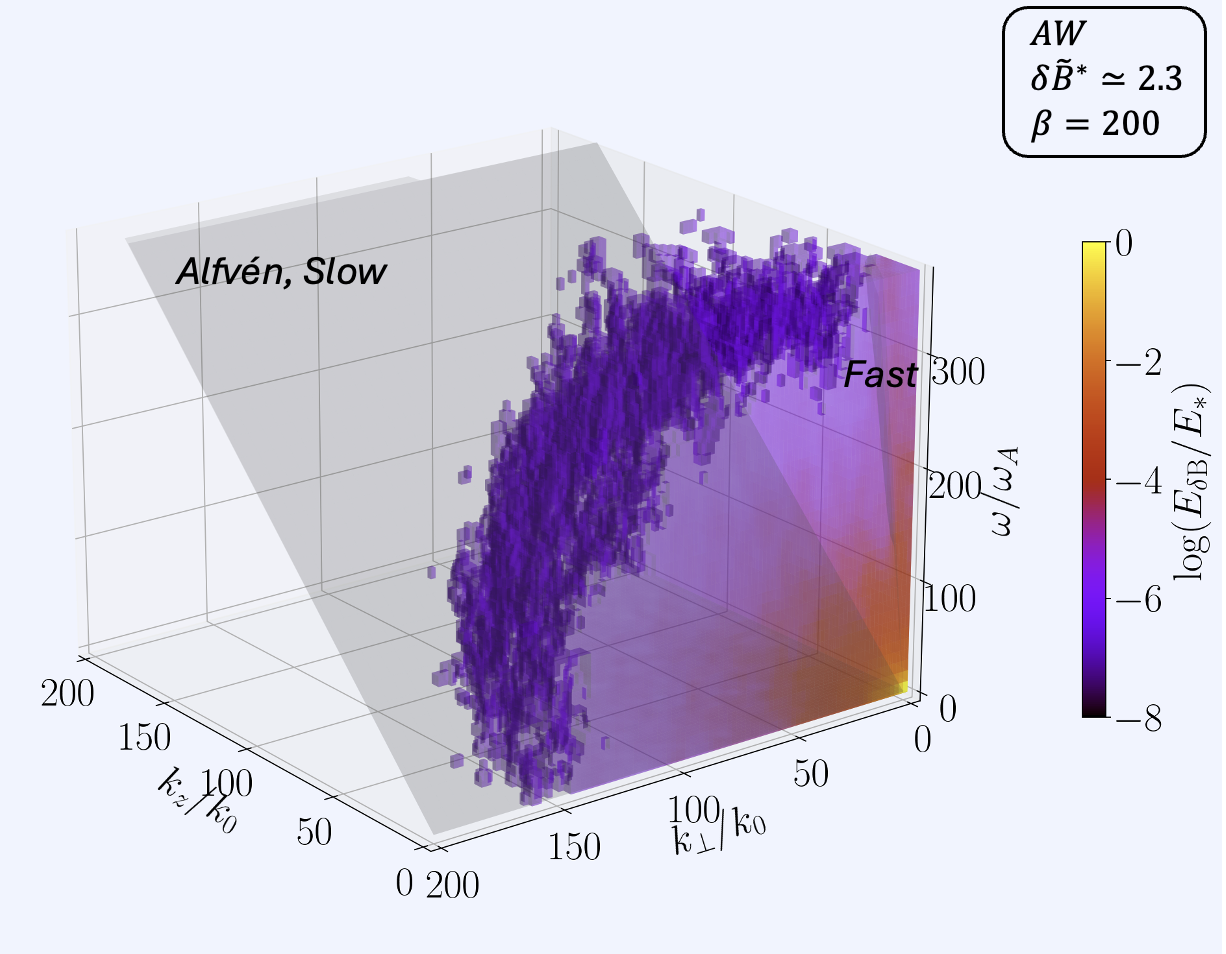}
    \includegraphics[width=0.49\linewidth]{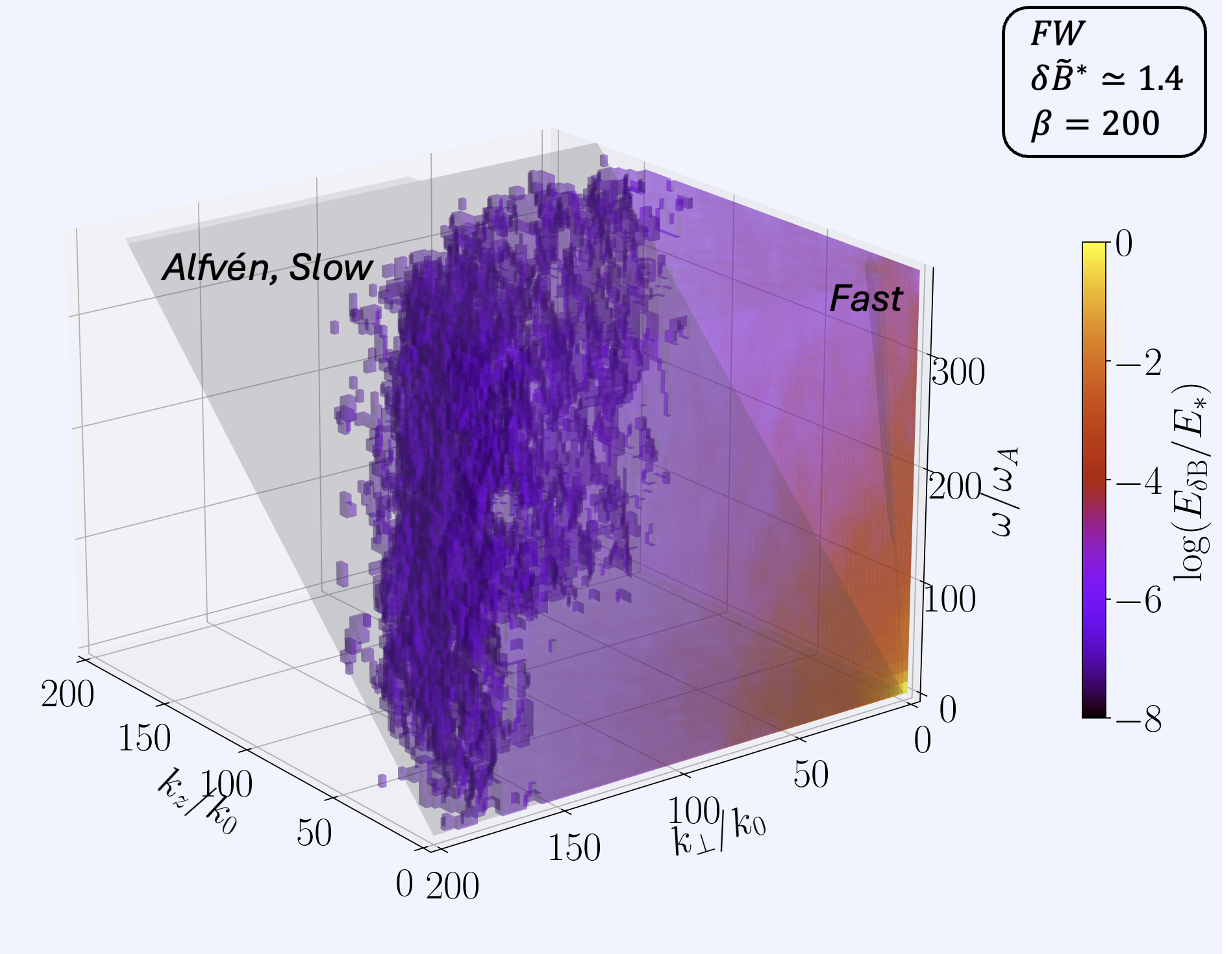}
    \caption{3D wavenumber-frequency spectrum of magnetic fluctuations, $E_{\delta B}(k_z, k_\perp, \omega)$, obtained at times $t/t^* \gtrsim 0.2$ and normalized to its peak value $E_*=\max(E_{\delta B})$. Wavenumbers and frequencies are normalized to $k_0=2\pi/L$ and to $\omega_A=k_0\,v_{\rm A}$, respectively. The semi-transparent gray surfaces represent the linear dispersion relations $\omega(k_z,k_\perp)$ of MHD modes.
    \textit{Top panels:} low-amplitude Alfvénic (left) and fast-magnetosonic (right) injection at $\beta=1$. \textit{Bottom panels:} large-amplitude Alfv\'enic (left) and fast-magnetosonic (right) injectiona at $\beta = 200$.}
    \label{fig:omegak}
\end{figure*}

The nature of turbulent fluctuations can be further investigated by looking at their power spectrum in wavenumber-frequency space~\citep[e.g.,][]{Gan2022}. To do so, we saved snapshots of the turbulent fluctuations $\delta\bb{B}$, $\delta\bb{v}$, and $\delta\rho$ on 2D slices of the simulation box at a high output cadence. These ($x_i$, $x_j$, $t$) cubes are Fourier transformed to obtain 3D spectra in ($k_i$, $k_j$, $\omega$) that can then be combined to obtain the averaged power spectrum of the fluctuations in ($k_z$, $k_\perp$, $\omega$) space.

Figure \ref{fig:omegak} shows the 3D wavenumber-frequency spectrum of magnetic-field fluctuations, $E_{\delta B}(k_z,k_\perp,\omega)$, obtained from the above-mentioned 2D slices of $\delta\bb{B}$ using times $t/t^* \gtrsim 0.2$. 
The spectrum $E_{\delta B}(k_z,k_\perp,\omega)$ is normalized to its peak value, the wavenumbers are normalized to the box wavenumber $k_0=2\pi/L$, and the frequency $\omega$ is normalized to the Alfvén frequency at the box scale $\omega_A=k_0v_{\rm A}$. The semi-transparent gray surfaces represent the linear dispersion relations $\omega(k_z,k_\perp)$ corresponding to the Alfvén, fast, and slow modes. 
The upper panels show $E_{\delta B}(k_z,k_\perp,\omega)$ for low-amplitude turbulence at $\beta=1$, developing from Alfvénic (left) or fast-magnetosonic (right) injection, while the lower panels show the corresponding 3D power spectra for the $\beta =200$ regime with $\delta B/B_0 > 1$ (in this high-$\beta$ regime, the Alfvén and slow branches become nearly degenerate and are represented by the same gray surface).

When small-amplitude AWs are initialized (upper left) most of the power (warm colors) is confined in the low-frequency part of the space, not only along the Alfv\'enic and slow-mode surface, but also distributed over a broad region characterized by very low frequency, small $k_z$ and large $k_\perp$ (i.e., $\omega/\omega_{\rm A}\lesssim 3$, $k_z/k_0\lesssim20$, and up to $k_\perp/k_0\sim100$). This may be due to the presence of coherent structures. Nevertheless, the spectrum shows some non-negligible amount of power near the fast-magnetosonic branch, extending to slightly higher frequency (i.e., up to $\omega/\omega_{\rm A}\sim10$). 
In contrast, for small-amplitude FWs initial condition (upper right), a significantly larger amount of higher-frequency fluctuations are produced (i.e., up to $\omega/\omega_{\rm A}\sim 20$). Moreover, most of this power is concentrated near the fast-magnetosonic branch, while the Alfvén and slow-magnetosonic dispersion surfaces remain poorly populated.
Some amount of very low frequency, quasi-perpendicular fluctuations that may denote coherent structures are still generated in this regime, although it represents a much smaller fraction of the total power than the case of Alfv\'enic injection (upper left). Furthermore, a non-negligible fraction of the fluctuation power reaches very high-$k$ and high-frequency regions ($k_z/k_0 \,,\, k_\perp/k_0 > 50$ and $\omega/\omega_{\rm A}> 20$), where it seems to cross different branches, possibly denoting efficient nonlinear coupling between the modes.

The injection of large-amplitude fluctuations at $\beta=200$ generates a significant amount of spectral power that extends up to very high frequencies, i.e., up to $\omega/\omega_{\rm A}\sim 200$ for Alfvénic injection (lower left) and up to $\omega/\omega_{\rm A}\sim 300$ for fast-magnetosonic injection (lower right). Moreover, a large fraction of such fluctuation energy populates a region of space that is enclosed by the fast-mode dispersion surface. In particular, the case of large-amplitude Alfvénic injection at high $\beta$ exhibits a substantially larger fraction of the total power along the fast-magnetosonic branch than in the corresponding $\beta=1$, low-amplitude case (upper left). Away from the fast-magnetosonic surface, the remaining power seems to be isotropically distributed in ($k_z$, $k_\perp$, $\omega$)-space and does not exhibit a clear accumulation around the degenerate Alfvén-slow plane. 
However, we stress that the definition of linear dispersion surfaces $\omega(\bb{k})$ and, in particular, identifying $k_z$ with $k_\parallel$, has a much weaker meaning in large-amplitude turbulence and this should not be over-interpreted in terms of the dispersion branches. Nevertheless, the evidence that large-amplitude turbulence inevitably produces very high-frequency fluctuations regardless of the nature of injected fluctuations does not depend on these assumptions, and points out that such a turbulent regime is never completely of ``Alfv\'enic nature'' in the standard sense (i.e., low-frequency turbulence).

\FloatBarrier 
\twocolumn
\FloatBarrier 
\clearpage

\end{document}